\documentclass[a4paper,12pt]{book}

\usepackage[frenchb,spanish,english]{babel}
\usepackage[utf8x]{inputenc}
\usepackage[T1]{fontenc}
\usepackage{ucs}
\usepackage{textcomp}
\usepackage{microtype}
\usepackage{lmodern}

\usepackage[small,labelfont={sf,bf},labelsep=quad]{caption}
\usepackage{geometry}
\usepackage{fancyhdr}
\usepackage{fancybox}
\usepackage{url}
\usepackage{hyperref}
\usepackage{cite}
\usepackage{enumerate}

\usepackage{xspace}
\usepackage{lastpage}
\usepackage{stmaryrd}

\usepackage{graphicx}
\usepackage{subcaption}
\usepackage[table]{xcolor}
\usepackage{rotating}
\usepackage{empheq}
\usepackage{ifthen}
\usepackage{tensor}
\usepackage{paralist}
\usepackage{wrapfig}
\usepackage{array}

\usepackage{amsmath}
\usepackage{amsfonts}
\usepackage{amssymb}
\usepackage{wasysym}
\usepackage{mathrsfs}
\usepackage{extarrows}
\usepackage{verbatim}
\usepackage{siunitx}
\usepackage{upgreek}
\usepackage{multirow}
\usepackage[overload]{abraces} 

\usepackage{makeidx}
\usepackage[sectionbib]{chapterbib}

\usepackage{lettrine}
\usepackage{minitoc}
\usepackage[section]{placeins}

\usepackage[sf,bf,newlinetospace]{titlesec}
\usepackage{titletoc}

\usepackage{ntheorem}
\theoremheaderfont{\sf\bfseries}
\theorembodyfont{\normalfont}
\theoremseparator{.}
\usepackage[framemethod=TikZ]{mdframed}

\usepackage{import}

\newcommand\ph{\ensuremath{\varphi}}
\newcommand\eps{\ensuremath{\varepsilon}}

\newcommand{\cc}{\text{c.c.}}
\newcommand{\cst}{\mathrm{cst}}

\newcommand\define{\equiv}

\newcommand\vect[1]{\overrightarrow{#1}}
\newcommand\fvect[1]{\boldsymbol{#1}}
\newcommand{\mat}[1]{\boldsymbol{#1}}

\newcommand\ex[1]{\mathrm{e}^{#1}}
\renewcommand\i{\ensuremath{\mathrm{i}}}

\newcommand\transpose[1]{#1^{\rm T}}
\newcommand{\tr}{\mathrm{tr}}

\newcommand\e[1]{_{\text{#1}}}
\newcommand\h[1]{^{\text{#1}}}
\newcommand\U[1]{\:\mathrm{#1}}

\newcommand{\dd}{\mathrm{d}}
\newcommand{\Dd}{\mathrm{D}}
\newcommand{\pd}[3][]{\frac{\partial^{#1} #2}{\partial {#3}^{#1}}}
\newcommand{\ddf}[3][]{\frac{\dd^{#1} #2}{\dd {#3}^{#1}}}
\newcommand{\Ddf}[3][]{\frac{\Dd^{#1} #2}{\dd {#3}^{#1}}}

\renewcommand\lim[2]{\underset{ #1 \rightarrow #2 }{ \mathrm{lim} } \,}

\newcommand{\delimiters}[4][]{
\ifthenelse{ \equal{#1}{1} }{  #2 #3 #4  }
					{ \ifthenelse{\equal{#1}{2}}{ \big#2 #3 \big#4 }
						{ \ifthenelse{\equal{#1}{3}}{ \Big#2 #3 \Big#4 }
							{ \ifthenelse{\equal{#1}{4}}{ \bigg#2 #3 \bigg#4 }
								{ \ifthenelse{\equal{#1}{5}}{ \Bigg#2 #3 \Bigg#4 }
									{ \left#2 #3 \right#4 }
								}
							}
						}
					}
													}

\newcommand{\pa}[2][]{\delimiters[#1]{(}{#2}{)}}
\newcommand{\pac}[2][]{\delimiters[#1]{[}{#2}{]}}
\newcommand{\paac}[2][]{\delimiters[#1]{\{}{#2}{\}}}
\newcommand{\abs}[2][]{\delimiters[#1]{|}{#2}{|}}
\newcommand{\norm}[2][]{\abs[#1]{\abs[#1]{#2}}}
\newcommand{\ev}[2][]{\delimiters[#1]{\langle}{#2}{\rangle}}

\newcommand{\wl}{\mathscr{L}}

\newenvironment{system}
{ \left\{ \begin{aligned} }
{ \end{aligned} \right. }

\newmdtheoremenv[%
outerlinewidth=1.5,%
roundcorner=5pt,%
backgroundcolor=black!4,%
ntheorem=true%
]{exercise}{Exercise}

\newlength{\boxtitlelength}
\newlength{\halfrulelength}
\newcommand{\boxtitle}[1]{\footnotesize\bf{\:#1\:}}

\definecolor{blue4}{RGB}{0,0,143}
\definecolor{red4}{RGB}{143,0,0}
\definecolor{orange}{RGB}{255,128,0}
\definecolor{darkcyan}{RGB}{0,128,128}
\definecolor{olive}{RGB}{0,128,0}
\definecolor{purple}{RGB}{128,0,128}
\definecolor{cyan2}{RGB}{0,255,255}
\definecolor{fushia}{RGB}{255,0,255}
\definecolor{mygray}{gray}{0.5}
\definecolor{lightgray}{gray}{0.85}

\input{style.tex}

\makeatletter
\renewcommand{\cleardoublepage}{%
  \clearpage\thispagestyle{empty}%
  \if@twoside
    \ifodd\c@page
    \else
      \hbox{}\newpage
      \if@twocolumn
         \hbox{}\newpage
      \fi
    \fi
  \fi}
\makeatother

\title{Gravitation: from Newton to Einstein}
\author{Pierre Fleury}

\begin{document}
\dominitoc
\frontmatter 



\begin{center}

\noindent
\begin{minipage}{8cm}
\includegraphics[width=\linewidth]{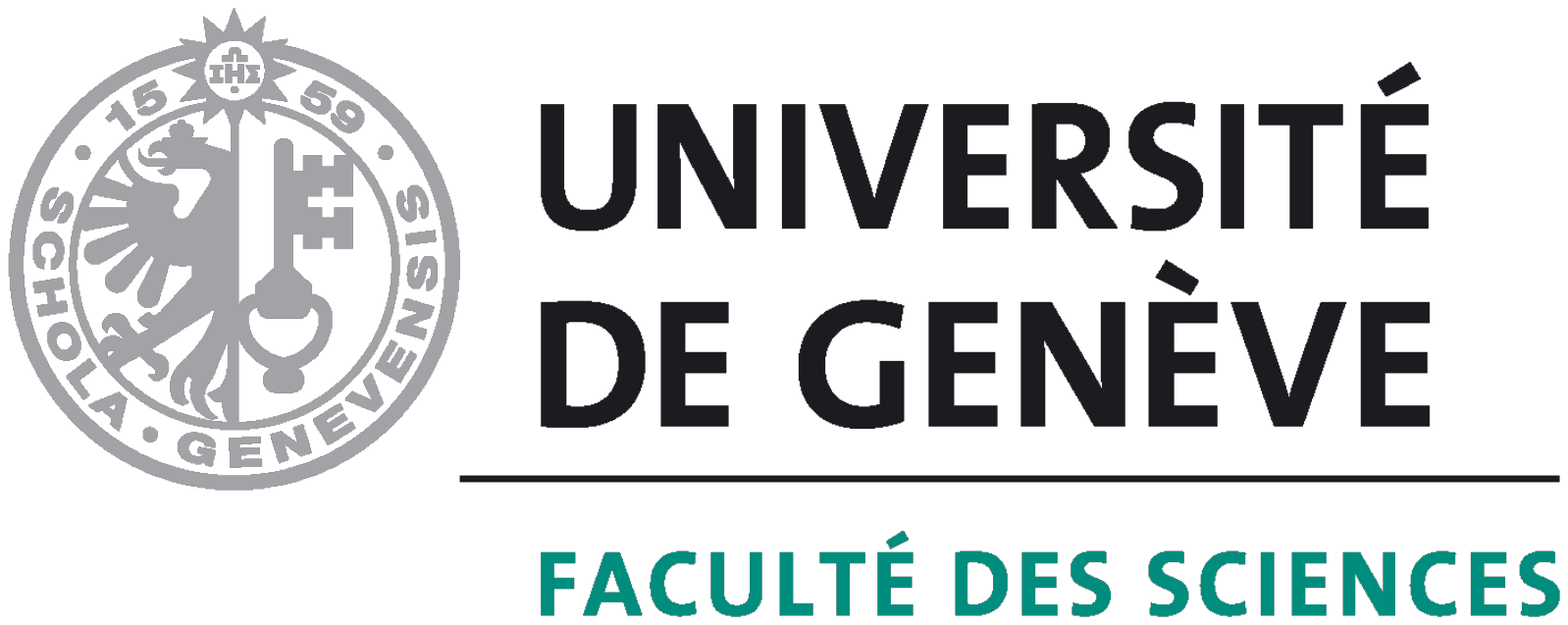}
\end{minipage}
\hfill
\begin{minipage}{4cm}
\includegraphics[width=\linewidth]{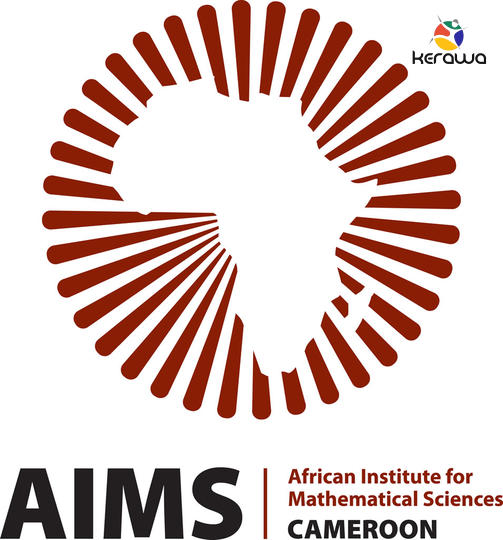}
\end{minipage}

\vspace*{4cm}

{%
\noindent\rule{\linewidth}{2pt}
{\Huge
\textbf{Gravitation:}\\[0.5cm]
\textbf{From Newton to Einstein}
}
\noindent\rule{\linewidth}{2pt}
}

\vspace*{1cm}

{\large
Lectures given at the African Institute for Mathematical Sciences, Cameroon (AIMS-Cameroon), in January 2018 and January 2019. 
}

\vspace*{2cm}

{\Large
\textsc{Pierre Fleury}\\[0.3cm]
\large
Département de Physique Théorique\\
Université de Genève, Switzerland \\
\texttt{pierre.fleury@unige.ch}
}

\vspace*{4cm}

Version: \today

\end{center}

\newpage


\cleardoublepage

\thispagestyle{empty}
\vspace*{\fill}
\begin{center}
{\LARGE\em To the students of Africa.}
\end{center}
\vspace*{\fill}

\chapter*{Foreword}

\lettrine{T}{he} African Institute for Mathematical Sciences (AIMS) is a pan-African non-profit educational organisation founded by the South African cosmologist Neil Turok, with the purpose of promoting mathematical sciences in Africa. It proposes an intensive one-year master-level programme for excellent and highly-motivated African students, with courses ranging from fundamental to applied mathematics, theoretical physics, and languages. The AIMS network consists of six centres in Cameroon, Ghana, Rwanda, Senegal, South Africa, and Tanzania. Each of them trains a cohort of about 50 students per year.

This document gathers the lecture notes of a course entitled \emph{Gravitation: from Newton to Einstein}, which I gave in January 2018 and January 2019 at AIMS-Cameroon. The course was initially designed to fit in thirty hours, each section corresponding to a two-hour lecture.
My main goal, in this course, was to propose a big picture of gravitation, where Einstein's theory of relativity arises as a natural increment to Newton's theory. The students are expected to be familiar with the fundamentals of Newton's mechanics and gravitation, for the first chapter to be a mere reformulation of known concepts. The second chapter then introduces special and general relativity at the same time, while the third chapter explores concrete manifestations of relativistic gravitation, notably gravitational waves and black holes.
The numerous exercises must be considered part of the course itself; they are intended to stimulate active reading.

\bigskip

\noindent \textbf{Acknowledgements.} I would not have had the opportunity to deliver this course without my mentor and friend Jean-Philippe Uzan, who both introduced me to the AIMS network and helped me designing the structure of the course itself. I also thank the academic director of AIMS-Cameroon, Marco Garuti, for his warm welcome and for having trusted me to take care of his students two years in a row. Many thanks to the tutors Peguy Kameni Ntseutse, Hans Fotsing and Pelerine Nyawo, for their daily assistance, and to my fellow lecturers, notably Patrice Takam, Charis Chanialidis, Jane Hutton, and Julia Mortera. Finally, I would like to express my sincere congratulations to the AIMS students for their remarkable attitude, dedication, and hard work.

\bigskip

\noindent \textbf{Influential references.} The organisation and content of this course, especially the first chapter, are partly inspired from \emph{Relativity in Modern Physics}~\cite{Deruelle:2018ltn} by Nathalie Deruelle and Jean-Philippe Uzan. They also reflect my personal approach to relativity and gravitation, which has been influenced by \emph{Special Relativity in General Frames}~\cite{Gourgoulhon:2013gua} by Eric Gourgoulhon, \emph{A Relativist's Toolkit}~\cite{Poisson:2009pwt} by Eric Poisson, and a remarkable doctoral course on general relativity that Gilles Esposito-Farèse gave at the Institut d'Astrophysique de Paris in 2013. I also used bits and pieces of a course given by my esteemed colleague Martin Kunz at the University of Geneva in 2017 and 2018, itself based on the very comprehensive \emph{General Relativity}~\cite{Straumann:2013spu} by Norbert Straumann.

\setcounter{tocdepth}{1}
\tableofcontents

\chapter*{Introduction}
\addstarredchapter{Introduction}

\hfill
\begin{minipage}{11.5cm}
\begin{otherlanguage}{french}
{\it
\begin{itemize}
\item Sur quelle plan\`{e}te suis-je tomb\'{e} ? demanda le petit prince.
\item Sur la Terre, en Afrique, r\'{e}pondit le serpent.
\item Ah ! \ldots Il n'y a donc personne sur la Terre ?
\item Ici c'est le d\'{e}sert. Il n'y a personne dans les d\'{e}serts. La Terre est grande, dit le serpent.
\end{itemize}
}
\begin{flushright}
Antoine de Saint-Exup\'{e}ry, \textit{Le Petit Prince}.
\end{flushright}
\end{otherlanguage}
\end{minipage}

\bigskip

\lettrine{G}{ravitation} surely is not the most appreciated of all forces. In its absence, my first steps as a child would have been far easier to achieve, and my clumsiness would have less practical consequences. This is unfair judgement though, for without gravitation there would be no one to enjoy floating around. Without gravitation, the splendid structures of our Universe could not have formed. Without gravitation, galaxies would not swirl and stars would not shine; planets would never have come to existence, and life would not be.

Because of its evidence and ubiquity in our daily experience of motion, it is not surprising that gravitation was the first physical interaction ever described within a solid scientific framework. Newton's theory of the universal attraction of massive bodies was, at the end of the 17\textsuperscript{th} century, a proper scientific revolution. It remains one of the best examples of conceptual unification -- how audacious was it to claim that objects falling on the ground and the orbits of celestial bodies are merely two facets of the same phenomenon?

Albeit unchallenged for more than two centuries, Newton's formulation of physics was only a prelude. In the early $20\textsuperscript{th}$ century, another revolution occurred, and dramatically changed our conception of the Universe. With the advent of Einstein's relativity, the hitherto distinct concepts of space and time merged into a hybrid structure called space-time. Furthermore, this space-time turned out to be somehow malleable, gravity being nothing but its geometry. This superb theory, formulated in 1915, was not less superbly confirmed, in 1919, by Eddington's measurement of the deflection of starlight by the Sun.

Besides light bending, relativity also predicted some exotic phenomena, among which gravitational waves and black holes. The first ones, which are to gravity what light is to electromagnetism, were first detected in 2015, that is, exactly one century after the formulation of the theory encompassing them. As these gravitational waves were produced by the collision of two black holes, they also provided indirect proof of their existence; and if this does not convince you, take a look at the 2019 photograph of the M87\textsuperscript{*} super-massive black hole! That picture, whose interest relies on the deviation of light by the black hole, remarkably marked the centenary of Eddington's observation.

Could there be a better occasion to start a journey around the world of gravity? Be careful though, as you may fall in love with it, just like I did.

\mainmatter 

\chapter{Newton's physics}
\label{chap:Newton}

\lettrine{I}{n} the somewhat legendary book \textit{Philosophiae naturalis principia mathematica}~\cite{Newton:1687eqk} (Mathematical principles of the natural philosophy), published in 1687, Isaac Newton set the fundamentals of modern physics, based on mathematics and calculus. His formulation of mechanics and gravitation remained unchallenged for more than two centuries.

\minitoc
\newpage

\section{Kinematics}

The term \emph{kinematics}, which comes from the French word \emph{cin\'{e}matique}, itself inspired from the Greek $\kappa\iota\nu\eta\mu\alpha$ (movement, motion), is the description of motion in physics. This first section deals with the fundamental postulates of Newtonian physics, namely the notions of time, space, and hence motion. It will be the opportunity to introduce notation and mathematical concepts that will be useful in all the remainder of this course.

\subsection{Time and space}

\paragraph{Absolute time} Newton's mechanics was probably the first consistent mathematical description of the world perceived by our senses. In this perception, there is a notion of time, which quantifies how things age, or change. Time is also tightly related to \emph{causality}, in that it classifies events depending on what can possibly be the cause or the consequence of what. As such, an essential property of time is that it allows events to be ordered, and the simplest mathematical tool for that purpose is a real number, denoted $t$. If two events $E_1$, $E_2$ are characterised by times $t_1, t_2$, then $t_1<t_2$ implies that $E_1$ can be the cause of $E_2$; if $t_1=t_2$, theses events are simultaneous, and cannot be causally connected.

Still in our sensitive experience, the way things age and change is \emph{absolute}. In other terms, the history of a given phenomenon depends neither on who observes it, nor on how, where, and when the observer performs the observation. Only at the beginning of the twentieth century was this intuitive framework challenged and finally proved wrong. We will nevertheless assume, in this first chapter, that it applies.

\paragraph{Spatial coordinates} Once the \emph{when} of an event is sorted, one also has to specify the \emph{where}. Contrary to time, a single number is not enough to characterise a position in space. Besides, space does not require any absolute ordering like time does. In our daily experience, space seems to have three dimensions, in the sense that the minimal structure that we need to locate points in space is a set of three numbers, called \emph{spatial coordinates}.

\begin{figure}[h!]
\centering
\input{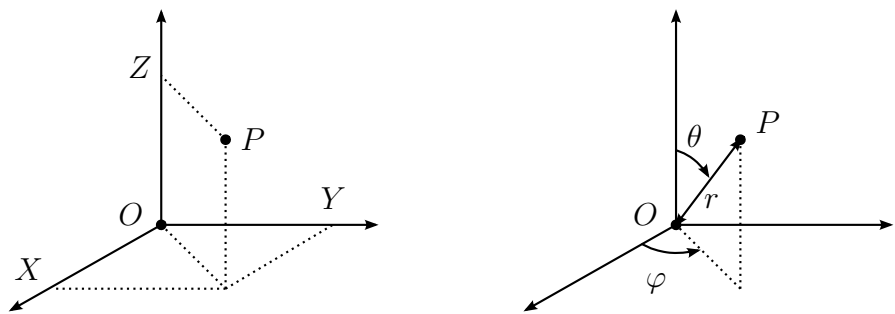}
\caption{Cartesian (left) and spherical (right) coordinates of a point $P$.}
\label{fig:coordinates}
\end{figure}

A fundamental example is the set of Cartesian (also called rectangular) coordinates $(X,Y,Z)$, which locate positions with respect to an arbitrary reference $O$ as depicted on the left of fig.~\ref{fig:coordinates}. Spherical coordinates $(r,\theta,\ph)$, on the right of fig.~\ref{fig:coordinates} are another important example.

\begin{exercise}\label{ex:coordinates}
Show that Cartesian and spherical coordinates are related by
\begin{align}
X &= r\sin\theta\cos\ph \\
Y &= r\sin\theta\sin\ph \\
Z &= r \cos\theta.
\end{align}
\vspace*{-0.5cm}
\end{exercise}

\paragraph{Notation} It is customary to denote coordinates in space with the abstract notation $(x^i)=(x^1, x^2, x^3)$, which can stand for any coordinate system. \emph{Beware!} the superscripts are indices, not exponents. For example, with spherical coordinates, $x^1=r$, $x^2=\theta, x^3=\ph$. This notation will allow us to write equations without having to specify the coordinate system that we are using.
We will keep the beginning of the alphabet ($a,b,c,\ldots$) for the Cartesian coordinates~$(X^a)=(X,Y,Z)$, which have a very special status.

\subsection{Metric}

When solving exercise~\ref{ex:coordinates}, you have certainly used the fact that $r^2=X^2+Y^2+Z^2$, that is to say the Pythagorean theorem of Euclidean geometry. More generally, you have used the fact that the distance $d_{AB}$ between two points $A, B$ reads, in Cartesian coordinates
\begin{align}
d_{AB}^2
&= (X_B-X_A)^2 + (Y_B-Y_A)^2 + (Z_B-Z_A)^2 \label{eq:distance_Cartesian_coordinates}\\
&= \sum_{a=1}^{3}\sum_{b=1}^3 \delta_{ab} (X^a_B-X^a_A) (X^b_B - X^b_A) \label{eq:introducing_Kronecker}\\
&\define \delta_{ab} (X^a_B-X^a_A) (X^b_B - X^b_A) \qquad \text{[Einstein's notation],} \label{eq:introducing_Einstein_summation}
\end{align}
where, in eq.~\eqref{eq:introducing_Kronecker} we introduced the \emph{Kr\"onecker symbol}
\begin{equation}
\delta_{ab} \define
\begin{cases}
1 & \text{if $a=b$}\\
0 & \text{if $a\not=b$}
\end{cases}
\end{equation}
and, in eq.~\eqref{eq:introducing_Einstein_summation} we used \emph{Einstein's convention for the summation over repeated indices}. This latter convention consists in implicitly summing over any repeated index in an expression, which highly alleviates notation. We will use it in the remainder of this course.

\paragraph{Euclidean metric} Clearly, for non-Cartesian coordinates, one cannot directly use the expression~\eqref{eq:distance_Cartesian_coordinates} to calculate $d_{AB}$. For example, with spherical coordinates
\begin{equation}
d^2_{AB} \not= (r_B-r_A)^2 + (\theta_B-\theta_A)^2 + (\ph_B-\ph_A)^2;
\end{equation}
the above expression is even dimensionally incorrect. In order to calculate distances with any coordinate system, consider two points $P,P'$ whose Cartesian coordinates are almost equal, $X_P^a$ and $X^a_{P'}=X_{P}^a+\dd X^a$. Then, applying eq.~\eqref{eq:introducing_Einstein_summation}, we have
\begin{equation}\label{eq:Cartesian_metric}
d_{PP'}^2 \define \dd\ell^2 = \delta_{ab} \, \dd X^a \dd X^b.
\end{equation}
This expression is now ready to be converted to any other coordinate system. Indeed, consider another coordinate system $(x^i)$; because $(X^a)$ and $(x^i)$ describe the same space, they are related by three functions $f^a$ such that $X^a=f^a(x^i)$. For example, if $(x^i)$ denote spherical coordinates, you have derived these functions in exercise~\ref{ex:coordinates}: $f^1(r,\theta,\ph)=r\sin\theta\cos\ph$, $f^2(r,\theta,\ph)=r\sin\theta\cos\ph$, $f^3(r,\theta,\ph)=r\cos\theta$.

Since the coordinates of the neighbouring points $P$ and $P'$ differ by $(\dd X^a)$, their other coordinates differ by $(\dd x^i)$, with
\begin{equation}
\dd X^a = \pd{f^a}{x^i} \, \dd x^i,
\end{equation}
(do not forget that there is summation over repeated indices). It is customary to replace the notation $f^a$ simply by $X^a$, and when this is inserted into the expression~\eqref{eq:Cartesian_metric}, we find
\begin{empheq}[box=\fbox]{equation}\label{eq:Euclidean_metric}
\dd\ell^2 = e_{ij} \,\dd x^i \dd x^j,
\qquad \text{with} \quad
e_{ij} \define \delta_{ab} \, \pd{X^a}{x^i} \pd{X^b}{x^j}.
\end{empheq}
The object formed by the set of coefficients~$e_{ij}=e_{ji}$, which can be thought of as a symmetric matrix, is called the \emph{metric tensor}. It is an example of tensor, a mathematical notion that will come back in the next chapter. For now, the important thing is that the metric is \emph{a machine that transforms coordinates into distances}.

\begin{exercise}
Show that, in spherical coordinates, the infinitesimal distance between two neighbouring points reads
\begin{equation}
\dd\ell^2 = \dd r^2 + r^2 \pa{\dd\theta^2 + \sin^2\theta \,\dd\ph^2},
\end{equation}
and give the associated metric coefficients $e_{rr}, e_{r\theta}$, etc.
\end{exercise}

\begin{exercise}\label{ex:inverse_metric}
Show that the inverse of the metric~\eqref{eq:Euclidean_metric}, in the sense of matrix inversion, denoted $e^{ij}$ and defined by the relation~$e^{ik}e_{kj}=\delta^i_j$, reads
\begin{equation}
e^{ij} = \delta^{ab} \, \pd{x^i}{X^a} \pd{x^j}{X^b}.
\end{equation}
\end{exercise}

\begin{figure}[h!]
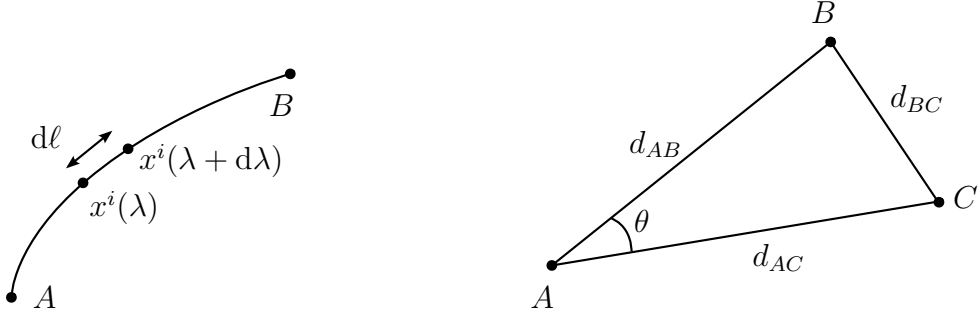

\centering
\input{curve.pdf_tex}
\input{angle.pdf_tex}
\caption{Left: parametrised curve between $A$ and $B$. Right: angles and distances}
\label{fig:curve_angle}
\end{figure}

\paragraph{Curvilinear distance} Let us draw a curve between two points $A$ and $B$, as in fig.~\ref{fig:curve_angle} (left). This curve can be parametrised by three functions~$x^i(\lambda)$, where $\lambda$ is an arbitrary parameter that allows one to move along the curve, assumed to be strictly increasing from $\lambda_A$ to $\lambda_B$ on the way from $A$ to $B$. The length of the curve is obtained by summing the lengths of every infinitesimal step from $A$ to $B$, that is
\begin{equation}\label{eq:length_curve}
\ell_{AB} = \int_A^B \dd\ell
= \int_A^B \sqrt{e_{ij} \dd x^i \dd x^j}
= \int_{\lambda_A}^{\lambda_B} \sqrt{e_{ij} \ddf{x^i}{\lambda} \ddf{x^j}{\lambda}} \; \dd\lambda.
\end{equation}
What we have called the distance~$d_{AB}$ between $A$ and $B$ is the \emph{shortest} length $\ell_{AB}$ among all possible curves connecting those two points. Such a curve is called a \emph{geodesic}. In Euclidean geometry and in the absence of constraints, it is simply a straight line; on the surface of a sphere, it is a great circle.

\subsection{Scalar product}

Finally, since we are able to compute distances in any coordinate system, we can also get \emph{angles}. Indeed, considering three points $A, B, C$, as depicted in fig.~\ref{fig:curve_angle} (right), we know that the angle~$\theta$ between $(AB)$ and $(AC)$ reads
\begin{equation}
 \cos\theta
 = \frac{d_{AB}^2+d_{AC}^2-d_{BC}^2}{2 d_{AB} d_{AC}} \ .
\end{equation}

\begin{exercise}
Assuming that $A$, $B$, and $C$ are separated by infinitesimal distances, show that the scalar product between the vectors $\vect{AB}$ and $\vect{AC}$ reads, in terms of arbitrary coordinates,
\begin{equation}\label{eq:scalar_product_3_points}
\vect{AB} \cdot \vect{AC} = e_{ij} (x^i_B-x^i_A)(x^j_C-x^j_A).
\end{equation}
\end{exercise}

It is customary to associate to any coordinate system $(x^i)=(x^1, x^2, x^3)$ a local basis $(\vec{\partial}_i)=(\vec{\partial}_1, \vec{\partial}_2, \vec{\partial}_3)$. This basis is defined so that if $A, A'$ have coordinates $x^i, x^i+\dd x^i$, where $\dd x^i$ is infinitesimal, then
\begin{equation}\label{eq:decomposition_vector_3D}
\vect{AA'} = \dd x^i \vec{\partial}_i \ .
\end{equation}
The use of the symbol~$\partial_i$, which is a short-hand notation for $\partial/\partial x^i$, is justified by its behaviour under coordinate transformations. Indeed, eq.~\eqref{eq:decomposition_vector_3D} holding in any coordinate system, we have $\vect{AA'}=\dd x^i\vec{\partial}_i=\dd X^a \vec{\partial}_a$, and hence the two bases are related as
\begin{equation}
\vec{\partial}_i = \pd{X^a}{x^i} \, \vec{\partial}_a \ ,
\end{equation}
which is reminiscent of the chain rule for partial derivatives. The decomposition~\eqref{eq:decomposition_vector_3D} actually applies to any vector~$\vec{u}$, which can be seen as the extension of an arrow connecting two neighbouring points $A, A'$. We thereby define the \emph{components} $u^i, u^a$ of this vector as
\begin{equation}
\vec{u} = u^a \vec{\partial}_a = u^i \vec{\partial}_i \ .
\end{equation}
This immediately implies the following transformation rule under $X^a \rightarrow x^i$,
\begin{equation}\label{eq:transformation_vector}
u^i = \pd{x^i}{X^a} u^a \ , \qquad
u^a = \pd{X^a}{x^i} u^i \ .
\end{equation}
Preserving the altitude of indices in eq.~\eqref{eq:transformation_vector} is useful trick to remember which Jacobian matrix ($\partial x^i/\partial X^a$ or $\partial X^a/\partial x^i$) must be used.

\begin{exercise}
Combining eqs.~\eqref{eq:scalar_product_3_points} and \eqref{eq:decomposition_vector_3D}, show that the metric components read
\begin{equation}\label{eq:scalar_product_basis_vectors}
e_{ij} = \vec{\partial}_i \cdot \vec{\partial}_j \ .
\end{equation}
Conclude that the metric gives the scalar product of any two vectors as
\begin{equation}
\vec{u} \cdot \vec{v} = e_{ij} u^i v^j ,
\end{equation}
and discuss the case of Cartesian coordinates: what is $\vec{\partial}_a\cdot\vec{\partial}_b$?
\end{exercise}

\noindent\textit{Remark.} Equation~\eqref{eq:scalar_product_basis_vectors} shows that the basis~$\vec{\partial}_i$ is \emph{not} orthonormal in general. For example, with spherical coordinates, $(\vec{\partial}_r,\vec{\partial}_\theta, \vec{\partial}_\ph)$ is different from the usual orthonormal basis~$(\vec{u}_r, \vec{u}_\theta, \vec{u}_\ph)$ because the latter is normalised. Both bases are related by
\begin{equation}\label{eq:normalised_basis}
\vec{u}_r
= \vec{\partial}_r \ ,
\qquad
\vec{u}_\theta
=\frac{1}{\sqrt{e_{\theta\theta}}} \, \vec{\partial}_\theta
=\frac{1}{r} \, \vec{\partial}_\theta \ ,
\qquad
\vec{u}_{\ph}
=\frac{1}{\sqrt{e_{\ph\ph}}} \, \vec{\partial}_{\ph}
=\frac{1}{r\sin\theta} \, \vec{\partial}_{\ph} \ .
\end{equation}

Summarising, the metric is not only as a machine to compute distances between points, but also scalar products between vectors. As such, it is the object that quantifies space. In Newtonian physics, space, just like time, is considered to be absolute, in the sense that the distances or angles between objects does not depend on who, how, and when they are observed. In other words, the metric is independent from the observer.

\subsection{Motion}

\paragraph{Velocity} Putting together the notions of time and space naturally leads to the concept of \emph{motion}, i.e. the change of position in space of an object as time passes. The trajectory of an object is characterised by a curve $x^i(t)$ parametrised with time. Its \emph{velocity} is the rate of change of its position, thus it is given by the vector~$\vec{v}$ with
\begin{equation}
v^i \define \ddf{x^i}{t} \define \dot{x}^i
\end{equation}
in any coordinate system. The \emph{speed}~$v$ of the object is the norm of its velocity, $v^2 = e_{ij} v^i v^j = \delta_{ab}v^a v^b$.

\paragraph{Acceleration} Similarly, the \emph{acceleration} $\vec{a}$ is the rate of change of the velocity. In Cartesian coordinates,
\begin{equation}
a^b = \dot{v}^b = \ddot{x}^b.
\end{equation}
Both $\vec{v}$ and $\vec{a}$ are vectors, hence their components change according to eq.~\eqref{eq:transformation_vector} under coordinate transformations. However, for an arbitrary coordinate system, $a^i \not= \dot{v}^i$. Let us show this explicitly:
\begin{align}
a^i
&= \pd{x^i}{X^b} \, a^b \\
&= \pd{x^i}{X^b} \ddf{v^b}{t} \\
&= \pd{x^i}{X^b} \,\ddf{}{t} \pa{ \pd{X^b}{x^j} v^j } \\
&= \pd{x^i}{X^b} \pa{\pd{X^b}{x^j}\ddf{v^j}{t}+v^j\ddf{}{t}\pd{X^b}{x^j}}
\label{eq:Christoffel_appears} \\
&= \pd{x^i}{X^b} \pd{X^b}{x^j} \ddf{v^j}{t}
	+ \pd{x^i}{X^b} v^j \, \ddf{x^k}{t}
		 \frac{\partial^2 X^b}{\partial x^k \partial x^j}\\
&=  \ddf{v^i}{t} +
		\pd{x^i}{X^b} \frac{\partial^2 X^b}{\partial x^k \partial x^j} \, v^j v^k,
\end{align}
which contains a new term, proportional to $\partial^2 X^b/\partial x^k \partial x^j$. We see that the key step that is responsible for this term is \eqref{eq:Christoffel_appears}; namely, the derivatives~$\partial X^b/\partial x^j$ are, in general, functions of $x^i$, which change as the object moves.

\paragraph{Covariant derivative} The above calculation reveals a crucial feature of general coordinate transformations: they change how derivatives act on vector fields. For a vector field~$u^i(x^j)$, we introduce the \emph{covariant derivative} of $\vec{u}$ in the $i$th direction as
\begin{empheq}[box=\fbox]{align}
\nabla_i u^k &\define \partial_i u^k + \Gamma\indices{^k_j_i} u^j \\
\text{with} \quad
\Gamma\indices{^k_j_i} &\define \frac{1}{2} e^{kl} \pa{ \partial_i e_{jl}
																				+ \partial_j e_{il}
																				- \partial_l e_{ij} } ,
\label{eq:Christoffel_def_Euclidean}
\end{empheq}
where $\Gamma\indices{^k_j_i}$ are called \emph{Christoffel symbols}, and $e^{ij}$ are the component of the inverse metric (see exercise~\ref{ex:inverse_metric}). This definition ensures that $\nabla_i \vec{u}=(\nabla_i u^j) \vec{\partial}_j$ is a vector, in the sense that it behaves correctly with respect to coordinate transformations:
\begin{equation}
\nabla_i u^j = \pd{x^j}{X^b} \, \nabla_i u^b \ .
\end{equation}

\begin{exercise}
Using the expression~\eqref{eq:Euclidean_metric} of the metric coefficients $e_{ij}$, show that the Christoffel symbols~\eqref{eq:Christoffel_def_Euclidean} also satisfy
\begin{equation}
\Gamma\indices{^i_j_k}
= \frac{\partial x^i}{\partial X^a}
	\frac{\partial^2 X^a}{\partial x^j \partial x^k} \ .
\end{equation}
Conclude that the acceleration in arbitrary coordinates reads
\begin{equation}
a^i = \Ddf{v^i}{t} \define \ddf{v^i}{t} + \Gamma\indices{^i_j_k} v^j v^k \ ,
\end{equation}
which we shall call the covariant derivative of $\vec{v}$ with respect to time.
\end{exercise}

\begin{exercise}\label{ex:acceleration_spherical}
Show that the acceleration in spherical coordinates reads
\begin{align}
a^r &= \ddot{r} - r\dot{\theta}^2 - r\sin^2\theta \, \dot{\ph}^2\\
a^\theta &= \ddot{\theta} + \frac{2\dot{r}\dot{\theta}}{r} - \sin\theta\cos\theta \, \dot{\ph}^2\\
a^\ph &= \ddot{\ph} + \frac{2\dot{r}\dot{\ph}}{r} + \frac{2\cos\theta}{\sin\theta} \, \dot{\theta}\dot{\ph} \ ,
\end{align}
and compare with the expression given in the literature (e.g. \href{https://en.wikipedia.org/wiki/Spherical_coordinate_system}{Wikipedia}). Explain the apparent difference in light of eq.~\eqref{eq:normalised_basis}.
\end{exercise}

\subsection{Reference frames}
\label{sec:frames}

\begin{figure}[h!]
\centering
\input{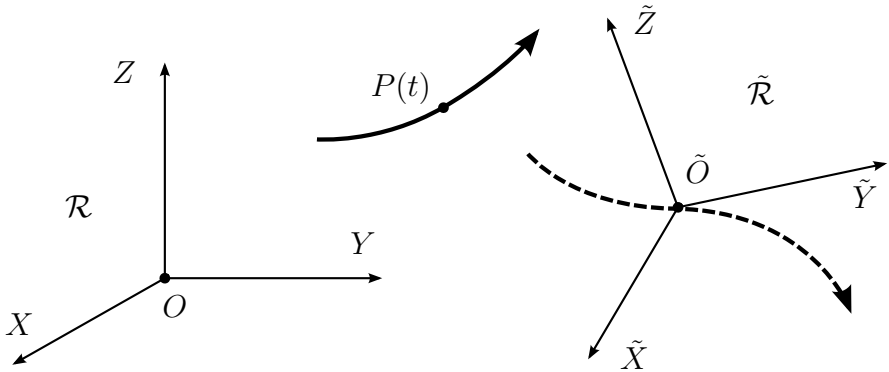}
\caption{The motion of a particle~$P(t)$ can be described relatively to the reference frames~$\mathcal{R}(X,Y,Z)$ and $\tilde{\mathcal{R}}(\tilde{X},\tilde{Y},\tilde{Z})$. The origin~$\tilde{O}$ and the axes of $\tilde{\mathcal{R}} $ are moving with respect to those of $\mathcal{R}$.}
\label{fig:frames}
\end{figure}

Contrary to time and space, velocity and acceleration are not independent from the observer, because they rely on a reference that might be moving itself. This is the obvious relativity of motion. A \emph{reference frame} formalises the intuitive notion of viewpoint; it is a particular Cartesian coordinate system, with respect to which one describes the motion of objects. Different reference frames may have origins and axes that move relative to each other (see fig.~\ref{fig:frames}).

For example, a corner of the room can be the origin of a reference frame $\mathcal{R}$, and the edges between the walls and the floor (or ceiling) can form its axes. It describes the point of view of someone who would be standing still at this corner. Another frame~$\tilde{\mathcal{R}}$ can be formed by you, walking in the room, holding your arms horizontally.

\paragraph{Change of frame} Changing the reference frame~$\mathcal{R}\rightarrow \tilde{\mathcal{R}}$ is a \emph{time-dependent} transformation from some Cartesian coordinates $(X^a)$ to other Cartesian coordinates~$(\tilde{X}^b)$,
\begin{equation}
X^a \rightarrow \tilde{X}^b(t,X^a).
\end{equation}
The condition that both systems are Cartesian is actually very restrictive. Only the transformations that preserve the Kr\"onecker form of the metric unchanged are allowed:
\begin{equation}
\dd\ell^2
= \delta_{ab} \, \dd X^a \dd X^b
= \delta_{cd} \, \dd\tilde{X}^c \dd\tilde{X}^d,
\qquad \text{i.e.} \quad
\delta_{ab} \, \frac{\partial X^a}{\partial\tilde{X}^c}
					\frac{\partial X^b}{\partial \tilde{X}^d}
= \delta_{cd} \ .
\end{equation}
These are called \emph{isometries}, they consist of translations and rotations. Thus, a change of frame must take the form
\begin{align}\label{eq:change_frame}
X^a(t,\tilde{X}^b) = X^a_{\tilde{O}}(t) + R\indices{^a_b}(t) \tilde{X}^b,
\end{align}
where $X^a_{\tilde{O}}(t)$ represents the trajectory of the origin~$\tilde{O}$ of $\tilde{\mathcal{R}}$ ($\tilde{X}^b=0$) as seen in $\mathcal{R}$, and $(R\indices{^a_b})$ are the components of a rotation matrix $\mat{R}(t)\in\mathrm{SO}(3)$, which encodes the rotation of the axes of $\tilde{\mathcal{R}}$ with respect to those of $\mathcal{R}$.

\paragraph{Composition of velocities and accelerations} Let us examine the consequences of $X^a\rightarrow \tilde{X}^a$ on kinematics. Taking the time derivative of eq.~\eqref{eq:change_frame}, we get
\begin{equation}
v^a = v^a_{\tilde{O}}
			+ \dot{R}\indices{^a_b} \tilde{X}^b
			+ R\indices{^a_b} \tilde{v}^b.
\end{equation}
The second term of the right-hand side, $\dot{R}\indices{^a_b}$, can be rewritten using the properties of $\mathrm{SO}(3)$. Taking the time derivative of the identity~$\transpose{\mat{R}}\mat{R}=\mat{1}_3$, where $\mat{1}_3$ is the $3\times 3$ unity matrix, we conclude that $\transpose{\mat{R}}\dot{\mat{R}}\define\mat{A}$ is an antisymmetric matrix. Thus, there exists a vector~$\vec{\Omega}$ such that
\begin{equation}
\dot{\mat{R}} =
\mat{R}\mat{A}
=
\mat{R}
\begin{bmatrix}
0 & - \Omega^3 & \Omega^2 \\
\Omega^3 & 0 & - \Omega^1 \\
-\Omega^2 & \Omega^1 & 0 
\end{bmatrix}.
\end{equation}
In terms of components and indices, this can be written
\begin{equation}\label{eq:Rdot_indices}
\dot{R}\indices{^a_b} = R\indices{^a_c}\eps\indices{^c_d_b} \Omega^d,
\end{equation}
where $\eps_{abc}$ denotes the \emph{Levi-Civita symbol}\footnote{The position of \emph{Cartesian} indices $a,b,c,\ldots$ does not really matter, $\eps\indices{^a_b_c}=\eps_{abc}$. Things are different for indices $i,j,k\ldots$ associated with arbitrary coordinates.}, such that
\begin{equation}
\eps_{abc} =
\begin{cases}
1 & \text{if $abc$ is an even permutation of 123}, \\
-1 & \text{if $abc$ is an odd permutation of 123}, \\
0 & \text{if any two indices are identical}.
\end{cases}
\end{equation}

\begin{exercise}
Check the relation~\eqref{eq:Rdot_indices}. Show that the Levi-Civita symbol gives the cross-product of two vectors; namely, if $\vec{w}=\vec{u}\times\vec{v}$, then
\begin{equation}
w^a = \eps\indices{^a_b_c} u^b v^c \ .
\end{equation}
\end{exercise}

Putting everything together, and changing some of the names of the indices that are summed over, we obtain the relation between the velocities in different frames
\begin{empheq}[box=\fbox]{equation}\label{eq:composition_velocities}
v^a = v^a_{\tilde{O}}
			+ R\indices{^a_b} \pa{ \tilde{v}^b
													+ \eps\indices{^b_c_d} \Omega^c \tilde{X}^d
												} ,
\end{empheq}
or, in a vector form,
\begin{equation}
\vec{v} = \vec{v}_{\tilde{O}} + \vec{\Omega} \times \vec{\tilde{X}} \ .
\end{equation}
While $\vec{v}_{\tilde{O}}$ represents the relative movements of the origins of $\mathcal{R}$ and $\tilde{\mathcal{R}}$, $\vec{\Omega}$ represents the instantaneous rotation velocity of their axes. More precisely, the direction of $\vec{\Omega}(t)$ is the axis of $\mat{R}(t)$, and its norm is the angular velocity of the rotation. 

\begin{exercise}
Taking the time derivative of eq.~\eqref{eq:composition_velocities}, show that the acceleration in $\mathcal{R}$ is related to the acceleration in $\tilde{\mathcal{R}}$ as
\begin{equation}\label{eq:composition_accelerations}
a^b
= a^b_{\tilde{O}} + R\indices{^b_c} 
								\pa{ 
									\tilde{a}^c
								+ \eps\indices{^c_d_e} \dot{\Omega}^d \tilde{X}^e
								+ \eps\indices{^c_d_e} \eps\indices{^e_f_g}
									\Omega^d \Omega^f \tilde{X}^g
								+ 2 \eps\indices{^c_d_e} \Omega^d \tilde{v}^e
								}.
\end{equation}
The third term on the right-hand side is sometimes called Euler acceleration, while the fourth is the centrifugal acceleration, and the fifth is the Coriolis acceleration.
\end{exercise}


\section{Dynamics}

Kinematics was the description of motion. In this section, we would like to analyse the causes of motion. \emph{Dynamics}, from the Greek word $\delta\upsilon\nu\alpha\mu o\varsigma$ (power), is the study of how forces affect the movement of objects.

\subsection{Newton's three laws of dynamics}


\paragraph{First law: inertia} We postulate the existence of a class of reference frames, called \emph{inertial}, or \emph{Galilean} frames, with respect to which any isolated body (i.e. undergoing no external forces) has a constant velocity, $v^a=\cst, a^b=0$. It thus follows a linear trajectory at constant speed. Any frame in constant-speed linear translation with respect to an inertial frame is, itself, inertial. In terms of the transformation~\eqref{eq:change_frame} of the previous section, it corresponds to $v_{\tilde{O}}^a=\cst$ and $\Omega^a=0$.

This Newtonian notion of inertial frame is quite theoretical. There actually exists no physical frame in the Universe that would be \emph{exactly} inertial. In practice, one has to rely on approximations: the less accelerated, the more inertial a frame is. For example, the Terrestrial frame (attached to the ground) is less inertial than the geocentric frame, because of the Earth's proper rotation, which is itself less inertial than the heliocentric frame, because of the Earth's revolution around the Sun, and so on.

\paragraph{Second law: dynamics} In an inertial frame, the time evolution of the \emph{momentum}~$\vec{p}$ of an object is driven by the sum of external forces~$\vec{F}$,
\begin{equation}
\ddf{p^a}{t} = F^a,
\qquad \text{with} \quad p^a \define m v^a,
\end{equation}
where $m$ is the \emph{inertial mass} of the object. This mass characterises the difficulty of an object to be moved, since the larger $m$, the smaller the acceleration for a given force. In an arbitrary coordinate system, this becomes
\begin{empheq}[box=\fbox]{equation}
\Ddf{p^i}{t} \define \ddf{p^i}{t} + \Gamma\indices{^i_j_k} p^j v^k = F^i.
\end{empheq}
If the mass of the object is constant, then Newton's second law reads $m a^i= F^i$, but its expression in terms of momentum is more general.

\begin{exercise}
Consider an object that progressively disintegrates into light, in such a way that its mass decreases proportionally to itself, $\dot{m}=-m/\tau$, where $\tau$ is a constant characteristic time. Show that this leads to an apparent force on the object, which can be compared with friction.
\end{exercise}

\paragraph{Third law: action and reaction} If an object $1$ exerts a force $\vec{F}_{1\rightarrow 2}$ on an object $2$, then $2$ exerts in return a force $\vec{F}_{2\rightarrow 1}=-\vec{F}_{1\rightarrow 2}$ on $1$. We experience this law every time we throw something heavy, and feel its recoil. It is also what makes sails and planes to work.

\subsection{Conserved quantities}
\label{sec:conserved_quantities}

Once one knows the forces applied to an object, Newton's laws allow one to predict its motion. In practice, one has to solve second-order differential equations for each individual situation that one studies. Nevertheless, Newton's laws also imply that some quantities related to the motion of isolated systems remain constant whatever happens to it. These are called \emph{integrals of motion}, or simply \emph{conserved quantities}.

\paragraph{Linear momentum} Consider an isolated particle, i.e. with no force acting on it. In an inertial frame, the second Newton's law implies that its momentum is conserved, $p^a=\cst$. If now we consider an isolated system of $N$ interacting particles, where the particle~$m$ exerts a force $\vec{F}_{m\rightarrow n}$ on the particle~$n$, then obviously the momentum of every particle is changing, since
\begin{equation}
\ddf{p^a_n}{t} = \sum_{m=1}^N F^a_{m\rightarrow n} \not= 0
\end{equation}
in general. However, the \emph{total} momentum of the whole system is conserved. Indeed,
\begin{equation}
\ddf{P^a}{t}
= \sum_{n=1}^N \ddf{p^a_n}{t}
= \sum_{n=1}^N \sum_{m=1}^N F^a_{m\rightarrow n}
= 0
\end{equation}
by virtue of the third Newton's law. This can be generalised to arbitrary coordinate systems by replacing the standard time derivative by a covariant derivative, $\Dd P^i/\dd t=0$.

\paragraph{Angular momentum} The angular momentum of a particle at $M(t)$ with respect to the origin~$O$ of the frame is defined as $\vec{L}\define \vect{OM}\times \vec{p}$. In terms of components in Cartesian coordinates, it reads
\begin{equation}
L^a \define \eps\indices{^a_b_c} X^b p^c.
\end{equation}
Newton's second law then implies
\begin{equation}
\ddf{L^a}{t}
= \eps\indices{^a_b_c} v^b p^c + \eps\indices{^a_b_c} X^b F^c
= \eps\indices{^a_b_c} X^b F^c,
\end{equation}
which is sometimes called the \emph{angular momentum theorem}. If the particle undergoes a \emph{central force}, i.e. a force always directed along $\vect{OM}$, then $\eps\indices{^a_b_c}X^b F^c=0$, and its angular momentum is conserved. Furthermore, just like linear momentum, the angular momentum of any isolated system of interacting particles is conserved.

\paragraph{Energy} Consider again an isolated particle. Taking the scalar product of Newton's second law with its momentum, we find that if the mass of the particle is conserved, then its \emph{kinetic energy}~$K$ is conserved,
\begin{equation}
\text{isolated particle:}
\quad
\ddf{K}{t} = 0,
\qquad \text{with} \quad K \define \frac{p^2}{2 m} = \frac{m v^2}{2}.
\end{equation}
Recall that, for arbitrary coordinates, $p^2 = e_{ij} p^i p^j$. So far, there is nothing more than a consequence of the conservation of momentum. Things become more interesting if the particle undergoes \emph{conservative forces}, i.e. forces that derive from a \emph{potential energy}~$U(X^a)$,
\begin{equation}
\vec{F} = -\vec{\nabla} U,
\end{equation}
where the \emph{gradient operator}~$\vec{\nabla}$ has Cartesian components $\partial^a U \define \delta^{ab} \partial_b U$.

\begin{exercise}
The expression of the gradient operator is more subtle with arbitrary coordinates. Assuming that $\vec{\nabla}U$ is a vector, in the sense that it behaves as eq.~\eqref{eq:transformation_vector} under coordinate transformations, show that
\begin{equation}
\partial^i U = e^{ij} \partial_j U,
\end{equation}
and deduce the expression of the gradient in spherical coordinates.
\end{exercise}

When the particle undergoes conservative forces, its kinetic energy is not conserved, but the \emph{total energy} $E\define K+U$ of the particle is conserved,
\begin{equation}\label{eq:cons_energy}
\ddf{E}{t} = \ddf{}{t} \pa{ K + U } = 0.
\end{equation}
This conservation law is then trivially generalised to a system of $N$ particles.

\begin{exercise}
Show that eq.~\eqref{eq:cons_energy} is not satisfied if the potential energy~$U$ explicitly depends on time, and must be replaced by
\begin{equation}
\ddf{E}{t} = \pd{U}{t}.
\end{equation}
\textit{Hint}: What is the time derivative of $U[t,X^a(t)]$? Give an example where this happens.
\end{exercise}

\subsection{Non-inertial frames}
\label{sec:non-inertial_frames}

A non-inertial frame is, by definition, a frame~$\tilde{\mathcal{R}}$ that is accelerated with respect to an inertial frame~$\mathcal{R}$, either because its origin~$\tilde{O}$ has a velocity that is not constant ($v^a_{\tilde{O}}\not=\cst$), or because its axes are rotating~($\Omega^a\not= 0$). When this is the case, Newton's second law does not apply, and fictitious forces appear.

In order to derive the generalised law of dynamics in non-inertial frames, one has to postulate that \emph{the forces applied to an object are frame-independent}. This seems perfectly reasonable in principle---if you are pulling a table, the force that you are producing should not depend on who measures it. Therefore, contrary to velocity and acceleration, the components~$\tilde{F}^b$ of a force in $\tilde{\mathcal{R}}$ are related to its components~$F^a$ in $\mathcal{R}$ as
\begin{equation}
F^a
= \frac{\partial X^a}{\partial \tilde{X}^b} \tilde{F}^b
= R\indices{^a_b} \tilde{F}^b \ .
\end{equation}
%

Applying Newton's second law in $\mathcal{R}$, replacing the expression of the acceleration and of the force in $\tilde{\mathcal{R}}$, and assuming that the mass of the object is constant, we find
\begin{equation}
m \tilde{a}^b = \tilde{F}^b
						\underbrace{
						- m R\indices{_c^b} a^c_{\tilde{O}} 
						- m \eps\indices{^b_c_d} \dot{\Omega}^c \tilde{X}^d
						- m\eps\indices{^b_c_d} \eps\indices{^d_e_f}
									\Omega^c \Omega^e \tilde{X}^f
						- 2m \eps\indices{^b_c_d} \Omega^c \tilde{v}^d
						}_{\text{fictitious forces}},
\end{equation}
where $R\indices{_c^b}=(\transpose{\mat{R}})\indices{^b_c}=(\mat{R}^{-1})\indices{^b_c}$ denote the components of the inverse of the matrix~$\mat{R}$.
The fictitious forces are naturally proportional to the inertial mass~$m$ of the object, as they come from its acceleration and not from any exterior phenomenon.

In the fictitious forces,
\begin{equation}
\tilde{F}^b\e{fic} = - m R\indices{_c^b} a^c_{\tilde{O}} 
						- m \eps\indices{^b_c_d} \dot{\Omega}^c \tilde{X}^d
						- m\eps\indices{^b_c_d} \eps\indices{^d_e_f}
									\Omega^c \Omega^e \tilde{X}^f
						- 2m \eps\indices{^b_c_d} \Omega^c \tilde{v}^d,
\end{equation}
the first term corresponds to the force that pushes one backwards in an accelerating car; the third one is the centrifugal force; and the last one is the so-called Coriolis force, which creates large-scale circular winds on the Earth due to its rotation. It is also the effect responsible for the precession of Foucault's pendulum.

\section{Lagrangian mechanics}
\label{sec:Lagrangian_mechanics}

Newton's second law can be reformulated in various ways. A particularly elegant one was developed at the end of the 18th century by Euler, Lagrange, and Hamilton. Lagrangian mechanics consists in defining a quantity called the \emph{action}, such that among all the possible trajectories that a particle could have between two points, the physical trajectory is the one that extremises the action. This principle turns out to be much more than a mere reformulation: it is the language in which modern physics is written.

\subsection{Euler-Lagrange equation}

As a first step, we show in this section that Newton's second law in arbitrary coordinates can be expressed in terms of the derivatives of a quantity called \emph{Lagrangian}. Such a reformulation, however, is only possible if all the forces applied to the object under study are conservative; we will therefore make this assumption for now on, and call $U$ the total potential energy. The Lagrangian is then defined simply as
\begin{equation}
L \define K - U .
\end{equation}
Note the minus sign in front of $U$, which makes $L$ differ from the total energy~$E=K+U$. The Lagrangian must actually be understood as a function on \emph{phase space}, that is, a function of six variables---position~$x^i$ and velocity~$v^i=\dot{x}^i$,
\begin{empheq}[box=]{equation}
L(t, x^i, \dot{x}^i) = \frac{m}{2} \, e_{ij}(x^k) \,\dot{x}^i \dot{x}^j - U(t,x^i) \ ,
\end{empheq}
where we considered an arbitrary coordinate system~$(x^i)$, and allowed the potential energy~$U$ to explicitly vary with time~$t$. We are now going to show that Newton's second law is equivalent to the \emph{Euler-Lagrange} equation
\begin{empheq}[box=\fbox]{equation}\label{eq:Euler-Lagrange}
\ddf{}{t}\pa{\frac{\partial L}{\partial \dot{x}^i}} - \pd{L}{x^i} = 0 .
\end{empheq}

Let us start with the first term:
\begin{align}
\ddf{}{t}\pa{\frac{\partial L}{\partial \dot{x}^i}}
&= \ddf{}{t} \pa{ m e_{ij} \dot{x}^j } \\
&= e_{ij}\dot{p}^j + \dot{e}_{ij} p^j \\
&= e_{ij} \dot{p}^j + e_{ij,k} v^j p^k ,
\end{align}
where a comma is a short-hand notation for partial derivatives~$e_{ij,k}\define\partial_k e_{ij}$. We can then deal with the second term
\begin{equation}
\pd{L}{x^i} 
= \frac{m}{2} e_{jk,i} \dot{x}^j \dot{x}^k - \partial_i U
= \frac{1}{2} e_{jk,i} v^j p^k - \partial_i U \ .
\end{equation}
Putting everything together, we find
\begin{align}
\ddf{}{t}\pa{\frac{\partial L}{\partial \dot{x}^i}} - \pd{L}{x^i}
&= e_{ij} \dot{p}^j 
	+ \frac{1}{2} \pa{ 2 e_{ij,k} - e_{jk,i} } v^j p^k
	+ \partial_i U
	\label{eq:Euler-Lagrange_calculation_1}\\
&= e_{ij} \dot{p}^j 
	+ \frac{1}{2} \pa{ e_{ij,k} + e_{ik,j}  - e_{jk,i} } v^j p^k
	+ \partial_i U
	\label{eq:Euler-Lagrange_calculation_2}\\
&= e_{il} \pa{ \dot{p}^l + \Gamma\indices{^l_j_k} v^j p^k } + \partial_i U \ .
	\label{eq:Euler-Lagrange_calculation_3}
\end{align}
To go from eq.~\eqref{eq:Euler-Lagrange_calculation_1} to eq.~\eqref{eq:Euler-Lagrange_calculation_2}, we renamed indices that are summed over:
\begin{equation}
e_{ik,j} v^j p^k
= m e_{ik,j} v^j v^k
= m e_{ij,k} v^k v^j
= m e_{ij,k} v^j p^k .
\end{equation}
Inside the parentheses of eq.~\eqref{eq:Euler-Lagrange_calculation_3}, we recognise the covariant time derivative of $p^l$. Multiplying the above expression by the inverse metric, we conclude that the Euler-Lagrange equation is equivalent to
\begin{equation}\label{eq:Newton_law_2_arbitrary_coordinates}
\Ddf{p^i}{t} = - e^{ij} \partial_j U  \ ,
\end{equation}
which is Newton's second law in arbitrary coordinates, when the forces derive from a (possibly time-dependent) potential $U$. Note the advantage of the Euler-Lagrange equation over the standard equation of motion~\eqref{eq:Newton_law_2_arbitrary_coordinates}, in that it directly gives the result in terms of arbitrary coordinates.

\begin{exercise}
Consider a particle with mass $m$ moving on a sphere of radius $R$, and described by spherical coordinates $\theta, \ph$. We assume that the particle is attached with an elastic to the top of the sphere, and submitted to gravity. Its Lagrangian is
\begin{equation}
L = \frac{1}{2} m R^2 
		\pa{ \dot{\theta}^2 + \sin^2\theta\dot{\ph}^2 } 
		- \frac{1}{2} k R^2 \theta^2 - m g R\cos\theta \ ,
\end{equation}
where $k, g$ are two constants. Using the Euler-Lagrange equation, show that the equations of motion of the particle are
\begin{align}
\ddot{\theta} - \cos\theta\sin\theta \, \dot{\ph}^2 
&= -\frac{k}{m} \, \theta + \frac{g}{R}\sin\theta \ , \\
\ddf{}{t}\pa{\sin^2\theta \dot{\ph} }
&= 0 \ .
\end{align}
\end{exercise}

\subsection{Variational calculus}

In order to perform the second step of the reformulation of Newton's second law towards the least action principle of Lagrangian mechanics, we have to introduce the notion of \emph{functional}, and \emph{variational calculus}.

\paragraph{Functionals} A functional~$\mathcal{F}$ is a function of functions, i.e., a function that eats a function and returns a number, assumed here to be real
\begin{equation}
\mathcal{F}:f\mapsto \mathcal{F}[f]\in\mathbb{R} .
\end{equation}
It is customary to denote the argument of functionals in square brackets $[\cdots]$ rather than in round brackets $(\cdots)$. For example, $\mathcal{F}_1$ could be the Dirac distribution, which to a function $x\mapsto f(x)$ associates its value at $x=0$, $\mathcal{F}_1[f]=f(0)$. Another example could be the functional that gives the mean square of a function between $a$ and $b>a$,
\begin{equation}
\mathcal{F}_2[f] = \frac{1}{b-a} \int_a^b f^2(x) \; \dd x .
\end{equation}

\paragraph{Functional derivation} We would like to build a notion of derivative for functionals, by analogy with the partial derivatives of functions of several variables. Suppose for simplicity that $\mathcal{F}[f]$ only depends on the values of $f$ in the interval $[a,b]$. Let us then split the interval $[a,b]$ in $N+1$ equal parts, defining
\begin{equation}
x_n \define \frac{n}{N}\,(b-a),
\end{equation}
so that $x_0=a$, $x_N=b$, and $\Delta x \define x_{n+1}-x_n = (b-a)/N$. The function $f$ can then be seen as the limit~$N\rightarrow\infty$ of a function that is constant on each interval~$[x_n,x_{n+1}]$, with $f_n=f(x_n)$. Therefore, $\mathcal{F}[f]$ can also be seen as a limit
\begin{equation}
\mathcal{F}[f] = \lim{N}{\infty} \mathcal{F}_N(f_0,f_1, f_2, \ldots, f_N),
\end{equation}
where $\mathcal{F}_N$ is not a functional, but simply a function of $N+1$ variables.

Now suppose that we slightly change the function~$f$ to $f+\delta f$. In general, this changes all the $f_n$ to $f_n+\delta f_n=f(x_n)+\delta f(x_n)$. The corresponding variation of $\mathcal{F}_N$ is
\begin{align}
\delta \mathcal{F}_N
&\define \mathcal{F}_N(f_0+\delta f_0,\ldots,f_n+\delta f_n) 
	- \mathcal{F}_N(f_0,\ldots,f_n) \\
&= \sum_{n=0}^{N} \pd{\mathcal{F}_N}{f_n} \, \delta f_n + \mathcal{O}(\delta f^2)\\
&= \sum_{n=0}^{N} \pac{\frac{1}{\Delta x} \pd{\mathcal{F}_N}{f(x_n)} } \delta f(x_n) \; \Delta x + \mathcal{O}(\delta f^2).\label{eq:construction_functional_derivative}
\end{align}
In the last equation, we have simply multiplied and divided by $\Delta x=(b-a)/N$. In the limit~$N\rightarrow\infty$, the sum turns into an integral, and we find
\begin{empheq}[box=\fbox]{equation}
\delta \mathcal{F}
= \int_a^b \frac{\delta \mathcal{F}}{\delta f(x)} \, \delta f(x) \; \dd x
+ \mathcal{O}(\delta f^2) ,
\end{empheq}
where the quantity~$\delta\mathcal{F}/\delta f(x)$ is called the \emph{functional derivative} of $\mathcal{F}$ at $f(x)$. We see that it is the limit of the term in brackets in eq.~\eqref{eq:construction_functional_derivative} as $N\rightarrow\infty$; as such, it must be understood as the generalisation of the notion of partial derivative: $\delta\mathcal{F}/\delta f(x)$ quantifies how much $\mathcal{F}$ varies as the value of $f$ at $x$ changes.

\begin{exercise}
Show that the functional derivatives of the two examples~$\mathcal{F}_1, \mathcal{F}_2$ given in the beginning of this section read
\begin{equation}
\frac{\delta \mathcal{F}_1}{\delta f(x)} = \delta\e{D}(x),
\qquad \text{and} \qquad
\frac{\delta \mathcal{F}_2}{\delta f(x)} = \frac{2f(x)}{b-a} \ ,
\end{equation}
where $\delta\e{D}(x)$ denotes the Dirac ``function''.
\end{exercise}

\subsection{Hamilton's least action principle}

\begin{figure}[h!]
\centering
\input{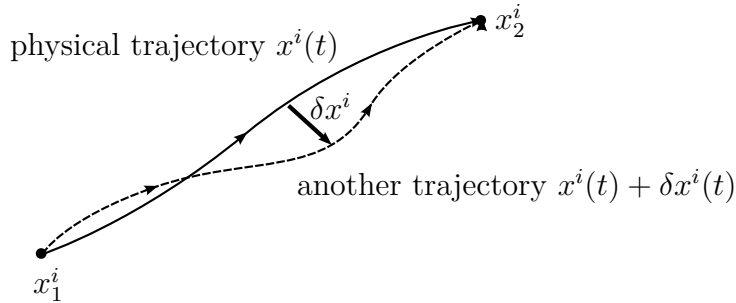}
\caption{The physical trajectory of a particle undergoing conservative forces is the one for which the action~$S$ is stationary.}
\label{fig:variational}
\end{figure}

We are now ready to express Newton's second law in terms of a variational principle. Consider a particle starting from coordinates $x^i_1$ at time $t_1$ and ending at $x^i_2$ at $t_2$. This particle could, in principle, follow any trajectory~$t\mapsto x^i(t)$ that interpolates between those two points (see fig.~\ref{fig:variational}). The \emph{action} of such a trajectory is defined as the integral of its Lagrangian over time,
\begin{equation}\label{eq:def_action}
S[x^i] \define \int_{t_1}^{t_2} L(x^i,\dot{x}^i) \; \dd t ,
\end{equation}
hence $S$ is a functional of the particle's trajectory. We are going to show that Newton's second law, or more precisely the Euler-Lagrange equation~\eqref{eq:Euler-Lagrange}, is equivalent to imposing that the physical trajectory between $(t_1,x^i_1)$ and $(t_2, x^i_2)$ is a stationary point of $S$, that is
\begin{empheq}[box=\fbox]{equation}\label{eq:Hamilton_principle}
\forall t \in [t_1, t_2] \qquad \frac{\delta S}{\delta x^i(t)} = 0.
\end{empheq}
This is known as \emph{Hamilton's least action principle}, because it turns out that this stationary point of $S$ is often a minimum: the physical trajectory minimises the action.

Let us now prove this statement. Consider two very close trajectories~$t\mapsto x^i(t)$ and $t\mapsto x^i(t)+\delta x^i(t)$, which connect at both ends~$(t_1,x^i_1)$ and $(t_2, x^i_2)$, that is $\delta x^i(t_1)=\delta x^i(t_2)=0$. The difference of the actions for those two trajectories is
\begin{align}
\delta S
&= S[x^i+\delta x^i] - S[x^i] \\
&= \int_{t_1}^{t_2} \pac{ L(x^i+\delta x^i,\dot{x}^i+\delta\dot{x}^i) 
											- L(x^i,\dot{x}^i)} \dd t \\
&=  \int_{t_1}^{t_2} \pac{ \pd{L}{x^i} \, \delta x^i 
										+ \frac{\partial L}{\partial \dot{x}^i} \, \delta \dot{x}^i
											}  \dd t \ .									
\end{align}
We can integrate the second term by parts,
\begin{equation}
\int_{t_1}^{t_2} \frac{\partial L}{\partial \dot{x}^i} \, \delta \dot{x}^i \; \dd t
= \int_{t_1}^{t_2} \frac{\partial L}{\partial \dot{x}^i} 
								\, \ddf{\delta x^i}{t} \; \dd t
= \underbrace{
			\pac{ \frac{\partial L}{\partial \dot{x}^i} \, \delta x^i }^{t_2}_{t_1}
			}_{0}
	- \int_{t_1}^{t_2} \ddf{}{t}\pa{\frac{\partial L}{\partial \dot{x}^i} }
		\delta x^i \; \dd t \ ,
\end{equation}
where we used that $\delta x^i(t_1)=\delta x^i(t_2)=0$. Therefore, the variation of the action reads
\begin{equation}
\delta S = \int_{t_1}^{t_2} \pac{ \pd{L}{x^i} 
											- \ddf{}{t}\pa{\frac{\partial L}{\partial \dot{x}^i}} 
												} \delta x^i(t) \; \dd t \ ,
\end{equation}
where we can directly read the functional derivative of $S$,
\begin{empheq}[box=\fbox]{equation}\label{eq:functional_derivative_action}
\frac{\delta S}{\delta x^i(t)}
= \pd{L}{x^i} - \ddf{}{t}\pa{\frac{\partial L}{\partial \dot{x}^i}}.
\end{empheq}
We recognise, in eq.~\eqref{eq:functional_derivative_action} the Euler-Lagrange term, which vanishes for the physical trajectory, as imposed by the laws of mechanics. This finally proves Hamilton's principle~\eqref{eq:Hamilton_principle}.

Note that eq.~\eqref{eq:functional_derivative_action} is true for any functional~$S$ that takes the form of \eqref{eq:def_action}, independently of the expression of the Lagrangian~$L$, provided it only depends on $x^i, \dot{x}^i$. In other words, the Euler-Lagrange equation can be applied to various situations where one has to extremise a functional, and not only in mechanics.

\begin{exercise}
Using variational calculus, show explicitly that the shortest-length curve between two points is a straight line.
\end{exercise}

\begin{exercise}
Consider a functional given by
\begin{equation}
\mathcal{F}[f] \define \int_a^b L(f,f',f'') \; \dd x,
\end{equation}
where the ``Lagrangian''~$L$ depends also on the second derivative of $f$. Show that
\begin{equation}
\frac{\delta \mathcal{F}}{\delta f(x)} = \pd{L}{f} - \ddf{}{x} \pa{\pd{L}{f'}}
																+ \ddf[2]{}{x} \pa{ \pd{L}{f''} },
\end{equation}
assuming that $\delta f$ and $\delta f'$ vanish at both $a$ and $b$. Generalise this to a Lagrangian that depends on the first $n$th derivatives of $f$, with the constraint that $\delta f$ and its first $n-1$ derivatives vanish at $a,b$.
\end{exercise}

\section{Gravitation}

Gravitation is the phenomenon that makes things fall. A key intellectual step was made by understanding that there is a unique cause for the falling of objects when we drop them, and for the orbit of planets in the Solar system. Newton was the first scientist to propose a mathematical description of gravitation that fitted in his formalism for mechanics.

\subsection{Universal gravity law}

The most striking property of gravitation is its universality: \emph{everything} falls, and, furthermore, everything falls \emph{the same way}. This universality of free fall was first emphasised by Galileo, and confirmed by many experiments over the years, in particular by E\"otv\"os in 1922~\cite{Eotvos:1922pb}. In December 2017, the French experiment MICROSCOPE compared the acceleration of cylinders made of Titanium and Platinum under the Earth's gravity, and concluded that they differed by less than two parts in $10^{14}$~\cite{Touboul:2017grn}.

\paragraph{Equivalence principle} The universality of free fall can be summarised as follows. Any object subject to gravity gets the same acceleration
\begin{equation}
\vec{a} = \vec{g},
\end{equation}
where $\vec{g}$ is naturally called the acceleration of gravitation. Multiplying the above relation by the mass $m$ of the object, $m\vec{a}=m\vec{g}$, and comparing with Newton's second law, we conclude that if gravitation is a force, then it has to read~$\vec{F}=m\vec{g}$. We see that the mass~$m$ intervenes here in two very different ways. On the one hand, in $m\vec{a}$, it quantifies inertia; on the other hand, in $m\vec{g}$, is quantifies how much an object feels gravity. Those two notions are sometimes explicitly distinguished by calling the former \emph{inertial mass}~$m\e{in}$, and the latter \emph{passive gravitational mass}~$m\e{pg}$. The universality of free fall is then expressed as the equivalence of those masses,
\begin{equation}
m\e{in} = m\e{pg},
\end{equation}
which is, therefore, called the \emph{equivalence principle}.

\paragraph{Gravitational force} If gravity is an interaction between objects, then it must satisfy Newton's third law of action and reaction. Hence, if an object $1$ exerts on an object $2$ the gravitational force $\vec{F}_{12}=m_2\vec{g}_1$, then $2$ exerts on $1$ the force $\vec{F}_{21}=m_1\vec{g}_2$, with
\begin{equation}
m_2 \vec{g}_1 = -m_1 \vec{g}_2 .
\end{equation}
Since this is true for any couple of objects, we conclude that $\vec{g}_1\propto m_1$ and $\vec{g}_2\propto m_2$, so that $\vec{F}_{12}\propto m_1 m_2$. This displays a third notion of mass, called \emph{active gravitational mass}~$m\e{ag}$, which now quantifies the capacity of objects to generate gravitation, instead of feeling it. The third Newton's law enforces the equality $m\e{ag}=m\e{pg}$.

Consider two objects in an otherwise empty Universe. Since there is no preferred direction apart from the line connecting these objects, the gravitational force between them must be aligned with it. Gravity being attractive, we have~$\vec{F}_{12}\propto - \vec{u}_{12}$, where
\begin{equation}
u^a_{12} = \frac{(X_2^a-X^a_1)}{\sqrt{\delta_{bc}(X_2^b-X_1^b)(X_2^c-X_1^c)}}
\end{equation}
is the unit vector directed from $1$ to $2$.

Finally, for reasons that will be clearer in the next section, for $\vec{F}_{12}$ to be independent from the size of the objects, it has to decrease with the square of the distance~$r$ between them. Therefore, the universal gravitational interaction must read
\begin{empheq}[box=\fbox]{equation}
\vec{F}_{12} = - \frac{G m_1 m_2}{r^2} \, \vec{u}_{12}
\ ,
\end{empheq}
that is, in terms of Cartesian components,
\begin{equation}
F^a_{12} = \frac{G m_1 m_2 \, (X_1^a-X^a_2)}
								{\pac{\delta_{bc}(X_2^b-X_1^b)(X_2^c-X_1^c)}^{3/2}} \ ,
\end{equation}
where $G=6.67408\times 10^{-11}\U{kg^{-1} m^3 s^{-2}}$ is Newton's gravitational constant.

\begin{exercise}
Show that the gravitational force is conservative, by checking that it derives from the potential energy
\begin{equation}
U = -\frac{G m_1 m_2}{r}.
\end{equation}
\end{exercise}

\subsection{Gravitational field}

In the previous paragraph, we introduced gravitation as an interaction between massive bodies. In this approach, the only physical objects are the massive bodies, while gravity is just a relation between them. However, it is possible to formulate an equivalent theory of gravity that is conceptually different. This formulation relies on the notion of \emph{gravitational field}, and consists in promoting the gravitational interaction into a proper physical object. This conceptual shift is comparable to the reformulation of electrostatics to electrodynamics. In the former, there is a force between electric charges; in the latter, there is an electromagnetic field that is affected by the existence and motion of charges, and affects in return the motion of charges.

\paragraph{Introducing the gravitational field} Let a set of $N$ masses~$m_1,\ldots, m_N$ be located at $\vec{X}_1, \ldots, \vec{X}_N$. Consider another mass~$m$ at $\vec{X}$; this mass feels the gravitational attraction of all the others
\begin{equation}
\vec{F}
= \sum_{n=1}^N \vec{F}_n
= - \sum_{n=1}^N \frac{G m m_n}{||\vec{X}-\vec{X}_n||^2}
= m \vec{g}(\vec{X}) \ ,
\end{equation}
where $\vec{g}$ is the gravitational field created by all the $N$ masses,
\begin{equation}\label{eq:grav_field_N_masses}
\vec{g}(\vec{X}) \define - \sum_{n=1}^N \frac{G m_n}{||\vec{X}-\vec{X}_n||^2} \ .
\end{equation}
The point of the notion of gravitational field is that it can be considered to exist independently of the mass~$m$ that may feel it. Similarly, one can introduce the \emph{gravitational potential}~$\Phi$, such that the potential energy of the mass~$m$ reads~$U= m\Phi$,
\begin{equation}\label{eq:grav_potential_N_masses}
\Phi(\vec{X}) = -\sum_{n=1}^N \frac{G m_n}{||\vec{X}-\vec{X}_n||} \ ,
\end{equation}
and we have the relation
\begin{equation}
\vec{g} = -\vec{\nabla} \Phi,
\end{equation}
that is $g^a = -\delta^{ab}\partial_b \Phi$, or $g^i = - e^{ij} \partial_j \Phi$ with an arbitrary coordinate system.

It is quite straightforward to generalise the expressions~\eqref{eq:grav_field_N_masses} and \eqref{eq:grav_potential_N_masses} for a continuous distribution of mass. If there is an amount of mass~$\dd m=\rho(\vec{Y}) \dd^3Y$ in the infinitesimal volume~$\dd^3Y$ about~$\vec{Y}$, where $\rho$ denotes the density field, then discrete sums can be turned into integrals, and we obtain
\begin{align}
\Phi(\vec{X})
&= -G \int_{\mathbb{R}^3} \frac{1}{||\vec{X}-\vec{Y}||}
				\; \rho(\vec{Y}) \,\dd^3Y,
\label{eq:grav_potential_density_field}\\
\vec{g}(\vec{X})
&= -G \int_{\mathbb{R}^3} \frac{\vec{X}-\vec{Y}}{||\vec{X}-\vec{Y}||^3}
																	\; \rho(\vec{Y})\,\dd^3Y	.
\label{eq:grav_field_density_field}			
\end{align}

\begin{exercise}
Check that eq.~\eqref{eq:grav_field_density_field} can be obtained from eq.~\eqref{eq:grav_potential_density_field} via $\vec{g}=-\vec{\nabla}\Phi$.
\end{exercise}

\paragraph{Poisson equation} Equation~\eqref{eq:grav_potential_density_field} can be seen as the solution of a second-order differential equation, called~\emph{Poisson equation},
\begin{empheq}[box=\fbox]{equation}\label{eq:Poisson}
\Delta \Phi = 4\pi G \rho,
\end{empheq}
where $\Delta$ denotes the Laplacian operator. It is defined as the divergence of the gradient, $\Delta \Phi \define \vec{\nabla}\cdot\vec{\nabla}\Phi$. In Cartesian coordinates, it is reads
\begin{equation}
\Delta \Phi = \delta^{ab} \partial_a \partial_b \Phi.
\end{equation}
The counterpart of eq.~\eqref{eq:grav_potential_density_field} with arbitrary coordinates is more complicated, as one would have to replace Cartesian distances by integrals involving the metric. However, the Poisson equation remains the same, except that the expression of the Laplacian is slightly different. Namely, since the divergence acts on a vector (the gradient), the simple partial derivatives must be replaced by covariant derivatives. For reasons that will become clearer in the next chapter, the result is
\begin{equation}
\Delta \Phi 
= e^{ij} \pa{\partial_i \partial_j \Phi - \Gamma\indices{^k_i_j} \partial_k\Phi}.
\end{equation}

\begin{exercise}
Solve the Poisson equation~\eqref{eq:Poisson} using a Green-function technique, and conclude that eq.~\eqref{eq:grav_potential_density_field} is indeed its solution.
\end{exercise}

\paragraph{Gauss's law} One can also write the Poisson equation~\eqref{eq:Poisson} in terms of the gravitational field, replacing~$\Delta \Phi =\vec{\nabla} \cdot \vec{\nabla}\Phi = -\vec{\nabla}\cdot \vec{g}$, which yields
\begin{equation}\label{eq:div_grav_field}
\vec{\nabla} \cdot \vec{g} = -4\pi G \rho .
\end{equation}

\begin{figure}[h!]
\centering
\input{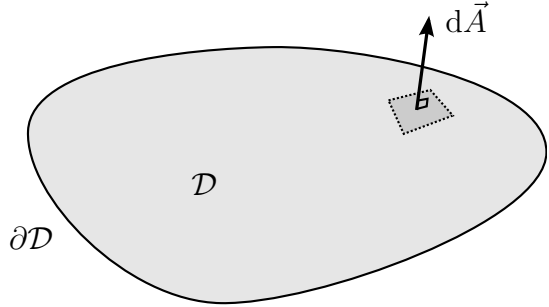}
\caption{A domain~$\mathcal{D}$, its boundary~$\partial\mathcal{D}$, and the normal area element vector~$\dd\vec{A}$.}
\label{fig:domain}
\end{figure}

Consider a closed domain~$\mathcal{D}$ of space. If we integrate eq.~\eqref{eq:div_grav_field} over this domain, the right-hand side is proportional to the total mass contained in $\mathcal{D}$,
\begin{equation}\label{eq:total_mass_D}
\int_{\mathcal{D}} \rho \; \dd V = M_{\mathcal{D}},
\end{equation}
where $\dd V$ denotes the infinitesimal element of volume. In Cartesian coordinates, it reads~$\dd V = \dd^3X \define \dd X \dd Y \dd Z$. With arbitrary coordinates, it involves the metric as
\begin{equation}
\dd V 
= \sqrt{\det \mat{e}} \, \dd^3 \vec{x} 
= \sqrt{\det \mat{e}} \, \dd x^1 \dd x^2 \dd x^3,
\end{equation}
where $\det \mat{e}$ denotes the determinant of the metric~$\mat{e}=[e_{ij}]$, seen as a matrix,
\begin{equation}
\det\mat{e} = \frac{1}{3!} \, \eps^{ijk} \eps^{lmn} e_{il} e_{jm} e_{kn}.
\end{equation}
\begin{exercise}
Show that, in spherical coordinates, $\dd V=r^2 \sin\theta \, \dd r \, \dd\theta \, \dd\ph$.
\end{exercise}
Besides, the left-hand side of eq.~\eqref{eq:div_grav_field}, once integrated over~$\mathcal{D}$, can be rewritten thanks to the Green-Ostrogradski divergence theorem,
\begin{equation}\label{eq:Green-Ostrogradski}
\int_{\mathcal{D}} \vec{\nabla} \cdot \vec{g} \; \dd V
= \int_{\partial \mathcal{D}} \vec{g} \cdot \dd\vec{A} \ ,
\end{equation}
where $\partial\mathcal{D}$ denotes the boundary of~$\mathcal{D}$, and $\dd \vec{A}$ is a vector that is locally normal to $\partial\mathcal{D}$, and whose norm is an infinitesimal area element of~$\partial\mathcal{D}$ (see fig.~\ref{fig:domain}). Just like the volume element~$\dd V$ in arbitrary coordinates, $\dd S$ is given by the determinant of the metric on~$\partial\mathcal{D}$. The right-hand side of eq.~\eqref{eq:Green-Ostrogradski} is called the flux of $\vec{g}$ through the surface~$\partial\mathcal{D}$. Combining eqs.~\eqref{eq:total_mass_D} and \eqref{eq:Green-Ostrogradski}, we finally find \emph{Gauss's law}
\begin{equation}
\int_{\partial \mathcal{D}} \vec{g} \cdot \dd\vec{A} = -4\pi G M_{\mathcal{D}} \ .
\end{equation}

\begin{exercise}
An important special case is when the distribution of mass is spherically symmetric. In spherical coordinates, this corresponds to~$\rho(r,\theta,\ph)=\rho(r)$. Argue that, in this case, the gravitational field~$\vec{g}$ is such that~$g^i = g(r) \delta^i_r$, and show that
\begin{equation}
g(r) = - \frac{G m(r)}{r^2} \ ,
\end{equation}
where $m(r)$ is the mass contained in the ball centred on $O$ and with radius~$r$. Is there a difference between the gravitational field generated by a ball of radius~$R<r$ and a point mass at $O$ with the same mass?
\end{exercise}


\subsection{Lagrangian formulation of Newton's gravity}
\label{sec:Lagrangian_Newton_gravity}

Just like Newton's second law, Poisson's equation can be reformulated as the consequence of a least action principle, similarly to what we have seen in \S~\ref{sec:Lagrangian_mechanics}. For the dynamics of a particle, the action~$S$ is stationary when the trajectory between two points is the physical trajectory of the particle, as determined by the equation of motion. In the case of gravitation, the action is stationary when the gravitational potential~$\Phi$ satisfies the Poisson equation~\eqref{eq:Poisson}.

\paragraph{Lagrangian density} As in \S~\ref{sec:Lagrangian_mechanics}, we proceed in two steps. We first define the \emph{Lagrangian density} of the gravitational field as
\begin{equation}
\mathcal{L}(\Phi,\vec{\nabla}\Phi)
\define -\frac{g^2}{2} - 4\pi G \rho (1+\Phi)
= -\frac{1}{2} \, \delta^{ab}\partial_a\Phi \partial_b\Phi 
	- 4\pi G\rho (1+\Phi),
\end{equation}
where we used Cartesian coordinates for simplicity; the calculation can also be done with arbitrary coordinates, but it is slightly more involved. From the above, it is straightforward to check that
\begin{equation}\label{eq:Euler-Lagrange_Poisson}
\partial_a \pac{\frac{\partial\mathcal{L}}
						{\partial(\partial_a\Phi)} }
- \frac{\partial \mathcal{L}}{\partial\Phi}
= -\Delta \Phi + 4\pi G \rho .
\end{equation}
so that Poisson's equation~\eqref{eq:Poisson} is equivalent to the Euler-Lagrange equation
\begin{equation}\label{eq:Euler-Lagrange_Newtonian_gravity}
\partial_a
 \pac{\frac{\partial\mathcal{L}}{\partial(\partial_a\Phi)} }
- \frac{\partial \mathcal{L}}{\partial\Phi}
= 0.
\end{equation}
Note the similarity with eq.~\eqref{eq:Euler-Lagrange} seen in \S~\ref{sec:Lagrangian_mechanics}. The difference, here, is that the trajectory~$x^i(t)$ is replaced with the Newtonian potential~$\Phi(X^a)$, and the time derivative~$\dd/\dd t$ is replaced with partial derivatives~$\partial_a$. Apart from those replacements, the structure of the Euler-Lagrange equation is the same.

\paragraph{Action of gravitation} Just like the action of classical mechanics is the time integral of the Lagrangian~$L$, the action of Newtonian gravitation is the spatial integral of the Lagrangian density~$\mathcal{L}$. More precisely, if $\mathcal{D}$ is a spatial domain, we define
\begin{equation}
S[\Phi] \define \int_{\mathcal{D}} \mathcal{L}(\Phi, \vec{\nabla}\Phi) \; \dd V,
\end{equation}
which is a functional of $\Phi$. We are now going to show that the Euler-Lagrange equation~\eqref{eq:Euler-Lagrange_Newtonian_gravity} is equivalent to imposing that $S$ is stationary.

Consider a variation~$\delta\Phi$ of the field, such that $\delta\Phi$ vanishes on the boundary~$\partial\mathcal{D}$ of $\mathcal{D}$. This requirement is similar to the~$\delta x^i(t_1)=\delta x^i(t_2)$ imposed in \S~\ref{sec:Lagrangian_mechanics}. The variation of the action implied by the variation of the field reads
\begin{equation}
\delta S
= \int_{\mathcal{D}}
		\pac{ \pd{\mathcal{L}}{\Phi} \, \delta \Phi 
				+ \pd{\mathcal{L}}{(\partial_a \Phi)}\, \partial_a \delta \Phi 
			} \, \dd V + \mathcal{O}(\delta\Phi^2) .
\end{equation}
The second term can be integrated by parts, as
\begin{align}
\int_{\mathcal{D}} \pd{\mathcal{L}}{(\partial_a \Phi)}
							\, \partial_a \delta \Phi \; \dd V
&= \int_{\mathcal{D}} \partial_a 
		\pac{ \pd{\mathcal{L}}{(\partial_a \Phi)} \,\delta\Phi } \, \dd V
	- \int_{\mathcal{D}}
		 \partial_a\pac{ \pd{\mathcal{L}}{(\partial_a \Phi)} }\delta\Phi \;\dd V \\
&= \int_{\partial\mathcal{D}} \pd{\mathcal{L}}{(\partial_a\Phi)} \,\delta\Phi 
			\; \dd A^a
	- \int_{\mathcal{D}}
		\partial_a\pac{ \pd{\mathcal{L}}{(\partial_a \Phi)} }\delta\Phi \;\dd V \\
&= - \int_{\mathcal{D}}
	\partial_a\pac{ \pd{\mathcal{L}}{(\partial_a \Phi)} }\delta\Phi \;\dd V \ ,
\end{align}
where we used the divergence theorem to get the second line, and $\delta\Phi|_{\partial\mathcal{D}}=0$ to get the third line. Therefore, we have obtained
\begin{equation}
\delta S
= \int_{\mathcal{D}}
	\underbrace{
					\paac{ \pd{\mathcal{L}}{\Phi}
								- \partial_a \pac{
															\pd{\mathcal{L}}{(\partial_a \Phi)}
															}
							}
					}_{\define \delta S/\delta \Phi}
	\delta \Phi \; \dd V + \mathcal{O}(\delta\Phi^2),
\end{equation}
and hence, combining with eq.~\eqref{eq:Euler-Lagrange_Poisson},
\begin{empheq}[box=\fbox]{equation}
\frac{\delta S}{\delta \Phi}
= \Delta\Phi - 4\pi G \rho.
\end{empheq}
Poisson's equation is thus equivalent to an action principle.

\section{Application to the Solar System}

Newton's theory has been very successful at explaining the dynamics of the Solar System. In this last section, we analyse its simplest aspects, namely the orbit of planets and tides.

\subsection{Orbits of planets}
\label{subsec:orbits_planets}

We consider here the simplified situation of a single planet $P$ orbiting around the Sun, i.e. we neglect the effect of the other planets on the system. Moreover, since the mass~$m$ of the planet is much smaller than the mass~$M$ of the Sun, we will neglect the effect of the planet on the Sun's motion, and assume that the heliocentric reference frame is \emph{inertial}.

\paragraph{Conservation of angular momentum} Let us pick the origin~$O$ of the coordinate system at the centre of the Sun. As the gravitational force of the Sun is \emph{central}, that is $\vec{F}\propto \vect{OP}$, we have seen in \S~\ref{sec:conserved_quantities} that the planet's angular momentum is conserved,
\begin{equation}
\vec{L} = \vect{OP} \times \vec{p} = \vect{\cst}.
\end{equation}
As a consequence, at any stage of the planet's motion, the vectors~$\vect{OP}$ and $\vec{p}$ belong to a unique plane, called \emph{ecliptic plane}, defined as the plane orthogonal to $\vec{L}$ and containing $O$. The trajectory of the planet thus belongs to this plane. In the following, we set the axes of the coordinate system such that the $Z$-axis is aligned with $\vec{L}$, then the trajectory satisfies $Z=0$, or $\theta=\pi/2$ in spherical coordinates.

\begin{exercise}
Show that the angular momentum reads
\begin{equation}
L^Z = -r L^\theta = m r^2 \dot{\ph}.
\end{equation}
Beware! For non-Cartesian coordinates the calculation of cross product is subtle. For two vectors~$\vec{u}, \vec{v}$ with components~$u^i, v^i$, we have
\begin{equation}
(\vec{u}\times\vec{v})^k
= \eps\indices{_i_j^k} u^i v^j
= \det(\mat{e}) \, e^{kl} [ijk] u^i v^j
\end{equation}
where $\det(\mat{e})$ is the determinant of $[e_{ij}]$, seen as a matrix, while $[ijk]$ denotes the permutation symbol, equal to $1$ if $(ijk)$ is an even permutation of $(123)$, $-1$ for an odd permutation, and $0$ otherwise. Finally, note that the spherical components of $\vect{OP}$ are simply $(r,0,0)$.
\end{exercise}

\begin{figure}[h!]
\centering
\input{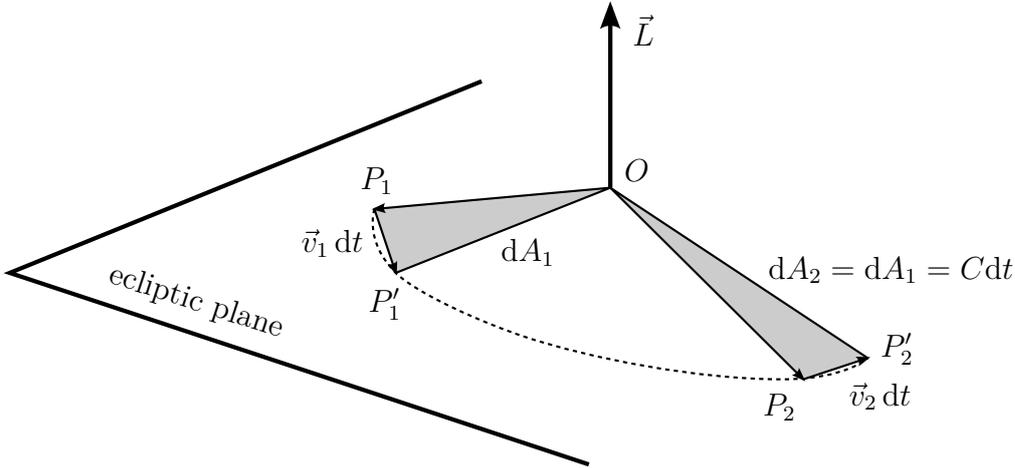}
\caption{Conservation of angular momentum and second Kepler's law.}
\label{fig:Kepler_2}
\end{figure}

An interesting consequence of the conservation of angular momentum is known as the second Kepler's law, and states that the area spanned by the segment $OP$ per unit time is always the same during the planet's motion (see fig.~\ref{fig:Kepler_2}). This can be explained as follows. Between $t$ and $t+\dd t$, the planet moves from $P$ to $P'$, and the area of the triangle~$OPP'$ is by definition
\begin{equation}
\dd A
= \frac{1}{2} \norm[2]{ \vect{OP} \times \vect{PP'} }
= \frac{1}{2} \norm[2]{ \vect{OP} \times \vec{v} \, \dd t }
= \frac{\norm[1]{\vec{L}}}{2 m} \, \dd t,
\end{equation}
and hence
\begin{equation}
\ddf{A}{t} = \frac{\norm[1]{\vec{L}}}{2 m} \define C  = \cst.
\end{equation}

\paragraph{Elliptical trajectory} Using the expression of the acceleration of the planet in spherical coordinates established in exercise~\ref{ex:acceleration_spherical}, with $\theta=\pi/2$, we find that the $r$-component of the planet's equation of motion reads
\begin{equation}
a^r = \ddot{r} - r \dot{\ph}^2 = - \frac{GM}{r^2}.
\end{equation}
Furthermore, we can substitute the constant $C=||\vec{L}||/(2m)=r^2\dot{\ph}/2$, which yields
\begin{equation}\label{eq:eom_Kepler_r}
\ddot{r} - \frac{4C^2}{r^3} = - \frac{GM}{r^2},
\end{equation}
that is a differential equation on the component $r$ only.

\begin{exercise}
Introducing Binet's variable~$u=1/r$, and parametrising the equation of motion with the angular component~$\ph$ instead of time~$t$, show that eq.~\eqref{eq:eom_Kepler_r} becomes
\begin{equation}\label{eq:eom_Kepler_u}
\ddf[2]{u}{\ph} + u = \frac{GM}{4 C^2}.
\end{equation}
\end{exercise}

The equation of motion~\eqref{eq:eom_Kepler_u} is much easier to solve than eq.~\eqref{eq:eom_Kepler_r}. With a suitable choice of the origin~$\ph=0$ of the polar angle, the solution reads
\begin{equation}
r(\ph) = \frac{1}{u(\ph)} = \frac{p}{1+e\cos\ph},
\end{equation}
which is the polar equation of a conic section (ellipse, parabola, or hyperbola) whose $O$ is a focus, with parameter~$p=4C^2/GM$ and eccentricity $e=p/r_0-1$. For planets, $e<1$, and the trajectory is therefore elliptical. This is known as the first Kepler's law, who established it empirically in 1608, along with the area law.

\paragraph{Third Kepler's law} Combining elliptical trajectories with the conservation of angular momentum leads to an interesting relation between the semi-major axis~$a$ of the orbit of planets and their sidereal period~$T$ (duration of one orbit). Namely, the ratio~$a^3/T^2$ is identical for all the planets of the Solar System. This observation was first established empirically by Kepler in 1618, and explained by Newton in 1687.

The proof is the following. Integrating the second Kepler's law~$\dd A/\dd t=C$ over a period~$T$ of the orbit, we first get
\begin{equation}\label{eq:Kepler_2_integrated}
\frac{\pi a b}{T} = C,
\end{equation}
where $a$ and $b$ are respectively the semi-major and semi-minor axes of the orbit.
\begin{exercise}
Show that the semi-major and semi-minor axes of an ellipse are related to its parameter via~$p=b^2/a$.
\end{exercise}
Then, combining this geometrical property with the expression~$p=4C^2/GM$ of the parameter, and with the square of eq.~\eqref{eq:Kepler_2_integrated}, we can eliminate $C$ and find
\begin{equation}
\frac{a^3}{T^2} = \frac{GM}{4\pi^2}.
\end{equation}
This ratio only depends on Newton's constant and the mass of the Sun, it is therefore the same for all the planets of the Solar System, which explains Kepler's third law.

\subsection{Tides}
\label{subsec:tides}

\paragraph{Removing gravity?} A very interesting property of the gravitational force, which will turn out to be crucial in the next chapter, is that it vanishes in a freely falling reference frame. For example, if you were in an elevator whose suspensions are cut, so that the elevator would fall freely in the gravitational field of the Earth, then you would feel as if there were no gravity at all. This is a direct consequence of the universality of free fall: the elevator and yourself undergo the same acceleration~$\vec{g}$ due to gravitation, and hence your relative motion discards gravity. Alternatively, in the elevator's frame, you feel a fictitious force
\begin{equation}
\vec{F}\e{fic} = - m\vec{a}\e{elev} = -m\vec{g} = - \vec{F}\e{grav}
\end{equation}
which exactly compensates the gravitational force.

In a similar manner, on Earth, we do not actually feel the gravitational attraction of the Sun (or the Moon), because the Earth itself is accelerated towards it as we are, and the resulting fictitious force exactly cancels the effect of Solar gravity. Well, in fact, not \emph{exactly}. There remains an effect due to the fact that the gravitational field of the celestial bodies is not homogeneous, and which is responsible for \emph{tides}.

\paragraph{Tidal field} Let us first consider the \{Sun, Earth\} system, leaving the Moon and the other celestial bodies aside for simplicity. Let an object~$M$ be on the surface of the Earth. In the geocentric frame, the sum of all forces applied to this object reads
\begin{equation}\label{eq:total_force_on_Earth}
\vec{F}\e{tot} = \vec{F}_\oplus + \vec{F}_\odot + \vec{F}\e{fic} + \vec{F}\e{other},
\end{equation}
where $\vec{F}_\oplus$ and $\vec{F}_\odot$ are the gravitational forces due to the Earth and the Sun,\footnote{$\oplus$ is the astronomical symbol of the Earth, while $\odot$ is the symbol of the Sun. All the planets of the Solar System have such a symbol, for example~$\mercury$ is Mercury, $\venus$ is Venus, and $\mars$ is Mars.} respectively; $\vec{F}\e{fic}$ are the fictitious forces due to the non-Galilean character of the geocentric reference frame; and $\vec{F}\e{other}$ regroups the other non-gravitational forces, like the reaction of the ground on the object, etc.

\begin{figure}[h!]
\centering
\input{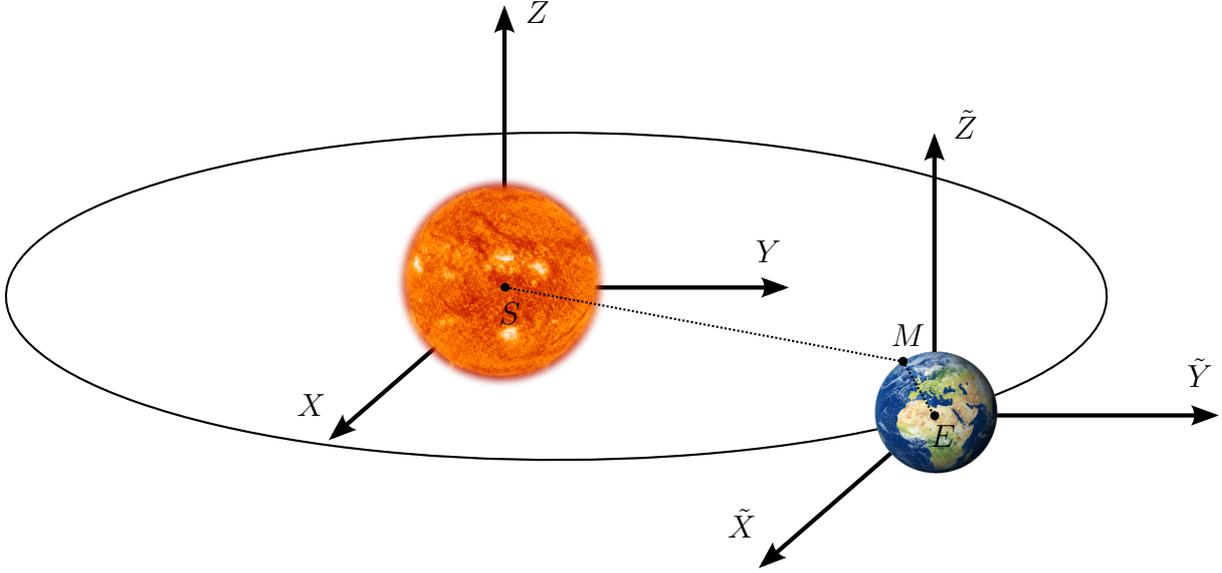}
\caption{Coordinates $(X^a)$ and $(\tilde{X}^a)$ of a point~$M$ at the surface of the Earth, in the heliocentric and geocentric frames.}
\label{fig:geocentric}
\end{figure}

Let us focus on the second and third terms, namely $\vec{F}_\odot + \vec{F}\e{fic}$. Assuming that the heliocentric frame~$\mathcal{R}_\odot$ is inertial, the only cause of non-inertiality of the geocentric frame~$\mathcal{R}_\oplus$ is the revolution of the Earth around the Sun. Recall that the geocentric frame is defined as the frame whose origin coincides with Earth's centre of mass, $E$, while its axes keep parallel to the axes of the heliocentric frame, thus
\begin{equation}
X^a = X^a_E + \tilde{X}^a ,
\end{equation}
where $(X^a)$ are the coordinates of $M$ in $\mathcal{R}_\odot$ while $(\tilde{X}^a)$ are its coordinates in $\mathcal{R}_\oplus$, as depicted in fig.~\ref{fig:geocentric}. In particular, there is no rotation, $\Omega^a=0$, between those frame. The fictitious forces derived in \S~\ref{sec:non-inertial_frames} then reduce to
\begin{equation}
\vec{F}\e{fic} = -m \vec{a}_E \ ,
\end{equation}
where $\vec{a}_E$ is the acceleration of $E$ in the heliocentric frame, and $m$ the mass of the object. Since $\vec{a}_E=\vec{g}_\odot(E)$, we have
\begin{equation}
\vec{F}_\odot + \vec{F}\e{fic}
= m \pac{ \vec{g}_\odot(M) - \vec{g}_\odot(E)}.
\end{equation}
If $M$ were at the Earth's centre of mass, then the above would be zero. Instead, here, there is a residual force~$m\vec{\gamma}_\odot$, with
\begin{align}
\gamma_\odot^a
&\define g^a_\odot(X^b) - g^a_\odot(X^b_E) \\
&= \tilde{X}^b \partial_b g^a_\odot(E) 
		+ \mathcal{O}\pa{|\tilde{X}^b|/D}^2 \\
&= - \tilde{X}^b  \partial_b \partial_a \Phi_\odot(E)
		+ \mathcal{O}\pa{|\tilde{X}^b|/D}^2,
\end{align}
where $D$ is the distance between the centres of the Earth and the Sun. The quantity~$\mat{T}_\odot$ with components $T_{ab}^\odot\define-\partial_a \partial_b \Phi_\odot(E)$ is called the \emph{tidal tensor} of the Sun at $E$, and $\vec{\gamma}_\odot$ is the associated \emph{tidal acceleration} exerted on the object.

\begin{exercise}
Show that the tidal tensor of the Sun on the Earth reads
\begin{equation}\label{eq:tidal_field}
T_{ab}^\odot = -\frac{GM_\odot}{D^3}
							\pa{ \delta_{ab} - 3 u_a u_b },
\end{equation}
where $D=|\vect{SE}|$ is the distance between the centre of the Earth~$E$ and the centre of the Sun~$S$, and $\vec{u}\define\vect{SE}/D$ is the unit vector in the direction of $\vect{SE}$. Note that the position of indices $a,b$ in eq.~\eqref{eq:tidal_field} does not matter, $u_a=\delta_{ab} u^b=u^a$.
\end{exercise}

\begin{figure}[h!]
\centering
\input{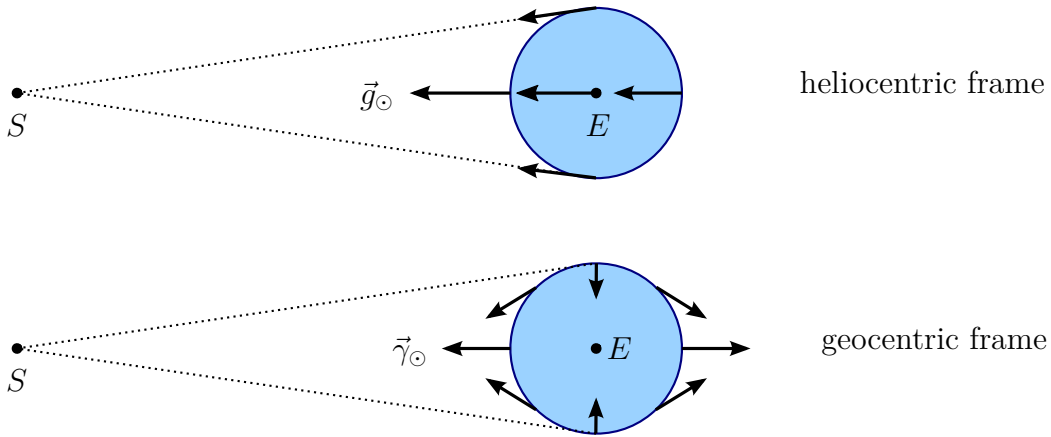}
\caption{Top: gravitational field~$\vec{g}_\odot$ generated of the Sun at different points of the Earth. Bottom: tidal acceleration field~$\vec{\gamma}_\odot\define \vec{g}_\odot-\vec{g}_\odot(E)$ at different points of the Earth.}
\label{fig:tidal_field}
\end{figure}

From the expression~\eqref{eq:tidal_field} of $T_{ab}^\odot$, we conclude that the tidal acceleration is
\begin{align}
\gamma^a_\odot &= -\frac{G M_\odot}{D^3}
									\pac{ \tilde{X}^a - 3 (u_b \tilde{X}^b) u^a }, \\
\text{i.e.} \quad
\vec{\gamma}_\odot &= -\frac{G M_\odot}{D^3}
					\pac{ \vec{\tilde{X}} - 3 \pa{\vec{u}\cdot\vec{\tilde{X}}}\vec{u} }.
\end{align}
The resulting acceleration field is depicted in the bottom panel of fig.~\ref{fig:tidal_field}. We see that it tends to elongate the Earth in the direction of the Sun, and to compress it in the orthogonal direction. This residual gravitational acceleration is responsible for slight deformations of the Earth's shape, but also for oceanic tides. Indeed, the mass of the oceans is more easily deformed by the tidal field than the ground.

\paragraph{Generalisation} It is easy to see that all the celestial bodies~$B$ of the Solar System---actually, of the entire Universe---generate a tidal field on the Earth. Indeed, we could have added to eq.~\eqref{eq:total_force_on_Earth} the gravitational force due to each body, and have combined it with the fictitious force that it also generates in the geocentric frame. The total tidal field on Earth is
\begin{equation}
\vec{\gamma}
= \sum_{B} \vec{\gamma}_B
= \sum_{B}	-\frac{G M_B}{D_{EB}^3}
	\pac{ \vec{\tilde{X}} - 3 \pa{\vec{u}_B\cdot\vec{\tilde{X}}}\vec{u}_B }.
\end{equation}
The amplitude of the tidal effect due to the body $B$ is set by the ratio~$G M_B/D_{EB}^3$, where $D_{EB}$ is the distance between the centre of the Earth and the centre of the body~$B$. The largest effect is actually due to the Moon; the second largest is due to the Sun, with approximately half the amplitude of the Moon's effect, while the effect of the other planets is essentially negligible.

\section*{Epilogue: when Newtonian physics fails}
\addcontentsline{toc}{section}{Epilogue: when Newtonian physics fails}

\paragraph{Precession of Mercury's perihelion} The laws of Newtonian mechanics and gravitation were very successful at explaining the observations of the Solar System, and astronomy in general, for more than two centuries. Only one measurement was in slight disagreement with its prediction: the precession rate of the orbit of Mercury.

Like the other planets of the Solar System, the axes of the elliptical trajectory of Mercury slowly rotate with time, with an angular velocity of $5600\U{arcsec/century}$. This is known as the \emph{precession of Mercury's perihelion}. Most of it ($5020\U{arcsec/century}$) is due to the fact that the Sun is not completely spherical, which affects the gravitational field that it generates. There is also the effect of the other planets of the Solar System (mostly Venus, Jupiter, and the Earth), responsible for $531\U{arcsec/century}$. But once those effects are taken into account, there are still $43\U{arcsec/century}$ that remain unexplained by Newtonian physics. This observation required Einstein’s theory of relativity to be fully understood.

\paragraph{If it had been measured in the past...} There are also facts that, if they had been observed in the past, would have disagreed with Newtonian physics. These include:
\begin{itemize}
\item Motion and interaction effectively change the mass of objects: a hot gas is heavier than a cold gas; a rotating gyroscope is heavier than a steady gyroscope; the set of two electrons gets heavier as they are closer. These cannot be explained by Newton's physics, where the mass of a system only depends on the amount of matter that constitutes it.
\item Light falls and attracts other objects, even though is has no mass.
\item Finally, time and distances are observer-dependent notions. Specifically, time ``slows down'' for observers who are moving, or who experience stronger gravitational fields.
\end{itemize}

The above facts represent the major differences between Newtonian gravitation and Einsteinian gravitation, which is the focus of the next chapter: the source of gravitation is not really mass, but rather any form of energy; and gravitation is not really a force, but rather a distortion of the geometry of space and time.

\chapter{Einstein's theory of relativity}
\label{chap:relativity}

\lettrine{I}{n} 1905, Einstein published three articles that dramatically changed our conception of physics. One of them introduced the \emph{special theory of relativity}~\cite{Einstein1905}, a new vision of space and time. It became the \emph{general theory of relativity}~\cite{Einstein:1915by} ten years later, in 1915, with the inclusion of gravity in this new framework. Although it is not the reason why Einstein earned a Nobel Prize, relativity is certainly the greatest achievement of his scientific career and, in my opinion, the most remarkable of all theories of physics. 

\minitoc
\newpage

\section{Space-time}

The first important conceptual step in the construction of the theory of relativity is the unification of the notions of time and space in a single, four-dimensional entity, called space-time. This section introduces the fundamentals of kinematics in four dimensions.

\subsection{Separation of two events}

Let $A, B$ be two events, respectively happening at times~$T_A, T_B$, and located at $(X_A,Y_A,Z_A)$, $(X_B, Y_B, Z_B)$ in a Cartesian coordinate system of an inertial frame\footnote{The importance of this assumption will be clearer in the following.}. Similarly to how we defined the Euclidean distance~$d_{AB}$, we introduce, as a postulate, the \emph{space-time separation} between those events as
\begin{align}
\Delta s^2_{AB}
&\define - c^2 (T_B-T_A)^2 + (X_B-X_A)^2 + (Y_B-Y_A)^2 + (Z_B-Z_A)^2 \\
&\define \eta_{\alpha\beta} (X^\alpha_B-X^\alpha_A)(X^\beta_B - X^\beta_A) \ ,
\end{align}
where $c$ denotes the speed of light. In the second line, we introduced new notation: Greek indices, contrary to Latin indices, are running from $0$ to $3$, $X^0\define cT$ being the temporal component of the four-dimensional coordinates of an event,
\begin{equation}
(X^\alpha) \define (X^0, X^a) = (cT, X^a) \ .
\end{equation}
Besides, the quantity $\eta_{\alpha\beta}$ is a particular 4-dimensional extension of the Kr\"onecker symbol, which can be written under a matrix form as
\begin{equation}
[\eta_{\alpha\beta}] =
\begin{bmatrix}
-1 & 0 & 0 & 0 \\
0 & 1 & 0 & 0 \\
0 & 0 & 1 & 0 \\
0 & 0 & 0 & 1
\end{bmatrix},
\qquad \text{that is} \quad
\eta_{\alpha\beta} =
\begin{cases}
-1 & \text{if } \alpha=\beta=0, \\
1 & \text{if } \alpha=\beta > 0, \\
0 & \text{if } \alpha\not=\beta. 
\end{cases}
\end{equation}

Note that, despite the ${}^2$ superscript, $\Delta s^2_{AB}$ is not necessarily a positive quantity. More precisely, the separation of the events $A$ and $B$ is said to be:
\begin{itemize}
\item \emph{Time-like} if $\Delta s^2_{AB}<0$, that is, if $c^2(T_B-T_A)^2> d_{AB}^2$. We will see, in \S~\ref{sec:physics_four_dimensions}, that such events can then be causally related, because information can travel from, say, $A$ to $B$ (assuming $T_A< T_B$) at a speed lower than the speed of light,
\begin{equation}
\frac{d_{AB}^2}{(T_B-T_A)^2} < c^2 \ .
\end{equation}
For instance, two events happening at the same place but at different times are separated by a time-like interval.
\item \emph{Null}, or sometimes \emph{light-like}, if $\Delta s^2_{AB}=0$. This typically corresponds to the case where $A$, for example, is the emission of a photon, and $B$ is its reception.
\item \emph{Space-like} if $\Delta s^2_{AB}>0$. In this case $A$ and $B$ cannot be causally related, because information should travel faster than light from $A$ to $B$. For example, two events happening simultaneously at different places are separated by a space-like interval.
\end{itemize}

\begin{figure}[h!]
\centering
\input{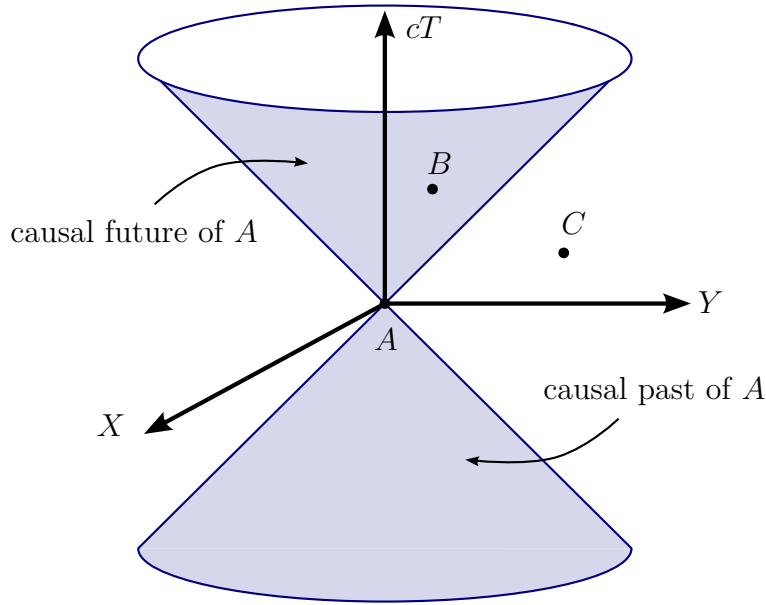}
\caption{Space-time diagram, where time is represented as the vertical axis, and two out of the three dimensions of space are represented as a horizontal plane. The light-cone of the event $A$, made of the set of events~$E$ with $\Delta s_{AE}^2=0$, is represented in blue. Event $B$ is located in the causal future of $A$: it can be the consequence of $A$. On the contrary, $C$ lies out of the light-cone of $A$, and hence it is causally disconnected from it.}
\label{fig:space-time}
\end{figure}

Those three cases are conveniently depicted in \emph{space-time diagrams}, where one represents time vertically, and two of the three dimensions of space as horizontal planes (see fig.~\ref{fig:space-time}). On this diagram, the events whose separation with an arbitrary event~$A$ are null form a cone, called the \emph{light-cone} of $A$. The events located inside the light-cone are time-like with respect to $A$, and hence can be a cause or a consequence of $A$. On the contrary, the events located outside the light-cone are space-like with respect to $A$, and hence causally disconnected from it.

\subsection{Minkowski metric and four-vectors}
\label{sec:Minkowski_metric}

In chapter~\ref{chap:Newton}, we have seen that the distance~$d_{AB}$ between two points $A$ and $B$ can be expressed in arbitrary coordinates, for which we had to introduce the notion of Euclidean metric. In a similar way, the space-time separation between two events can also be expressed in terms of arbitrary four-dimensional coordinates~$(x^\mu)\define(x^0,x^1,x^2,x^3)$. We will keep Greek indices of the beginning of the alphabet ($\alpha, \beta, \gamma,\ldots$) for the extension of Cartesian coordinates $(X^\alpha)=(cT, X^a)$, while the middle of the alphabet ($\mu, \nu, \rho, \ldots$) will correspond to arbitrary coordinates.

\paragraph{Minkowski metric} Consider two infinitesimally close events~$E, E'$, respectively associated with coordinates~$X^\alpha, X^\alpha+\dd X^\alpha$, or $x^\mu, x^\mu + \dd x^\mu$. The space-time interval between those events can then be written as
\begin{empheq}[box=\fbox]{equation}
\dd s^2
= \eta_{\alpha\beta} \, \dd X^\alpha \dd X^\beta
\define f_{\mu\nu} \dd x^\mu \dd x^\nu \ ,
\end{empheq}
where we introduced the \emph{Minkowski metric}~$\mat{f}$,\footnote{The symbol $f$ stands for ``flat''.} with components
\begin{equation}\label{eq:Minkowski_explicit}
f_{\mu\nu} = \eta_{\alpha\beta} \pd{X^\alpha}{x^\mu} \pd{X^\beta}{x^\nu}
\end{equation}
in arbitrary coordinates~$(x^\mu)$, which can be seen as a four-dimensional extension of the Euclidean metric. In the following, we will call \emph{inertial Cartesian coordinates} (ICCs) the class of coordinate systems $(X^\alpha)$ such that the Minkowski metric has components~$\eta_{\alpha\beta}$.

A key advantage of working directly in four dimensions is that there is no fundamental difference between a coordinate transformation and a change of reference frame. Indeed, we have seen in sec.~\ref{sec:frames} that a change of frame is just a time-dependent coordinate transformation~$x^i(t,X^a)$. This is just another way of writing $x^\mu(X^\alpha)$, with $x^0=X^0=ct$.

\begin{exercise}\label{ex:rotating_coordinates}
Consider the coordinate transformation $(X^\alpha)\rightarrow (x^\mu)=(ct,r,\theta,\ph)$,
\begin{align}
T &= t \\
X &= r \sin\theta\cos(\ph-\Omega t) \\
Y &= r \sin\theta\sin(\ph-\Omega t) \\
Z &= r \cos\theta,
\end{align}
where $\Omega$ is a constant. What is the physical meaning of this coordinate transformation? Show that the Minkowski metric reads, in this coordinate system,
\begin{equation}
\dd s^2 = (-1+\Omega^2 r^2\sin^2\theta) c^2\dd t^2
				- 2 \Omega r^2 \sin^2\theta \, \dd t \dd\ph
				+ \dd r^2
				+ r^2 \pa{ \dd\theta^2 + \sin^2\theta \, \dd\ph^2 } .
\end{equation}
\end{exercise}

\paragraph{Four-vectors} The four-dimensional analogue of a vector~$\vec{u}$ is called a four-vector, and is denoted with a bold symbol $\fvect{u}$. Just like three-vectors, four-vectors can be decomposed over the coordinate basis $(\fvect{\partial}_\alpha)$ for ICCs, and $(\fvect{\partial}_\mu)$ for arbitrary coordinates, with
\begin{equation}
\fvect{u} = u^\alpha \fvect{\partial}_\alpha = u^\mu \fvect{\partial}_\mu \ .
\end{equation}
The relations between components~$u^\alpha, u^\mu$ are, therefore,
\begin{equation}
u^\mu = \pd{x^\mu}{X^\alpha} \, u^\alpha ,
\qquad
u^\alpha = \pd{X^\alpha}{x^\mu} \, u^\mu .
\end{equation}

\paragraph{Minkowski product} The Minkowski metric defines a notion of product between four-vectors. Just like in three dimensions with the Euclidean metric, we have
\begin{equation}
\fvect{\partial}_\mu \cdot \fvect{\partial}_\nu = f_{\mu\nu}
\end{equation}
in general, and hence $\fvect{\partial}_\alpha\cdot\fvect{\partial}_\beta=\eta_{\alpha\beta}$ for ICCs. The scalar product of any two four-vectors $\fvect{u}$ and $\fvect{v}$ is then
\begin{empheq}[box=\fbox]{equation}
\fvect{u}\cdot\fvect{v}
\define \eta_{\alpha\beta} u^\alpha v^\beta
= f_{\mu\nu} u^\mu v^\nu .
\end{empheq}

Note that the Minkowski product is not exactly a scalar product in the pre-Hilbertian sense; namely, it is not positive definite. The sign of the Minkowskian self-product of a four-vector dictates its nature: $\fvect{u}$ is said to be space-like, null, time-like if, respectively, $\fvect{u}\cdot\fvect{u}>0,=0,<0$. This terminology is the same as the separation of events, because $\fvect{u}$ can be seen as an arrow linking two events.

\paragraph{Covariant or contravariant components} We have seen in chapter~\ref{chap:Newton}, with the example of the gradient of a function~$\vec{\nabla}U$, that the position (up or down) of an index can matter, when working in arbitrary coordinates; e.g., we had defined $\partial^i U=e^{ij} \partial_j U$, where the (inverse) Euclidean metric $e^{ij}$ appeared as a tool to \emph{raise indices}. For Cartesian coordinates, the position of indices did not matter, because they were raised and lowered with Kr\"onecker symbols, which do not change the components.

Things are slightly different with the Minkowski structure. The natural components of a vector~$\fvect{u}$ are the components with upper indices, $u^\alpha$; they are called \emph{contravariant components}, because the way they transform under coordinate transformations is contrary to the way the vector basis~$(\fvect{\partial}_\alpha)$ changes. But one can also introduce components with lower indices, $u_\alpha$,  called \emph{covariant components}, with
\begin{equation}
u_\alpha \define \eta_{\alpha\beta} u^\beta \ ,
\end{equation}
so that~$(u_\alpha)=(u_0,u_1,u_2,u_3)=(-u^0, u^1, u^2, u^3)$. We see that, even for the four-dimensional analogue of Cartesian coordinates, the position of indices does matter, because $u_0=-u^0$.

More generally, with arbitrary coordinates, we lower the index of a vector with the Minkowski metric
\begin{equation}
u_\mu \define f_{\mu\nu} u^\nu \ .
\end{equation}
Finally, these relations can be inverted using the inverse metric $f^{\mu\nu}$, defined just as in the three-dimensional case, in terms of matrix inversion,
\begin{equation}
f^{\mu\rho} f_{\rho\nu} = \delta^\mu_\nu \ .
\end{equation}
We then have $u^\mu = f^{\mu\nu} u_\nu$, so that $f_{\mu\nu}$ and $f^{\mu\nu}$ are objects that lower and raise the indices of vectors, respectively. Note finally that the Minkowskian product between two four-vectors $\fvect{u}, \fvect{v}$ can be seen as the \emph{contraction} of their covariant and contravariant components,
\begin{equation}
\fvect{u}\cdot\fvect{v} = f_{\mu\nu} u^\mu v^\nu = u_\mu v^\mu = u^\mu v_\mu \ .
\end{equation}

\begin{exercise}
Check that, for ICCs, the inverse metric is simply $\eta^{\alpha\beta}=\eta_{\alpha\beta}$.
\end{exercise}

\subsection{Relativity of time and space}

Like Cartesian coordinates for three-dimensional Euclidean geometry, ICCs are very special in Minkowskian geometry. They represent the class of coordinates such that~$f_{\alpha\beta}=\eta_{\alpha\beta}$. We can therefore wonder which class of coordinate transformations preserves that form of the Minkowski metric, i.e. the transformations~$X^\alpha \rightarrow \tilde{X}^\beta(X^\alpha)$ such that, for any two events~$A, B$,
\begin{equation}
\Delta s^2_{AB}
= \eta_{\alpha\beta} \, (X^\alpha_B-X^\alpha_A)(X^\beta_B-X^\beta_A)
= \eta_{\gamma\delta} \,
	(\tilde{X}^\gamma_B-\tilde{X}^\gamma_A)
	 (\tilde{X}^\delta_B-\tilde{X}^\delta_A) \ ,
\end{equation}
and in particular
\begin{equation}\label{eq:preserve_Minkowski}
\dd s^2 = \eta_{\alpha\beta} \, \dd X^\alpha \dd X^\beta 
= \eta_{\gamma\delta} \, \dd\tilde{X}^\gamma \dd\tilde{X}^\delta \ .
\end{equation}

\paragraph{Poincaré transformations} Transformations satisfying eq.~\eqref{eq:preserve_Minkowski} are called \emph{Poincaré transformations}; they form a group made of space-time translations (shift of the origin of time and space) plus the so-called \emph{Lorentz transformations}. Let us elaborate on the latter. Lorentz transformations are linear coordinate transformations, usually denoted
\begin{equation}
\tilde{X}^\alpha = \Lambda\indices{^\alpha_\beta} X^\beta \ ,
\end{equation}
and such that
\begin{equation}
\eta_{\gamma\delta} \Lambda\indices{^\alpha_\gamma} \Lambda\indices{^\beta_\delta} = \eta_{\alpha\beta} \ .
\end{equation}
As such, Lorentz transformations can be considered the generalisation of rotations in four dimensions, in a Minkowskian geometry\footnote{They differ from $\mathrm{SO}(4)$, which would generalise rotations to the four-dimensional Euclidean geometry, where we would replace $\eta_{\alpha\beta}$ by $\delta_{\alpha\beta}$.}. Any Lorentz transformation can be written as
\begin{equation}
\Lambda\indices{^\alpha_\beta}
= R\indices{^\alpha_\gamma} B\indices{^\gamma_\beta} \ ,
\end{equation}
where $[R\indices{^\alpha_\beta}]$ is a spatial rotation (leaving the time coordinate unchanged)
\begin{equation}
[R\indices{^\alpha_\gamma}] =
\begin{bmatrix}
1 & 0 \\
0 & [R\indices{^a_b}]
\end{bmatrix},
\end{equation}
with $[R\indices{^a_b}]\in \mathrm{SO}(3)$; while $[B\indices{^\gamma_\beta}]$ is called a \emph{Lorentz boost}.

\begin{figure}[h!]
\centering
\input{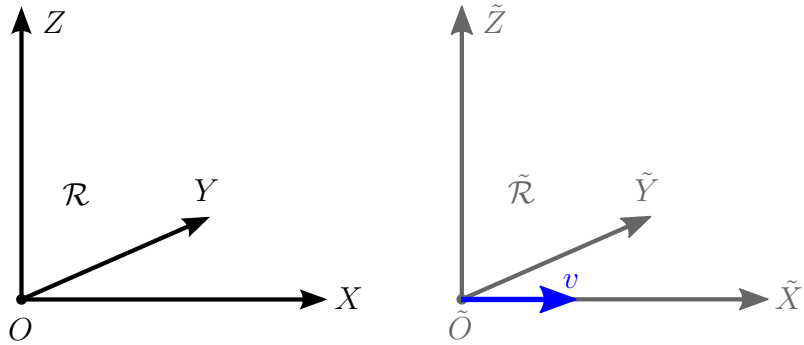}
\caption{Boost from an inertial frame $\mathcal{R}$ to another inertial frame~$\tilde{\mathcal{R}}$, in translation with respect to $\mathcal{R}$ at constant velocity $v$ in the direction $X$.}
\label{fig:Lorentz_boost}
\end{figure}

\paragraph{Lorentz boosts} \emph{Lorentz boosts are changes of inertial reference frames}. In Newtonian physics, according to Newton's first law, two inertial reference frames must be in constant-velocity translation with respect to each other. For example, if~$\tilde{\mathcal{R}}$ has the same axes as $\mathcal{R}$, while its origin~$\tilde{O}$ moves at constant velocity~$v$ in the $X$-direction with respect to $\mathcal{R}$ (see fig.~\ref{fig:Lorentz_boost}), then we expect to have~$\tilde{X}^\alpha=G\indices{^\alpha_\beta} X^\beta$, with
\begin{equation}\label{eq:Galilean_transformation}
[G\indices{^\alpha_\beta}]
=
\begin{bmatrix}
1 & 0 & 0 & 0 \\
-v/c & 1 & 0 & 0 \\
0 & 0 & 1 & 0 \\
0 & 0 & 0 & 1
\end{bmatrix},
\qquad \text{that is} \quad
\begin{system}
\tilde{T} &= T \\
\tilde{X} &= X - v T \\
\tilde{Y} &= Y \\
\tilde{Z} &= Z.
\end{system}
\end{equation}
The above transformation is called a Galilean transformation, but it turns out that \emph{it does not preserve the $\eta_{\alpha\beta}$ form of the Minkowski metric}. On the contrary, the Lorentz boost
\begin{empheq}[box=\fbox]{equation}\label{eq:Lorentz_boost}
[B\indices{^\alpha_\beta}]
=
\begin{bmatrix}
\gamma & -\gamma \beta & 0 & 0 \\
-\gamma \beta & \gamma & 0 & 0 \\
0 & 0 & 1 & 0 \\
0 & 0 & 0 & 1
\end{bmatrix},
\qquad \text{that is} \quad
\begin{system}
c\tilde{T} &= \gamma(cT-\beta X) \\
\tilde{X} &= \gamma(X - vT) \\
\tilde{Y} &= Y \\
\tilde{Z} &= Z,
\end{system}
\end{empheq}
where
\begin{equation}
\beta\define \frac{v}{c}, \qquad
\text{and} \quad
\gamma \define \frac{1}{\sqrt{1-\beta^2}} \geq 1
\end{equation}
is called the Lorentz factor, preserves the $\eta$-form of the Minkowski metric.

\begin{exercise}
Check that the Galilean transformation~\eqref{eq:Galilean_transformation} does not preserve the special $\eta$-form of the Minkowski metric, while the Lorentz boost~\eqref{eq:Lorentz_boost} does,
\begin{align}
\eta_{\gamma\delta} 
G\indices{^\gamma_\alpha} G\indices{^\delta_\beta}
&\not= \eta_{\alpha\beta}, \\
\eta_{\gamma\delta} B\indices{^\gamma_\alpha} B\indices{^\delta_\beta}
&= \eta_{\alpha\beta} \ .
\end{align}
\end{exercise}

\begin{exercise}
Show that the inverse transformation of \eqref{eq:Lorentz_boost} reads
\begin{equation}\label{eq:inverse_Lorentz_boost}
\begin{system}
cT &= \gamma(c\tilde{T}+\beta\tilde{X}) \\
X &= \gamma(\tilde{X} + \beta c\tilde{T}) \\
T &= \tilde{Y} \\
Z &= \tilde{Z} \ ,
\end{system}
\end{equation}
which, thus, simply consists in turning $v$ into $-v$.
\end{exercise}

\begin{exercise}
Generalise eq.~\eqref{eq:Lorentz_boost} by showing that, if the translation between $\mathcal{R}$ and $\tilde{\mathcal{R}}$ occurs in an arbitrary direction set by the unit vector~$\vec{e}$, then the components of the boost transformation read
\begin{align}
B\indices{^0_0} &= \gamma \\
B\indices{^a_0} &= -\gamma \beta e^a \\
B\indices{^a_b} &= \delta^a_b + (\gamma-1) e^a e_b \ .
\end{align}
\textit{Hint}: use rotation matrices.
\end{exercise}

\paragraph{Relativity of time} A key difference between the Galilean transformations of Newtonian physics and Lorentz boosts is that the latter do not leave time unchanged. To be more specific, consider two events $A, B$ that, in $\mathcal{R}$, happen at the same place $X_A^a=X^a_B$, and at times~$T_A=T, T_B=T+\Delta T$. In the frame~$\tilde{\mathcal{R}}$, however, those events happen at times
\begin{equation}
\begin{system}
c\tilde{T}_A &= \gamma (c T_A - \beta X_A) \\
c \tilde{T}_B &= \gamma (c T_B - \beta X_B)
\end{system}
\qquad
\text{whence}
\quad
\boxed{ \Delta \tilde{T}  = \gamma \Delta T \geq \Delta T \ .}
\end{equation}
The duration between the events $A$ and $B$ is therefore longer in $\tilde{\mathcal{R}}$ than in $\mathcal{R}$. The fact that time is not longer absolute, but rather relative to the state of motion of who measures it, is the reason that gave its name to \emph{relativity}.

\begin{exercise}
Show that, for any pair of events~$A$ and $B$ separated by a time-like interval, there exists an inertial frame in which those events happen at the same place.
\end{exercise}

From the above, we conclude that the reference frame in which the events occur at the same place is also the frame in which the duration between them is the shortest. In any other frame, the amount of time is dilated by the factor $\gamma$. For example, suppose that I clap my hands once, wait $\Delta T=1\U{s}$, and clap a second time, if you are moving with respect to me at $75\%$ of the speed of light, then you will measure, with your own clock, a duration
\begin{equation}
\Delta \tilde{T}
= \gamma \Delta T
= \frac{\Delta T}{\sqrt{1-\beta^2}}
= \frac{1\U{s}}{\sqrt{1-(3/4)^2}}
\approx 1.5\U{s}
\end{equation}
between the claps. This phenomenon is known as relativistic \emph{time dilation}.

\begin{exercise}
What is the Lorentz factor for $v=100\U{m/s}$? Recall that, in the international system of units, the speed of light is $c=3 \times 10^{8}\U{m/s}$. Why do not we notice time dilation in our daily life?
\end{exercise}

\begin{exercise}
Show that the notion of \emph{simultaneity} of two events is also relative: if two events happen at the same time in one frame, they do not in another frame.
\end{exercise}

\paragraph{Relativity of distances} Consider an object, say a ruler, and assume that~$\mathcal{R}$ is its \emph{rest frame}, i.e. the frame in which the ruler is at rest. In this frame, the coordinates of the ends of the ruler are, for example, $(X_1,Y_1,Z_1)=(0,0,0)$, and $(X_2,Y_2,Z_2)=(\ell,0,0)$. In other words, the length of the ruler is $\ell$, and it is aligned with the $X$ direction.

Now suppose that an observer in $\tilde{\mathcal{R}}$ measures the length of this ruler. In $\tilde{\mathcal{R}}$, the ruler moves, so it is essential that its length is measured by comparing the positions~$\tilde{X}_1, \tilde{X}_2$ of its ends at the same time~$\tilde{T}$,
\begin{equation}
\tilde{\ell} \define \tilde{X}_2(\tilde{T}) - \tilde{X}_1(\tilde{T}) \ .
\end{equation}
Using the inverse Lorentz boost~\eqref{eq:inverse_Lorentz_boost}, we find that the coordinates of the events corresponding to such measurement events read
\begin{equation}
\begin{system}
X_1 &= \gamma(\tilde{X}_1 + v\tilde{T}), \\
X_2 &= \gamma(\tilde{X}_2 + v\tilde{T}),
\end{system}
\qquad \text{whence} \quad
\boxed{
\tilde{\ell} = \frac{\ell}{\gamma}<\ell \ .
}
\end{equation}
The length of an object is therefore always smaller, when measured in a frame when it is moving, compared to the frame where it is at rest. This is called the relativistic \emph{contraction of lengths}. The size of an object as measured in its rest frame is called the \emph{proper size}.


\begin{exercise}
Show that, for any pair of events~$A$ and $B$ separated by a space-like interval, there exists an inertial frame in which those events happen at the same time.
\end{exercise}

\section{Physics in four dimensions}
\label{sec:physics_four_dimensions}

Now that we have set the structure of the four-dimensional space-time of the theory of relativity, let us review how Newton's mechanics can be extended to fit in this new picture. We will also mention, in \S~\ref{sec:Nordstrom}, an important historical attempt to include gravitation in the relativistic framework. This will lead us to the general theory of relativity at the end of this chapter.

\subsection{Motion and frames in relativity}

\begin{figure}[h!]
\centering
\input{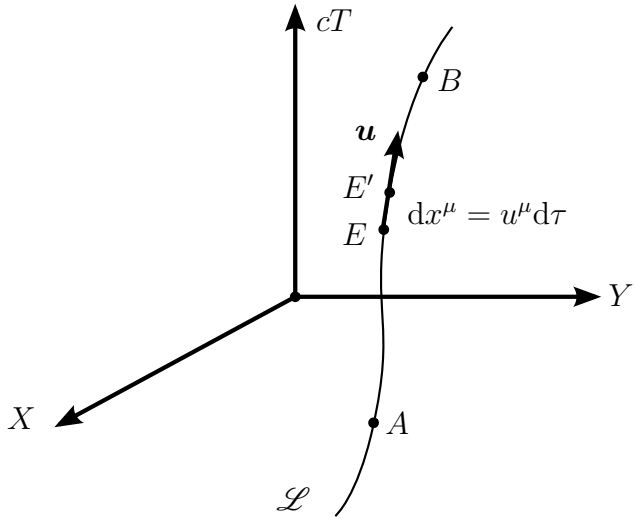}
\caption{World-line~$\wl$ of a particle. Between the events~$E,E'\in\wl$, separated by $\dd x^\mu$ in an arbitrary coordinate system, an observer sitting on the particle would measure a time interval~$\dd \tau$. The four-velocity~$\fvect{u}$ of the particle is the tangent vector to $\wl$, parametrised by $\tau$.}
\label{fig:world-line}
\end{figure}

\paragraph{World-lines and proper time} Consider a particle in an arbitrary state of motion. Instead of seeing this motion as a point in space which moves with time, we can consider it as a curve in the four-dimensional space-time (see fig.~\ref{fig:world-line}). This curve is called \emph{the world-line}~$\wl$ of the particle, and represents the whole history and future of its motion.

The world-line~$\wl$ of a particle defines a particular notion of time, which is the time measured by an observer~$\mathcal{O}$ who would be sitting on this particle. Let $E, E'$ be two events on $\wl$ separated by $\dd x^\mu$. $EE'$ is a time-like interval; indeed, by definition, there exists a frame in which those events happen at the same place: the rest frame of $\mathcal{O}$. Let us call $(X^\alpha)$ the coordinate system corresponding to an inertial frame that locally coincides with the observer's motion. By definition, in that frame, $(\dd X^\alpha)=(c\,\dd T,0,0,0)$, and hence
\begin{equation}
\dd s^2
= f_{\mu\nu} \dd x^\mu \dd x^\nu
= \eta_{\alpha\beta} \dd X^\alpha \dd X^\beta
= - c^2 \dd T^2 .
\end{equation}
The time interval~$\dd T$ is called the \emph{proper time} interval between $E$, $E'$, and it is more commonly denoted $\dd\tau$. Thus, we have, in general
\begin{equation}
\dd \tau = \frac{1}{c} \sqrt{-\dd s^2} \ .
\end{equation}
%

Now consider again two events $A$ and $B$ on $\wl$, but not necessarily separated by an infinitesimal interval. Denote~$x^\mu_A, x^\mu_B$ their respective coordinates, and let us parametrise $\wl$ with an arbitrary parameter~$\lambda$, as $x^\mu(\lambda)$. The proper time measured by $\mathcal{O}$ between those events is then
\begin{equation}\label{eq:proper_time_finite}
\tau_B - \tau_A
= \int_A^B \dd \tau
= \int_A^B \sqrt{- f_{\mu\nu} \dd x^\mu \dd x^\nu}
= \frac{1}{c} 
	\int_{\lambda_A}^{\lambda_B}
	\sqrt{-f_{\mu\nu} \ddf{x^\mu}{\lambda} \ddf{x^\nu}{\lambda} } 
		\; \dd\lambda \ ,
\end{equation}
where one can note the similarity with the length of a curve~\eqref{eq:length_curve} in three dimensions.

\paragraph{Four-velocity} In eq.~\eqref{eq:proper_time_finite}, there naturally appears in the integral a quantity~$\dd x^\mu/\dd\lambda$. This is nothing but the tangent vector of $\wl$, parametrised by $\lambda$. There is clearly a preferred parameter for this curve: its proper time. We call the \emph{four-velocity}~$\fvect{u}$ of a particle~$P$ the tangent vector to its world-line parametrised by proper time
\begin{empheq}[box=\fbox]{equation}
u^\mu \define \ddf{x^\mu}{\tau} \ .
\end{empheq}

\begin{exercise}
Show that~$\fvect{u}\cdot\fvect{u} = f_{\mu\nu} u^\mu u^\nu = -c^2$. As expected, it is time-like.
\end{exercise}

The four-velocity has a very specific form in inertial frames. Consider some ICCs $(X^\alpha)=(cT, X^a)$, attached to an inertial frame~$\mathcal{R}$. We can write
\begin{equation}
u^\alpha = \ddf{X^\alpha}{\tau} = \ddf{T}{\tau} \ddf{X^\alpha}{T} \ ,
\qquad \text{whence} \quad
(u^\alpha) = \ddf{T}{\tau}(c,v^a) \ ,
\end{equation}
where $v^a \define \dd X^a/\dd T$ in the velocity of $P$ as measured in $\mathcal{R}$.

\begin{exercise}
Check that the normalisation~$\fvect{u}\cdot\fvect{u}=-c^2$ of the four-velocity implies
\begin{equation}
\ddf{T}{\tau} = \frac{1}{\sqrt{1-\beta^2}} \define \gamma \ ,
\qquad \text{with} \quad
\beta^2 = \frac{\delta_{ab} v^a v^b}{c^2} \ ,
\end{equation}
so that $(u^\alpha) = (\gamma c, \gamma v^a)$.
\end{exercise}
\paragraph{Local space} The local space of an observer, at a point $A$ of its world-line, is defined as the hyperplane that is orthogonal to its four-velocity at this point, in the sense of Minkowski. It is therefore made of the events such that
\begin{equation}
0 = \fvect{u}\cdot\fvect{AE}
\define \eta_{\alpha\beta} u^\alpha (X_E^\beta-X_A^\beta) \ .
\end{equation}

\begin{exercise}
Show that, in the rest frame of the observer, these events $E$ are then all \emph{simultaneous}. This justifies the denomination of \emph{space} (the set of all events happening at the same time) for this hyperplane.
\end{exercise}

\paragraph{Four-acceleration} We define the four-acceleration of a particle as the derivative of its four-velocity with respect to proper time. With ICCs, this reads
\begin{equation}
a^\alpha \define \ddf{u^\alpha}{\tau} \ .
\end{equation}
In arbitrary coordinates, just like the Euclidean case, the simple derivative has to be replaced with a covariant derivative,
\begin{empheq}[box=\fbox]{equation}
a^\mu \define \Ddf{u^\mu}{\tau}
= \ddf{u^\mu}{\tau} + \Gamma\indices{^\mu_\nu_\rho} u^\nu u^\rho \ ,
\end{empheq}
where the Christoffel symbols of the Minkowski metric are defined in the same way as in the Euclidean case,
\begin{equation}
\Gamma\indices{^\rho_\mu_\nu}
= \frac{1}{2} f^{\rho\sigma}
	\pa{ f_{\sigma\mu,\nu} + f_{\sigma\nu,\mu} - f_{\mu\nu,\sigma}}.
\end{equation}

\paragraph{Changing frame} In the previous chapter, there was an important difference between a coordinate transformation, say~$X^a\rightarrow x^i(X^a)$, and changing the frame~$X^a\rightarrow\tilde{X}^b(t,X^a)$. In particular, for the latter, we have seen in \S~\ref{sec:frames} that the presence of time implies complicated transformations for velocity and acceleration when going from one frame~($\tilde{X}^b$) to the other~($X^a$). In four dimensions, things are much simpler.

\begin{exercise}\label{ex:changing_frame_relativity}
Show that $\fvect{u}$ and $\fvect{a}$ are four-vectors, in the sense that their components transform as
\begin{equation}
u^\mu = \pd{x^\mu}{X^\alpha} \, u^\alpha \ ,
\qquad
a^\mu = \pd{x^\mu}{X^\alpha} \, a^\alpha
\end{equation}
under any coordinate transformation~$(X^\alpha) \rightarrow (x^\mu)$.
\end{exercise}

The result of exercise~\ref{ex:changing_frame_relativity} is essential, because it describes both three-dimensional coordinate transformations \emph{and} changes of frame with a unique formula. For example, consider a particle with four-velocity~$(u^\alpha)=(\gamma c, \gamma v, 0, 0)$, that is, moving at velocity $v$ in the direction~$X^1$ in an ICC system $(X^\alpha)$. Suppose that we want to evaluate this velocity in another ICC system~$(\tilde{X}^\beta=B\indices{^\beta_\alpha} X^\alpha)$, moving at velocity~$v'$ in the same direction $X^1$ with respect to $(X^\alpha)$. Then we have
\begin{equation}
\tilde{u}^\beta
= \frac{\partial \tilde{X}^\beta}{\partial X^\alpha} u^\alpha
= B\indices{^\beta_\alpha} u^\alpha
\qquad
\text{hence}
\quad
\begin{system}
\tilde{u}^0 &= \gamma'\gamma c (1-\beta'\beta)\\ 
\tilde{u}^1 &= -\gamma'\gamma c (\beta-\beta')\\
\tilde{u}^2 &= \tilde{u}^3 = 0 \ .
\end{system}
\end{equation}
Therefore, if we write~$(\tilde{u}^\beta)=(\tilde{\gamma}, \tilde{\gamma} \tilde{v})$, we find the relativistic \emph{composition of velocities}
\begin{equation}
\tilde{v} = \frac{v-v'}{1-\frac{vv'}{c^2}} \ .
\end{equation}
Note the difference with Newtonian kinematics (and our intuition), in which~$\tilde{v}=v-v'$. The latter is approximately valid when~$v,v'\ll c$. On the contrary, if the particle is a photon, moving at $v=c$, then $\tilde{v}=c$ whatever the velocity~$\tilde{v}$ of the frame in which it is evaluated. This is the very important \emph{frame-independence of the speed of light in relativity}.

\subsection{Relativistic dynamics}

We now review the extension of the laws of mechanics in a relativistic context.

\paragraph{Four-momentum} We define the four-momentum of a particle with mass~$m\not=0$ as
\begin{equation}\label{eq:definition_four_momentum}
\fvect{p}=m\fvect{u}.
\end{equation}
With ICCs, this reads $(p^\alpha) = (\gamma m c, \gamma m \vec{v})$. The temporal component, $p^0$, is associated with the energy~$E\e{free}$ of the particle, that is, its energy when no forces are applied on it (free particle). More precisely, $p^0 c$ is the sum of the kinetic energy and \emph{rest-mass energy}~$m c^2$ of the particle. The usual expression of kinetic energy is recovered in the \emph{non-relativistic regime}, that is, when the particle moves slowly compared to the speed of light ($v\ll c$),
\begin{equation}
E\e{free} \define p^0 c
= \gamma m c^2 
= \frac{m c^2}{\sqrt{1-\pa{\frac{v}{c}}^2}}
= m c^2 + \frac{1}{2} m v^2 + \mathcal{O}\pa{\frac{v}{c}}^4.
\end{equation}

\begin{exercise}
Using the identification $E\e{free}\define p^0 c$ and the normalisation of the four-velocity, $\fvect{u}\cdot\fvect{u}=-1$, show that
\begin{equation}\label{eq:Pythagore-Einstein}
E\e{free}^2 = (m c^2)^2 + p^2 c^2 \ , 
\end{equation}
where $p^2\define \delta_{ab} p^a p^b$ is the norm of the spatial part of $\fvect{p}$.
\end{exercise}

While eq.~\eqref{eq:definition_four_momentum} cannot be applied for mass-less particles ($m=0$), like photons, eq.~\eqref{eq:Pythagore-Einstein} holds, in which case we have $E\e{free}=pc$. For example, a photon of frequency~$\omega$ and wave-vector~$\vec{k}$, with $k=\omega/c$, is associated with a four-momentum~$(p^\alpha)=\hbar(\omega/c,\vec{k})$. In this case, $\fvect{p}\cdot\fvect{p}=0$, so that $\fvect{p}$ is a null vector. Instead of eq.~\eqref{eq:definition_four_momentum}, we write~$\fvect{p}=\hbar \fvect{k}$, where $\fvect{k}$ is the wave-four vector of the photon and plays the role of its four-velocity.

\paragraph{Equation of motion} The relativistic generalisation of Newton's second law for a point particle is, in arbitrary coordinates,
\begin{empheq}[box=\fbox]{equation}\label{eq:relativistic_Newton_2}
\Ddf{p^\mu}{\tau}
\define \ddf{p^\mu}{\tau} + \Gamma\indices{^\mu_\nu_\rho} p^\nu u^\rho
= F^\mu \ ,
\end{empheq}
where $\tau$ is the particle's proper time, and $\fvect{F}$ is called the \emph{four-force} applied on the particle. Its spatial part~$F^i$ is the three-dimensional force, while its temporal component is the power of that force (work per unit time). When $m=\cst$, the above relation is just~$m a^\mu = F^\mu$. We will restrict to that case in the remainder of the course.

Contrary to classical mechanics in three dimensions, we do not need to make any assumption about the nature (inertial or not) of the frame. The equation of motion~\eqref{eq:relativistic_Newton_2} is \emph{valid in any frame}, because it is valid for any four-dimensional coordinate system. The fictitious forces appearing in non-inertial frames are, here, contained in the Christoffel symbols~$\Gamma\indices{^\mu_\nu_\rho}$ of the Minkowski metric, which are zero in ICCs, but non-zero in general.

\begin{exercise}
Calculate the Christoffel symbols of the Minkowski metric in the rotating coordinates of exercise~\ref{ex:rotating_coordinates}, and show that the centrifugal and Coriolis forces naturally appear in the equation of motion.
\end{exercise}

An interesting case, which illustrates the properties of relativistic dynamics, is when the four-force derives from a potential energy~$U(x^\mu)$. Its expression is, then,
\begin{equation}\label{eq:four-force_potential}
F^\mu =
- \pa{f^{\mu\nu}
+ \frac{u^\mu u^\nu}{c^2} } \frac{\partial_\nu U}{1+ U/m c^2} \ ,
\end{equation}
where $\fvect{u}$ is the four-velocity of the particle. The above expression can seem quite complicated at first sight. For example, one could wonder why it involves~$(f^{\mu\nu} + c^{-2}u^\mu u^\nu)$. This operator is the projector onto the particle's local space. In other words, it imposes $\fvect{F}\cdot\fvect{u}=0$, so that, in the particle's rest frame, $\fvect{F}$ is purely spatial. This projection is essential, because it ensures that the condition~$\fvect{p}\cdot\fvect{p}=-m^2 = \cst$ remains true along the particle's world-line. The role of the denominator~$1+U/mc^2$ in eq.~\eqref{eq:four-force_potential} is more elegantly understood as follows: first multiply the equation of motion by $1+U/mc^2$, and then use
\begin{equation}
\ddf{U}{\tau}
= \ddf{}{\tau} \, U[x^\mu(\tau)]
= \ddf{x^\mu}{\tau} \, \partial_\mu U
= u^\mu \partial_\mu U \ ;
\end{equation}
the result is
\begin{empheq}[box=\fbox]{equation}\label{eq:relativistic_dynamics_potential}
\Ddf{}{\tau}\pac{(mc^2 + U) u^\mu} = - c^2\partial^\mu U \ .
\end{empheq}
Let us clarify the physical meaning of this equation with the following exercise.

\begin{exercise}
With ICCs~$(X^\alpha)$ eq.~\eqref{eq:relativistic_dynamics_potential} simply becomes
\begin{equation}
\ddf{}{\tau} \pac{(mc^2 + U) u^\alpha} 
= -c^2 \partial^\alpha U \ .
\end{equation}
Separating the temporal part ($\alpha=0$) and the spatial part~$(\alpha=a)$, show that
\begin{align}
\ddf{E}{T} &= \frac{1}{\gamma} \pd{U}{T} \ , \\
\pa{ m+\frac{U}{c^2} } \ddf{v^a}{T}
&= - \frac{1}{\gamma^2} \pac{\partial^a U 
		+ \pa{\frac{v^a}{c^2}}\partial_T{U}},
\label{eq:relativistic_dynamics}
\end{align}
where we have defined the \emph{total energy} of the particle as $E=E\e{free}+\gamma U=\gamma(mc^2+U)$. Check that we recover Newtonian dynamics in the non-relativistic regime ($v\ll c$).
\end{exercise}

\paragraph{Light-speed cannot be exceeded} Another interesting limit of eq.~\eqref{eq:relativistic_dynamics} is the \emph{ultra-relativistic regime}, which corresponds to $v\rightarrow c$. In this case,
\begin{equation}
\gamma = \frac{1}{\sqrt{1-(v/c)^2}} \rightarrow\infty \ ,
\end{equation}
and hence
\begin{equation}
\ddf{v^a}{T} 
= - \frac{1}{\gamma^2} \frac{c^2}{mc^2+U}
		\pac{\partial^a U + \pa{\frac{v^a}{c^2}}\partial_T{U}}\rightarrow 0 \ ,
\end{equation}
even if a force keeps being applied to the particle. This shows that a massive particle can never reach the speed of light, even if it is constantly accelerated. The speed of light appears as the asymptotic velocity of a particle that would be constantly accelerated during an infinite amount of time, giving it infinite energy.

This fact can be interpreted as follows. Let us multiply eq.~\eqref{eq:relativistic_dynamics} by $\gamma^2$, then
\begin{equation}
\frac{E}{c^2} \, \ddf{v^a}{\tau} = - \pac{\partial^a U + \pa{\frac{v^a}{c}}\partial_T{U}} .
\end{equation}
This equation is very analogous to Newton's second law, except from the fact that the equivalent of inertial mass~$m$ is now the energy~$E/c^2$. This will turn out to be a generic fact in relativity: inertia and gravitation are not ruled by mass, but energy.

\paragraph{Lagrangian formulation} Just like in classical mechanics, the relativistic  equation of motion for a point particle in a potential~$U$ can be obtained from an action principle. Consider a particle evolving between events~$A$ and $B$, the corresponding action can be written as
\begin{empheq}[box=\fbox]{equation}\label{eq:action_relativistic_dynamics}
S[x^\mu] = - \int_A^B \pa{m c^2 + U} \dd\tau \ .
\end{empheq}
Note that we do recover the Lagrangian $K-U$ of Newtonian dynamics in the non-relativistic regime. Indeed, for an inertial frame such that~$v\ll c$,
\begin{align}
-\pa{m c^2 + U} \dd\tau
&= -\pa{m c^2 + U} \sqrt{1- \frac{v^2}{c^2}} \, \dd T \\
&= \pac{- m c^2 + \frac{m v^2 }{2} - U + \mathcal{O}\pa{\frac{v}{c}}^4 } \dd T \ ,
\end{align}
which is $(K-U)\,\dd T$, modulo the constant term~$m c^2$ which does not change the dynamics.

In order to recover the equation of motion from the action~\eqref{eq:action_relativistic_dynamics}, one has to rely on a trick which consists in artificially introducing an arbitrary parameter~$\lambda$ along the world-line of the particle:
\begin{equation}\label{eq:trick_relativistic_action}
S[x^\mu] =
- \int_{\lambda_A}^{\lambda_B} (mc^2 + U)
\sqrt{-f_{\mu\nu} \ddf{x^\mu}{\lambda} \ddf{x^\nu}{\lambda}} \; \dd\lambda \ .
\end{equation}
Indeed, with this notation, the relativistic Lagrangian becomes a function of $x^\mu$ and $\dd x^\mu/\dd \lambda$. We can then apply the usual techniques of variational calculus.

\begin{exercise}
Show that the functional derivative of $S$ reads
\begin{equation}
\frac{\delta S}{\delta x^\mu}
= \frac{\partial L}{\partial x^\mu}
	- \ddf{}{\lambda} \pa{ \frac{\partial L}{\partial\dot{x}^\mu} },
\end{equation}
where $L$ is the integrand of eq.~\eqref{eq:trick_relativistic_action}, and $\dot{x}^\mu \define \dd x^\mu / \dd\lambda$ here. Calculate the above explicitly, and, at the very end of the calculation, replace the arbitrary parameter~$\lambda$ by proper time. Conclude that
\begin{equation}
\frac{\delta S}{\delta x^\mu} = 0
\Longleftrightarrow
\Ddf{}{\tau} \pac{ (m c^2 + U) u^\mu } = -c^2 \partial^\mu U.
\end{equation}
\end{exercise}

\subsection{Nordstr\"{o}m's theory of gravity}
\label{sec:Nordstrom}

In 1912, the Finnish physicist Gunnar Nordstr\"{o}m proposed a theory of gravity within the framework of Einstein's special theory of relativity~\cite{Nordstrom1912}. Its reformulation~\cite{1914AnP...349..321E}, in 1914, by Einstein and Fokker, paved the way towards the general theory of relativity, published a year later.

\paragraph{Attempt for scalar gravity} The initial idea of Nordstr\"{o}m was to cure the \emph{instantaneous} character of Newtonian gravitation. Indeed, as we have seen in the previous chapter, the solutions of the Poisson equation,
\begin{equation}
\Delta \Phi = 4\pi G \rho \ ,
\end{equation}
allow information to propagate instantaneously---if $\rho$ changes somewhere at time $t$, then the gravitational potential $\Phi$ feels directly this change at the same time $t$, whatever its distance to the change of $\rho$. This is in contradiction with the relativistic idea that nothing can propagate quicker than the speed of light.

The simplest modification of the Poisson equation that satisfies this principle consists in turning the Laplace operator~$\Delta=\delta^{ab}\partial_a\partial_b$ into a d'Alembertian operator~$\Box=\eta^{\alpha\beta}\partial_\alpha\partial
_\beta$,
\begin{equation}\label{eq:modified_Poisson}
\Box \Phi = 4\pi G \rho \ ,
\end{equation}
which is similar to the equation for the electromagnetic potentials~$(V,\vec{A})$ in the Lorenz\footnote{The Danish physicist Ludvig Lorenz [1829-1891] must be distinguished from the Dutch physicist Hendrik Lorentz [1853-1928]; they differed by one letter and a couple of decades.} gauge. Just like in electrodynamics, the hyperbolic character of the modified Poisson equation~\eqref{eq:modified_Poisson} implies that its solutions can be expressed as retarded potentials,
\begin{equation}\label{eq:Nordstrom_field}
\Phi(T, \vec{X}) 
= - G \int_{\mathcal{D}} \frac{\rho(T-||\vec{X}-\vec{Y}||/c,\vec{Y})}{||\vec{X}-\vec{Y}||} \; \dd^3 Y \ ,
\end{equation}
ensuring that the gravitational information propagates at the speed of light.

\paragraph{Nordstr\"{o}m action} Consider a system of $N$ particles in gravitational interaction. An action that produces a field equation of the form \eqref{eq:modified_Poisson} is
\begin{equation}
S =  - \frac{1}{8\pi G c} 
			\int \eta^{\alpha\beta} \partial_\alpha\Phi \partial_\beta \Phi \; \dd^4 X
		- \sum_{p=1}^N m_p c^2 \int \pa{1+\frac{\Phi}{c^2}}\; \dd \tau_p \ ,
\end{equation}
where $m_p$ denotes the mass of the particle~$p$, while $\tau_p$ is its proper time. The first term is usually called the kinetic term of the field~$\Phi$. It is a straightforward generalisation of Newton's action seen in \S~\ref{sec:Lagrangian_Newton_gravity} and it will yield the d'Alembertian~$\Box\Phi$. The second term is the sum of individual actions of the form~\eqref{eq:action_relativistic_dynamics}, with $U_p=m_p \Phi$ for each particle~$p$. Thus, we already know that its variation with respect to $x^\mu_p$ produces
\begin{equation}
\forall p\in\{1,\ldots,N\} \qquad
\ddf{}{\tau_p} \pac{ \pa{1 + \frac{\Phi}{c^2}} u^\alpha_p }
= -\partial^\alpha \Phi \ .
\end{equation}
In the non-relativistic regimes, this simply becomes $\vec{a}_p = -\vec{\nabla}\Phi$.

The sum of the actions of all the particles~$p$ can, besides, be rewritten as
\begin{equation}
\sum_{p=1}^N m_p c^2 \int\pa{1+\frac{\Phi}{c^2}}\; \dd \tau_p
= \frac{1}{c} \int (\rho c^2 - 3P) \pa{1+\frac{\Phi}{c^2}}\; \dd^4 X \ ,
\end{equation}
where $\rho$ is the mass density and $P$ is the \emph{kinetic pressure} of the system of $N$ particles. We will, for the moment, accept this result with no proof, and come back to it in the last section of this chapter.

\begin{exercise}
Considering the action
\begin{equation}
S = - \frac{1}{c} \int
			\pac{ \frac{1}{8\pi G}\,\eta^{\alpha\beta} \partial_\alpha\Phi \partial_\beta \Phi
						+ (\rho c^2 - P) \pa{1+\frac{\Phi}{c^2}}	
					} \, \dd^4 X \ ,
\end{equation}
show that the field equation for $\Phi$, obtained by imposing~$\delta S/\delta\Phi=0$, reads
\begin{equation}
\Box\Phi = 4\pi G \pa{ \rho - \frac{3P}{c^2} } \ ,
\end{equation}
which the modified Poisson equation~\eqref{eq:modified_Poisson}, modulo the pressure term.
\end{exercise}

\paragraph{Einstein-Fokker reformulation} The key discovery of Einstein and Fokker in 1914 was to notice that the action of a point particle coupled to Nordstr\"{o}m's field,
\begin{equation}
S = - m c^2 \int\pa{ 1 + \frac{\Phi}{c^2} } \, \dd \tau \ ,
\end{equation}
is equivalent to the action of a free particle,
\begin{equation}
S = - m c^2 \int\dd \hat{\tau} \ ,
\end{equation}
if one replaces the Minkowski metric~$f_{\mu\nu}$ by $g_{\mu\nu} = (1+\Phi/c^2)^2 f_{\mu\nu}$. Indeed, with the $g_{\mu\nu}$ metric, the proper time interval between two events separated by~$\dd x^\mu$ along the particle's world-line reads
\begin{equation}
\dd\hat{\tau}^2
\define - g_{\mu\nu} \dd x^\mu \dd x^\nu 
= -\pa{1+\frac{\Phi}{c^2}}^2 f_{\mu\nu} \dd x^\mu \dd x^\nu 
= \pac{ \pa{1+\frac{\Phi}{c^2}} \dd\tau }^2.
\end{equation}
In this language, the gravitational field~$\Phi$ is absorbed in the metric of space-time, instead of being a force applied on a particle in Minkowski space-time. Moreover, because $S$ is now proportional to the proper time of the particle, $\delta S/\delta x^\mu=0$ imposes that its trajectory is a \emph{geodesic} of space-time with metric~$\mat{g}$ (see next section).

Furthermore, Nordstr\"{o}m's field equation can be rewritten, in this framework, as
\begin{equation}
g^{\mu\nu}R_{\mu\nu} = 24 \pi G\, g^{\mu\nu}T_{\mu\nu} \ ,
\end{equation}
where $R_{\mu\nu}$ is called the Ricci curvature of the space-time metric~$g_{\mu\nu}$, while $T_{\mu\nu}$ is the energy-momentum tensor of matter. We will explain the meaning of those quantities in the next sections. For now, the important thing is to realise the change of paradigm that we are about to make: instead of viewing gravity as a force, we consider the possibility that it can be the \emph{curvature of space-time}. This curvature is the reason why trajectories of particles in a gravity field are not straight lines, while the \emph{energy and momentum} of matter would generate it.

\paragraph{Towards general relativity} Nordstr\"{o}m's theory turns out to be wrong: it does not agree with experiments. In particular, it does not predict the right trajectory for Mercury around the Sun, and its does not predict any deflection of light by massive bodies. However, the Einstein-Fokker formulation shows that it is possible to encode gravitational phenomena in the geometry of space-time, through a metric~$g_{\mu\nu}$ which is not the Minkowski metric. This opens the door to the theory of general relativity, hereafter abbreviated GR.

\section{Differential geometry tool kit}
\label{sec:differential_geometry}

Before entering into the details of GR, we need to introduce the main tools of differential geometry, which is the language of that theory. This section is a crash course aiming to introduce those in roughly two hours. We will, therefore, adopt a very utilitarian approach, introducing mathematical objects \textit{à la physicienne}, without proper definitions, but rather as a set of intuitions, recipes, and calculation rules. The interested reader is encouraged to refer to more rigorous presentations; I personally find \textit{Gauge fields, knots, and gravity}, by John Baez \& Javier Muniain~\cite{Baez:1995sj}, very well written. For French speakers, the lecture notes \href{http://science.thilucmic.fr/TELECHARGEMENTS/LECTURES/coursgeodiff-2x1.pdf}{\it Géométrie différentielle, groupes et algèbres de Lie, fibrés et connexions}, by Thierry Masson, are also very good and thorough.

\subsection{Tensors}

\paragraph{Space-time manifold} The mathematical structure of a space-time is a four-dimensional \emph{manifold}~$\mathcal{M}$. This is just the name for a topological space, i.e., a space in which we are told which points can be linked by a curve, which curves can be continuously deformed to a point, etc. Here we will assume that our space-time has a trivial topology, that is, the same topology as $\mathbb{R}^4$. On this space-time, we can define a coordinate system, or chart, $(x^\mu)$, which allows us to locate points.

\paragraph{Scalars} Functions~$f:\mathcal{M}\rightarrow \mathbb{R}$, i.e., that take a point of space-time and return a number, are called \emph{scalar fields}, or simply \emph{scalars}. They trivially change under coordinate transformations.\footnote{In this section, for notational ease, we will use Greek indices of the beginning of the alphabet $(\alpha,\beta,\gamma,\ldots)$ similarly to indices of middle of the alphabet $(\mu,\nu,\rho,\ldots)$; they will also refer to arbitrary coordinates, and not necessarily to ICCs.} For $(x^\mu) \rightarrow (y^\alpha)$, we have $f\rightarrow \tilde{f}$, with
\begin{equation}\label{eq:transformation_scalar}
\tilde{f}(y^\alpha) = f[x^\mu(y^\alpha)] \ .
\end{equation}
Although $y^\alpha \mapsto \tilde{f}(y^\alpha)$ and $x^\mu \mapsto f(x^\mu)$ are, analytically speaking, different functions, it is customary to denote them with the same symbol~$f$. The reason is that, in physics, we care more about the physical meaning of $f$ (like temperature, gravitational potential, etc.) than the mathematical function of the coordinates that is used to represent it. For example, Nordstr\"{o}m's field $\Phi$ is a scalar, and we write~$\Phi(y^\alpha)=\Phi[x^\mu(y^\alpha)]$.

\begin{figure}[h!]
\centering
\input{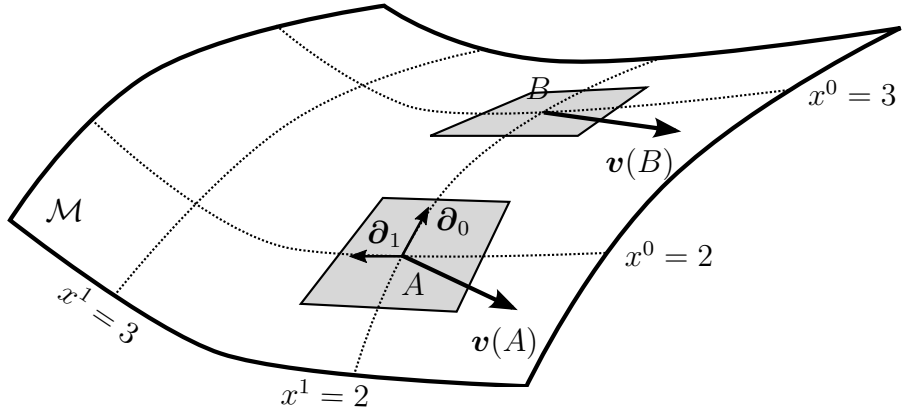}
\caption{A vector field $\fvect{v}$ evaluated at two points $A, B$ of the manifold~$\mathcal{M}$.}
\label{fig:tangent}
\end{figure}

\paragraph{Vectors} The notion of vector was extensively used in the previous sections. Slightly more mathematically, the idea is that, at each point $P$ of the space-time manifold, one can define a flat tangent space-time. This notion is quite intuitive (see fig.~\ref{fig:tangent}); if space-time were a sphere, the tangent space at a point of the sphere would be the plane that is tangent to the sphere at that point. This tangent space-time is where four-vectors live. A \emph{four-vector field}~$\fvect{v}$ is a function which, to each point $x^\mu$ associates a four-vector~$\fvect{v}(x^\mu)$.

The coordinate system~$(x^\mu)$ on $\mathcal{M}$ generates a basis $(\fvect{\partial}_\mu)$ for each of its tangent spaces. These vectors are constructed as follows: let two events $E, E'$ have the same coordinates, apart from, e.g., $x^1$ which differs by $\dd x^1$ from $E$ to $E'$; then $\fvect{\partial}_1=\fvect{EE'}/\dd x^1$. Any four-vector field (we will simply say four-vector, or vector, for short) $\fvect{v}$ can be decomposed over this basis as $\fvect{v}=v^\mu \fvect{\partial}_\mu$. Under coordinate transformation~$(x^\mu)\rightarrow(y^\alpha)$, the basis vectors and the vector components over it change according to
\begin{equation}\label{eq:transformation_four-vector}
\fvect{\partial}_\alpha = \pd{x^\mu}{y^\alpha} \, \fvect{\partial}_\mu \ ,
\qquad
v^\alpha = \pd{y^\alpha}{x^\mu} \, v^\mu \ ,
\end{equation}
where we now omit to specify where the quantities are evaluated---it is understood that, like scalars, they are taken at the same event, described by $y^\alpha$ in one coordinate system, and $x^\mu(y^\alpha)$ in the other.

\paragraph{Forms} A differential form, or one-form, or co-vector, $\fvect{\omega}$, is a linear map that, at each point of space-time, takes a vector and returns a number. In other words, it takes a vector field and returns a scalar field. In this course, we will be mostly interested in manipulating the components of forms, defined through their effect on the vector basis as
\begin{equation}
\omega_\mu \define \fvect{\omega}(\fvect{\partial}_\mu) \ .
\end{equation}

\begin{exercise}
Using the linearity of $\fvect{\omega}$, show that its components transform as 
\begin{equation}
\omega_\alpha = \pd{x^\mu}{y^\alpha} \, \omega_\mu
\end{equation}
under a coordinate transformation $(x^\mu)\rightarrow(y^\alpha)$. Besides, show that the action of $\fvect{\omega}$ on any vector is given by the \emph{contraction} of their components, $\fvect{\omega}(\fvect{v}) = \omega_\mu v^\mu$.
\end{exercise}

\paragraph{Tensors} The combination of an arbitrary number of forms and vectors, i.e., a multi-linear map that takes several vectors and returns several other vectors, is called a \emph{tensor}. Let us take the example of a tensor~$\fvect{T}$ that takes two vectors and returns one other vector. Its components are defined through its effect on the vector basis as
\begin{equation}
\fvect{T}(\fvect{\partial}_\mu, \fvect{\partial}_\nu)
= T\indices{_\mu_\nu^\rho} \fvect{\partial}_\rho \ .
\end{equation}
Under a coordinate transformation $(x^\mu)\rightarrow(y^\alpha)$, these components change according to
\begin{empheq}[box=\fbox]{equation}\label{eq:transformation_tensor}
T\indices{_\alpha_\beta^\gamma}
=  \pd{x^\mu}{y^\alpha} \pd{x^\nu}{y^\beta} \pd{y^\gamma}{x^\rho} \,
	T\indices{_\mu_\nu^\rho} \ .
\end{empheq}
The Jacobian matrices $\partial x^\mu / \partial y^\alpha$ and $\partial y^\alpha/\partial x^\mu$ are used so as to \emph{preserve the altitude of indices}; namely, two members of a sum or an equality involving free indices must have those indices at the same altitude. Dummy indices must have different altitudes, e.g. $\omega_\mu v^\mu$.

\paragraph{Terminology} It is customary, in physics, to neglect the ontological distinction between a tensor and its components. The transformation rule~\eqref{eq:transformation_tensor} may then be considered the
definition of a tensor: it is a prescription for deciding whether a quantity with multiple indices does or does not represent the components of tensor (see exercise~\ref{ex:tensor_or_not}). In that language, the contraction of a pair of indices in a tensor leads to a quantity that is still a tensor. For instance, starting from a tensor~$T\indices{_\mu_\nu^\rho}$, the quantity~$T\indices{_\mu_\nu^\nu}$ represents the components of another tensor---in this case, it is a form.

\subsection{Metric}

We have already introduced the concept of metric in the previous sections. We have understood that it is a tool that allows one to compute distances, times, vector products, and also to lower and raise indices.

\paragraph{Definition} A metric~$\mat{g}$ is a symmetric tensor defining the scalar product of vectors. Its components dictate the scalar product of basis vectors as
\begin{equation}
\fvect{\partial}_\mu \cdot \fvect{\partial}_\nu
\define \fvect{g}(\fvect{\partial}_\mu , \fvect{\partial}_\nu)
= g_{\mu\nu} \ .
\end{equation}
By bi-linearity, the scalar product of any two vectors~$\fvect{u}$, $\fvect{v}$ then reads $\fvect{u}\cdot\fvect{v}=g_{\mu\nu} u^\mu v^\nu$. If $\fvect{u}=\fvect{v}$ connects two neighbouring events $E, E'$ with coordinates $x^\mu, x^\mu+\dd x^\mu$, then $\fvect{u}\cdot\fvect{u}$ represents the space-time interval between those events,
\begin{equation}
\dd s^2 = g_{\mu\nu} \dd x^\mu \dd x^\nu \ .
\end{equation}

\paragraph{What is different now?} In chapter~\ref{chap:Newton} and in the beginning of the present chapter, we have used two very particular metrics, namely the Euclidean metric in three dimensions, and the Minkowski metric in four dimensions. The latter, for example, is characterised by the fact that there existence a particular class of coordinate systems $(X^\alpha)$, which we called ICC, such that $f_{\alpha\beta} = \eta_{\alpha\beta}$ over the whole space-time. This property does not hold for a general metric tensor~$\mat{g}$, in particular,
\begin{equation}
g_{\mu\nu} \not= \eta_{\alpha\beta} 
								\pd{X^\alpha}{x^\mu}\pd{X^\beta}{x^\nu} \ .
\end{equation}

\paragraph{Signature} What is not globally true remains, however \emph{locally} true. Namely, at any event~$E$, one can always find a particular coordinate system such that
\begin{equation}
g_{\alpha\beta}(E) = \eta_{\alpha\beta} \ ,
\qquad
\text{but} \quad g_{\alpha\beta}(E'\not=E) \not= \eta_{\alpha\beta} \ ,
\end{equation}
the metric can be turned into $\eta_{\alpha\beta}$ anywhere, but not everywhere at the same time.

This allows us to define the \emph{signature} of a metric: as $g_{\mu\nu}$ locally corresponds to the matrix $\mathrm{diag}(-1,1,1,1)$, we say that its signature is $(-+++)$, which is called a \emph{Lorentzian} signature. A manifold equipped with such a metric is then called a Lorentzian manifold. Note that some authors, mostly in particle physics, use the opposite signature $(+---)$, which distributes minus signs here and there in the equations. In contrast, a \emph{Riemannian} manifold would be equipped with a metric with signature $(++++)$.

\paragraph{Lowering and raising indices} In \S~\ref{sec:Minkowski_metric}, we mentioned that the metric could be used to lower indices, while its inverse raises indices. Now that the notion of form has been presented, we can understand why. Indeed, starting from a vector field~$\fvect{u}$ and a scalar product~$\fvect{g}$, we can naturally define a form~$\fvect{\Upsilon}$, which takes any vector~$\fvect{v}$ and returns its scalar product with $\fvect{u}$,
\begin{equation}
\fvect{\Upsilon}(\fvect{v})
\define \fvect{u}\cdot\fvect{v}
= g_{\mu\nu} u^\mu v^\nu \ .
\end{equation}
The components of $\fvect{\Upsilon}$ are therefore $\Upsilon_\nu=g_{\mu\nu}u^\mu$; because there is a one-to-one relation between $\fvect{\Upsilon}$ and $\fvect{u}$, we decide to use the same symbol for their components, and just write $u_\mu \define\Upsilon_\mu$. Thus, in that sense, $g_{\mu\nu}$ lowers indices as $u_\nu = g_{\mu\nu}u^\mu$.

The above was about turning vectors into forms. The reverse process uses the inverse metric, with components~$g^{\mu\nu}$ such that
\begin{equation}
g^{\mu\rho} g_{\rho\nu} = \delta^\mu_\nu \ ,
\end{equation}
we then have~$u^\mu = g^{\mu\nu} u_\nu$. This can be generalised to any index of any tensor, for example,
\begin{equation}
T\indices{_\lambda^\nu^\sigma}
= g_{\mu\lambda} g^{\rho\sigma} T\indices{^\mu^\nu_\rho} \ .
\end{equation}

\subsection{Connection}

We have already met the notion of covariant derivative in the previous sections. It appeared naturally as a way to properly take derivatives of components of vectors, by taking into account the spurious changes of the coordinate system when one moves from one point to another. The underlying mathematical structure is called a \emph{connection}, and, more specifically here, the Levi-Civita connection associated with the space-time metric.

\paragraph{Covariant derivative} The covariant derivative can be seen as a generalisation of the partial derivative. Its effect depends on the object it is applied to. First of all, the covariant derivative of a scalar in the $\mu$th direction, i.e. the direction of the basis vector $\fvect{\partial}_\mu$, denoted~$\nabla_\mu$, is simply
\begin{equation}
\nabla_\mu f \define \partial_\mu f \ .
\end{equation}

The covariant derivative of a vector~$\fvect{v}$ is another vector~$\nabla_\mu \fvect{v}$, whose components are
\begin{equation}
\nabla_\mu v^\nu 
\define
v\indices{^\nu_;_\mu}
= v\indices{^\nu_,_\mu} + \Gamma\indices{^\nu_\rho_\mu} v^\rho \ .
\end{equation}
The semicolon~``;'' serves as a short-hand notation for the covariant derivative, and the Christoffel symbols~$\Gamma\indices{^\nu_\rho_\mu}$, also called connection coefficients, are
\begin{empheq}[box=\fbox]{equation}
\Gamma\indices{^\nu_\rho_\mu}
= \frac{1}{2} \, g^{\nu\sigma}
	\pa{ g_{\sigma\rho,\mu} + g_{\sigma\mu,\rho} - g_{\mu\rho,\sigma}}.
\end{empheq}
Note that the Christoffel symbols are symmetric in their last indices: $\Gamma\indices{^\nu_\rho_\mu}=\Gamma\indices{^\nu_\mu_\rho}$. It is common to introduce the notation
\begin{equation}
\Gamma_{\sigma\rho\mu}
=\frac{1}{2}\pa{ g_{\sigma\rho,\mu} + g_{\sigma\mu,\rho} - g_{\mu\rho,\sigma}}
= g_{\sigma\nu} \Gamma\indices{^\nu_\rho_\mu} \ .
\end{equation}

\begin{exercise}
Show that $g_{\mu\nu,\rho} = 2\pa{\Gamma_{\mu\nu\rho}+\Gamma_{\nu\rho\mu}}$.
\end{exercise}

\begin{exercise}\label{ex:tensor_or_not}
By performing a general coordinate transformation, show that:
\begin{enumerate}
\item $\partial_\mu f$ are the components of a vector; while
\item $\partial_\mu v^\nu$ are not the component of a tensor; and
\item $\Gamma\indices{^\nu_\rho_\mu}$ are not the components of a tensor; but
\item $\nabla_\mu v^\nu $ are the components of a tensor.
\end{enumerate}
\end{exercise}

One can also define the covariant derivative~$\nabla_\mu\fvect{\omega}$ of a form~$\fvect{\omega}$, which is a form, with
\begin{equation}
\nabla_\mu \omega_\nu
\define \omega_{\nu;\mu}
= \partial_\mu \omega_\nu
	- \Gamma\indices{^\rho_\mu_\nu} \omega_\rho \ .
\end{equation}
More generally, the covariant derivative of a tensor is a tensor with components
\begin{empheq}[box=\fbox]{align}
T\indices{^{\mu_1\ldots\mu_n}_{\nu_1\ldots\nu_m;\rho}}
&\define
T\indices{^{\mu_1\ldots\mu_n}_{\nu_1\ldots\nu_m,\rho}}
+ \Gamma\indices{^{\mu_1}_{\sigma\rho}} T\indices{^{\sigma\ldots\mu_n}_{\nu_1\ldots\nu_m}}
+\ldots
+ \Gamma\indices{^{\mu_n}_{\sigma\rho}} T\indices{^{\mu_1\ldots\sigma}_{\nu_1\ldots\nu_m}}\nonumber\\
&\hspace{3.2cm}
- \Gamma\indices{^\sigma_{\nu_1\rho}} T\indices{^{\mu_1\ldots\mu_n}_{\sigma\ldots\nu_m}}
-\ldots
- \Gamma\indices{^\sigma_{\nu_m\rho}} T\indices{^{\mu_1\ldots\mu_n}_{\nu_1\ldots\sigma}}.
\end{empheq}
The structure is: there is a Christoffel symbol for each index of the tensor, with a plus sign if the index is upstairs (like  vectors), and a minus sign if the index is downstairs (like forms). One cannot mess up with the position of indices if one respects the rule of the preservation of index altitude.

\paragraph{Leibniz rule} Just like partial derivatives, covariant derivatives are subject to the Leibniz rule with respect to multiplication. An example tells everything:
\begin{equation}
\nabla_\mu\pa{ T^{\nu\rho} v_{\sigma} }
= (\nabla_\mu T^{\nu\rho}) v_{\sigma} + T^{\nu\rho} (\nabla_\mu v_\sigma) \ .
\end{equation}
In particular, for the scalar product of two vectors, we have
\begin{equation}
\partial_\mu (\fvect{u}\cdot\fvect{v})
= \partial_\mu(u^\nu v_\nu)
= \nabla_\mu(u^\nu v_\nu)
= v_\nu \nabla_\mu u^\nu + u^\nu \nabla_\mu v_\nu \ .
\end{equation}

\paragraph{Metric preservation} Last, but not least, we have
\begin{empheq}[box=\fbox]{equation}\label{eq:metric_preservation}
\nabla_\rho g_{\mu\nu} = 0 = \nabla_\rho g^{\mu\nu} \ ,
\end{empheq}
a property called metric-preservation by $\nabla$. Combined with the Leibniz rule, this means that whenever the metric appears in a covariant derivative, it can freely be taken in or out. A particular consequence is that indices can be freely raised and lowered when they are inside a covariant derivative. This property is not true for simple partial derivatives.

\begin{exercise}
Demonstrate equation~\eqref{eq:metric_preservation}, using that $\fvect{g}$ is a tensor.
\end{exercise}

\paragraph{Parallel transport} The covariant derivative of any tensor~$\mat{T}$ in the direction of a vector~$\fvect{u}$ is defined as
\begin{equation}
\nabla_{\fvect{u}} \mat{T} \define u^\mu \nabla_\mu \fvect{T} \ .
\end{equation}
Now consider a curve~$\mathscr{C}$ in space-time, parametrised by $\lambda$. The tangent vector to this curve has components~$t^\mu\define \dd x^\mu/\dd \lambda$. The covariant derivative of $\mat{T}$ with respect to $\lambda$ is then defined as
\begin{equation}
\Ddf{\mat{T}}{\lambda}
\define \nabla_{\fvect{t}}\fvect{T}
= t^\mu \nabla_\mu \mat{T} \ .
\end{equation}
The tensor~$\mat{T}$ is said to be \emph{parallely transported} along the curve~$\mathscr{C}$ if $\Dd \fvect{T}/\dd \lambda=\fvect{0}$ along~$\mathscr{C}$.

\subsection{Geodesics}
\label{sec:geodesics}
\newcommand{\geo}{\ensuremath{\mathscr{G}}}

There are two equivalent definition of a geodesic in Lorentzian geometry:
\begin{enumerate}
\item A geodesic is an extremal curve~$\mathscr{C}$. More precisely, for two events~$A$ and $B$ in space-time, the length or time between $A$ and $B$ along $\mathscr{C}$ must be stationary with respect to infinitesimal variations:
\begin{equation}
\frac{\delta s}{\delta x^\mu}=0
\qquad \text{with} \quad
s
= \int_A^B \dd s
= \int_A^B 
	\sqrt{\abs{g_{\mu\nu }\ddf{x^\mu}{\lambda}\ddf{x^\nu}{\lambda}}}
	\; \dd\lambda \ ,
\end{equation}
where $\lambda$ is any parameter on $\mathscr{C}$. The absolute value in the square-root is here to account for the time-like case. In that case, $s$ is usually denoted~$\tau$: it is the proper time between $A$ and $B$.

\item A geodesic is a \emph{self-parallel} curve, i.e., a curve whose tangent vector~$\fvect{t}$ satisfies~$\nabla_{\fvect{t}}\fvect{t}=\kappa\fvect{t}$, where $\kappa$ is any scalar function. In terms of components, this reads
\begin{equation}\label{eq:geodesic_equation}
\Ddf{t^\nu}{\lambda}
= \ddf{t^\nu}{\lambda} + \Gamma\indices{^\nu_\mu_\rho} t^\mu t^\rho 
= \kappa t^\nu \ .
\end{equation}
Equation~\eqref{eq:geodesic_equation} is called the \emph{geodesic equation}.
\end{enumerate}

Three categories of geodesics can be distinguished, depending on the nature of the tangent vector~$\fvect{t}$: it is time-like, null, or space-like if $\fvect{t}\cdot \fvect{t}$ is negative, zero, or positive.

\begin{exercise}
Show the equivalence of the above two definitions of a geodesic.
\end{exercise}

\begin{exercise}
Show that, if $\geo$\xspace is a geodesic described by eq.~\eqref{eq:geodesic_equation} then the norm of the tangent vector, $N\define\fvect{t}\cdot\fvect{t}=t^\mu t_\mu$, with $t^\mu=\dd x^\mu/\dd\lambda$, reads
\begin{equation}
\frac{1}{N}\ddf{N}{\lambda} = 2\kappa.
\end{equation}
Conclude that there exists a suitable choice for $\lambda$, called~\emph{affine parameter}, such that the geodesic equation has no right-hand side, that is, $\kappa=0$. Check that, in the time-like case, proper time~$\tau$ is such a parameter.
\end{exercise}

\subsection{Curvature}

\paragraph{Riemann tensor} There are various ways of introducing the curvature of a manifold. One that I particularly like is based on the so-called \emph{geodesic deviation equation}. If $\geo_1$ and $\geo_2$ are two very close geodesics, affinely parametrised by $s$, and if we call~$\xi^\mu(s)=x^\mu_2(s)-x^\mu_1(s)$ their separation vector, then
\begin{equation}\label{eq:geodesic_deviation_equation}
\Ddf[2]{\xi^\mu}{s} =
R\indices{^\mu_\nu_\rho_\sigma} t^\nu t^\rho \xi^\sigma \ ,
\end{equation}
where $t^\mu\define \dd x^\mu/\dd s$ is the tangent vector of one of the geodesics, and the four-index quantity $R\indices{^\mu_\nu_\rho_\sigma}$ represents the components of the \emph{Riemann curvature tensor}. Before we give their expression, let us discuss the geometrical meaning of eq.~\eqref{eq:geodesic_deviation_equation}. The left-hand side can be understood as a relative ``acceleration'' between the two geodesics, as one moves along them. In a \emph{flat} geometry, geodesics are straight lines, and therefore their relative distance changes at a constant rate as we move along them, $\xi^\mu\propto s$. This is the case of the Euclidean and Minkowski geometries, for which the Riemann tensor is zero. In a curved space, or space-time, things are different: two neighbouring geodesics can, for instance, start diverging and end up converging, like great circles on a sphere.

\begin{figure}[h!]
\centering
\input{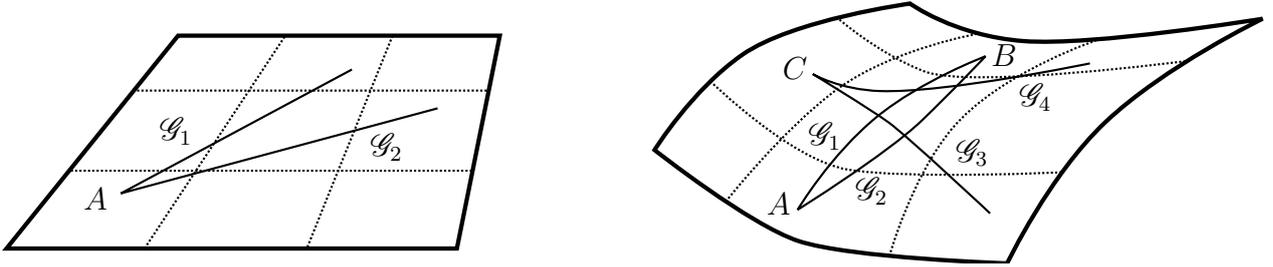}
\caption{\textit{Left:} two geodesics $\geo_1$ and $\geo_2$ in a flat space, diverging linearly from a point~$A$. \textit{Right}: geodesic deviation in a curved space; geodesics~$\geo_1$ and $\geo_2$ start diverging from $A$, and then converge again towards $B$; geodesics $\geo_3$ and $\geo_4$ diverge from $C$ quicker than linearly.}
\label{fig:geodesic_deviation}
\end{figure}

The Riemann tensor can also be defined by its effect on a vector~$\fvect{v}$,
\begin{empheq}[box=\fbox]{equation}
(\nabla_\mu \nabla_\nu - \nabla_\nu \nabla_\mu) v^\sigma
= R\indices{^\sigma_\rho_\mu_\nu} v^\rho \ ,
\end{empheq}
from which we can deduce the expression of its components.

\begin{exercise}
Show that the components of the Riemann tensor read
\begin{equation}
R\indices{^\sigma_\rho_\mu_\nu} 
=
\partial_\mu \Gamma\indices{^\sigma_\rho_\nu} - \partial_\nu \Gamma\indices{^\sigma_\rho_\mu}
+ \Gamma\indices{^\sigma_\lambda_\mu} \Gamma\indices{^\lambda_\rho_\nu}
- \Gamma\indices{^\sigma_\lambda_\nu} \Gamma\indices{^\lambda_\rho_\mu} \ .
\end{equation}
Mind that you only know how to apply covariant derivative to tensors. In particular, you should avoid to have terms like $\nabla_\mu\Gamma\indices{^\sigma_\nu_\rho}$ in your calculation. Justify that the Minkowski metric has a zero Riemann tensor.
\end{exercise}

\paragraph{Identities of the Riemann tensor} Although the Riemann tensor has, in four dimensions, $4^4=256$ possible combinations of indices, it enjoys a number of symmetries and identities that make this number fall to $20$. We give them here without proof:
\begin{align}
R_{\mu\nu\rho\sigma} &= -R_{\nu\mu\rho\sigma} \ ,\\
R_{\mu\nu\rho\sigma} &= -R_{\mu\nu\sigma\rho} \ ,\\
R_{\mu[\nu\rho\sigma]} &= 0 \ .
\end{align}
In the last line, $[\nu\rho\sigma]$ corresponds to a sum over all the permutations of $(\nu,\rho,\sigma)$, with a plus sign if the permutation is even, that is, if it corresponds to an even number of transpositions, and a minus sign if it is odd. Explicitly, we have
\begin{align}
R_{\mu[\nu\rho\sigma]}
&\define \frac{1}{3!}
\pa{ R_{\mu\nu\rho\sigma} + R_{\mu\rho\sigma\nu} + R_{\mu\sigma\nu\rho}
	- R_{\mu\nu\sigma\rho} - R_{\mu\rho\nu\sigma} - R_{\mu\sigma\rho\nu} }\\
&= \frac{1}{3} \pa{ R_{\mu\nu\rho\sigma} + R_{\mu\rho\sigma\nu} + R_{\mu\sigma\nu\rho} },
\end{align}
where the second line is obtained using the anti-symmetry of the last pair of indices. The above relations can also be combined to show that the components of the Riemann tensor are invariant under the exchange of the first pair and second pair of indices,
\begin{equation}
R_{\mu\nu\rho\sigma} = R_{\rho\sigma\mu\nu}.
\end{equation}

Finally, the covariant derivative of the Riemann tensor satisfies the \emph{Bianchi identity}
\begin{equation}\label{eq:Bianchi}
R_{\mu[\nu\rho\sigma;\lambda]}=0,
\end{equation}
where, again, $[\nu\rho\sigma;\lambda]$ corresponds to a full anti-symmetrisation over the indices $(\nu,\rho,\sigma,\lambda)$, that is, a sum over all permutations with a plus sign for even permutations and a minus sign for odd permutations.\footnote{Beware! An even permutation of four indices is \emph{not} a circular permutation. In general, an even (resp. odd) permutation is a permutation made of an even (resp. odd) number of \emph{transpositions}. A transposition is the exchange of two indices.}

\paragraph{Ricci tensor} The Ricci tensor~$R_{\mu\nu}$ is defined as a sort of trace of the Riemann tensor, in the sense that its components are
\begin{equation}
R_{\mu\nu} \define R\indices{^\rho_\mu_\rho_\nu} \ ,
\end{equation}
where we contracted the first and third indices.
\begin{exercise}
Using that the symmetries of the Riemann tensor, show that the Ricci tensor is symmetric, i.e. $R_{\mu\nu}=R_{\nu\mu}$.
\end{exercise}
Finally, we call \emph{Ricci scalar} the trace of the Ricci tensor,
%
$
R \define R^\mu_\mu = g^{\mu\nu} R_{\mu\nu} .
$

\section{Space-time tells matter how to fall}

As John A. Wheeler famously wrote in \textit{Geons, Black Holes, and Quantum Foam}~\cite{Wheeler:1998vs}, general relativity can be summarised in one sentence: ``Space-time tells matter how to move; matter tells space-time how to curve''. Equipped with our brand new tool kit, we are ready to successively explore both aspects of this sentence.

\subsection{Equivalence principles}

If one had to pick axioms, or fundamental principles, on which the general theory of relativity is built, the first one would certainly be the equivalence principle. There are three versions of it, which we will state from the weakest to the strongest, that is, from the easiest to the hardest to satisfy. This paragraph is inspired from the excellent presentation of Clifford Will in \textit{The Confrontation between General Relativity with Experiment}~\cite{Will:2014kxa}.

\paragraph{Weak equivalence principle} The weak equivalence principle is the universality of free fall. It states that any massive object has the same motion under an external gravity field, regardless of its mass or composition. To be precise, this applies to \emph{test bodies}. A test body is defined such that
\begin{enumerate}
\item no force apart from gravity act upon it (free fall);
\item the object is small enough not to experience tidal forces;
\item the object is light enough not to affect the geometry of space-time.
\end{enumerate}

The weak equivalence principle is easy to satisfy, in the sense that it is not too hard to cook up a theory of gravity in which the above is true. In Newtonian gravity, it is ensured by the equality between inertial mass and passive gravitational mass.

As already mentioned in the previous chapter, the universality of free fall is now tested at an exquisite level of precision. The E\"otv\"os ratio~$\eta$, defined as the relative acceleration of two bodies~$1$ and $2$ in a gravity field, has been constrained to be
\begin{equation}
\eta \define 2\, \frac{|\vec{a}_1-\vec{a}_2|}{|\vec{a}_1+\vec{a}_2|} < 1.3\times 10^{-14}
\end{equation}
for $(1, 2) = (\text{Pt, Ti})$, by the MICROSCOPE experiment~\cite{Touboul:2017grn}.

\paragraph{Einstein equivalence principle} This is the heart of the philosophy of general relativity. Given the universality of free fall, if I am freely falling myself, then any other freely falling body near me will have, in my own frame, a linear trajectory with constant velocity. For this reason, we can call \emph{inertial frame} any non-rotating freely-falling frame. Indeed, this definition fits with the one given by Newton's first law. The important difference is that, now, inertial frames are not a just a conceptual notion: they really exist in nature.

This reasoning applies to the motion of test bodies, but Einstein generalised it to any physical phenomenon. The \emph{Einstein equivalence principle} states that the outcome of any non-gravitational experiment (like an electromagnetic phenomenon) performed in any freely-falling frame is identical to its outcome in the absence of gravity.

A refined version of this principle can be formulated as:
\begin{enumerate}
\item The weak equivalence principle is valid.
\item The outcome of a non-gravitational experiment is independent of the velocity of the freely-falling frame in which it is performed; this is called \emph{local Lorentz invariance}.
\item The outcome of a non-gravitational experiment is independent of the location, in space-time, of the freely-falling frame in which it is performed; this is called \emph{local position invariance}.
\end{enumerate}

The Einstein equivalence principle is the reason why differential geometry is the natural language of general relativity. Indeed, if gravity is encoded in the geometry of space-time, then one should see a correspondence between the equivalence principle and the property of \emph{local flatness} of Lorentzian manifolds, that is, the fact that any manifold locally coincides with its tangent space-time at any point. For that reason, the Einstein equivalence principle is also relatively easy to satisfy; thanks to local flatness, it can be incorporated in any theory where gravity is encoded in space-time geometry, independently of this geometry and how it is produced.

\paragraph{Strong equivalence principle} The strong equivalence principle is the extension of the Einstein equivalence principle to all experiments, \emph{including gravitational experiments}. For example, this means that the attraction between the Sun and the Earth does not depend on the external (e.g. galactic)  gravitational field in which they are placed. Another important consequence is that, within all the forms of energies responsible for the inertia and gravity created by a physical system, gravitational binding energy contributes, too.

Contrary to the weak and Einstein equivalence principles, the strong equivalence principle is hard to satisfy. To date, general relativity (along with, to some extent, Norstr\"{o}m's gravity) is the only known theory that satisfies it.

\subsection{Geodesic motion}
\label{subsec:geodesic_motion}

\paragraph{Massive particles} Inspired from the Einstein-Fokker formulation of Nordstr\"{o}m's gravity, we postulate that the action of a massive test body, between two events $A$, $B$ of its world-line, reads
\begin{equation}\label{eq:action_free_particle_GR}
S[x^\mu] = -m \int_A^B \dd \tau \ , 
\end{equation}
where $m$ is the mass of the particle, and $\tau$ denotes the proper time measured along the particle's world-line, defined exactly like in special relativity, but with a general metric $g_{\mu\nu}$ instead of $f_{\mu\nu}$,
\begin{equation}
\dd \tau^2 = - \dd s^2 = - g_{\mu\nu} \dd x^\mu \dd x^\nu \ .
\end{equation}

Note that, in the expression of $S$, we have now dropped the factor $c^2$. Indeed, given the ubiquity of $c$ in relativity, it can be tedious to write it all the time. Thus, it is customary to work in a system of units such that $c=1$. For instance, if one uses the second as a time unit, the corresponding unit of distance has to be the light-second, i.e. the distance travelled by light during one second. In this case, one can consider that times and distances have the same dimension. We will adopt this convention in the remainder of the course.

The action principle $\delta S/\delta x^\mu = 0$ then means that the particle follows a \emph{time-like geodesic}. The corresponding geodesic equation can be derived easily using the following trick. The four-velocity of the particle satisfies~$\fvect{u}\cdot\fvect{u}=-1$, indeed, along the world-line,
\begin{equation}
\dd \tau^2
= -g_{\mu\nu} \dd x^\mu \dd x^\nu
= -g_{\mu\nu} (u^\mu \dd\tau) (u^\nu \dd\tau)
= (-g_{\mu\nu} u^\mu u^\nu) \dd\tau^2 \ .
\end{equation}
We can then rewrite the action as follows,
\begin{equation}
-\frac{S}{m}
= \int_A^B \pa{ -g_{\mu\nu} \ddf{x^\mu}{\tau} \ddf{x^\nu}{\tau} } \dd\tau \ ,
\end{equation}
where we just multiplied the integrand by $\sqrt{-g_{\mu\nu}u^\mu u^\nu}=1$. Calling~$L$ this new integrand, we can apply the Euler-Lagrange equation as
\begin{align}
-\frac{1}{m}\frac{\delta S}{\delta x^\mu}
&= \ddf{}{\tau}\pa{ \frac{\partial L}{\partial \dot{x}^\mu} } - \pd{L}{x^\mu} \\
&= \ddf{}{\tau} \pa{ 2 g_{\mu\nu} u^\nu } - g_{\nu\rho,\mu} u^\nu u^\rho \\
&= 2 \pa{ \ddf{u^\mu}{\tau} 
	+ \Gamma\indices{^\mu_\nu_\rho} u^\nu u^\rho } .
\end{align}
Hence, the equation of motion of the test particle is
\begin{empheq}[box=\fbox]{equation}
\Ddf{u^\mu}{\tau} = 0 ,
\end{empheq}
with
\begin{equation}
u^\nu \nabla_\nu u^\mu
=
\Ddf{u^\mu}{\tau}
=
\ddf{u^\mu}{\tau} 
	+ \Gamma\indices{^\mu_\nu_\rho} u^\nu u^\rho 
=
\ddf[2]{x^\mu}{\tau} 
+ \Gamma\indices{^\mu_\nu_\rho}\ddf{x^\nu}{\tau} \ddf{x^\rho}{\tau} \ ,
\end{equation}
from which we conclude that $\tau$ is an affine parameter (see \S~\ref{sec:geodesics}). Here, the Christoffel symbols not only contain the effect of a static change of coordinates, like in Newtonian physics, or the fictitious forces related to a change of frame, like in special relativity, they also contain the gravitational force.

\begin{exercise}
Let the space-time metric take the form
\begin{equation}
\dd s^2 = - \ex{2\Phi} \dd t^2 + \ex{-2\Phi} \delta_{ij} \dd x^i \dd x^j .
\end{equation}
From a variational approach, show that the geodesic equation reads
\begin{align}
0 &= \ddot{t}
		+ 2 \partial_i\Phi \dot{t} \dot{x}^i
		- \partial_t\Phi \ex{-4\Phi}
		 \delta_{ij} \dot{x}^i \dot{x}^j \ ,\\
0 &= \ddot{x}^i
		+ \ex{4\Phi} \delta^{ij} \partial_j\Phi \dot{t}^2
		- 2 \partial_t\Phi \dot{t} \dot{x}^i
		- ( \delta^i_k \delta^j_l + \delta^i_l \delta^j_k
			- \delta_{kl} \delta^{ij}  ) 
			\partial_j\Phi \dot{x}^k \dot{x}^l \ ,
\end{align}
where a dot denotes a derivative with respect to $\tau$. Deduce the expression of the Christoffel symbols. This can be remembered as a quick method to compute Christoffel symbols, especially when the metric is diagonal.
\end{exercise}

\paragraph{Fermi normal coordinates} The Einstein equivalence principle states that, in a freely falling frame, the laws of physics are the same as in an inertial frame in the absence of gravitation. We mentioned that this property is tightly related to the local flatness of Lorentzian manifolds. Here is the mathematical explanation.

Consider an observer~$\mathcal{O}$ in free fall, so that his world-line $\wl$ is a time-like geodesic. In this condition, one can show\footnote{The proof is not too hard, but a bit long. We will therefore admit this result here. The interested reader is referred to, e.g., the excellent \emph{A Relativist's Toolkit}~\cite{Poisson:2009pwt}, by Eric Poisson, for more details.} that there always exists a system of coordinates~$(X^\alpha)=(\tau, X^a)$, called \emph{Fermi normal coordinates} (FNCs), where $\tau$ is the observer's proper time, $X^a=0$ on $\wl$ (the spatial origin coincides with the observer), and such that the metric reads
\begin{empheq}[box=\fbox]{align}
g_{00} &=  -1 - R_{0a0b}(\tau, \vec{0}) X^a X^b + \mathcal{O}(\vec{X})^3 \\
g_{0a} &=  -\frac{2}{3} R_{0bac}(\tau, \vec{0}) X^b X^c 
					+ \mathcal{O}(\vec{X})^3 \\
g_{ab} &= \delta_{ab} - \frac{1}{3} R_{acbd}(\tau, \vec{0}) X^c X^d
					+\mathcal{O}(\vec{X})^3 .
\end{empheq}
In other words,
\begin{equation}
\forall \tau \qquad
\dd s^2 = \pac{\eta_{\alpha\beta} + \mathcal{O}(\vec{X})^2}
					\dd X^\alpha \dd X^\beta \ .
\end{equation}
We have, in particular, $\Gamma\indices{^\alpha_\beta_\gamma}(\tau,\vec{0})=0$, i.e. everywhere on $\wl$. FNCs are the local version of ICCs for any metric~$g_{\mu\nu}$. If you are freely falling, equipped with a clock and three rigid rulers, orthogonal to each other, then $\tau$ is the time that you measure with the clock, and $X^a$ are the distances that you measure with the rulers.

The distance from which $g_{\alpha\beta}$ starts to deviate significantly from $\eta_{\alpha\beta}$, i.e., from which the effects of gravity cannot be neglected any more, are set by the Riemann curvature of space-time. Curvature corresponds to the tidal effects mentioned at the end of chapter~\ref{chap:Newton}. Just like tidal forces cannot be eliminated in a freely-falling frame, curvature cannot be eliminated by picking inertial coordinates.

Remember that what was globally true for Minkowski is only \emph{locally} valid in general. While we could impose $f_{\alpha\beta}=\eta_{\alpha\beta}$ everywhere with a single coordinate transformation, we have $g_{\alpha\beta}=\eta_{\alpha\beta}$ only in the vicinity of a single time-like geodesic. This means that two freely-falling observers at a distance do not measure the same times and distances.

\paragraph{Mass-less particles} The action of a particle with no mass cannot be expressed as in eq.~\eqref{eq:action_free_particle_GR}, not only because $m=0$, but also because such a particle moves at the speed of light, i.e. along a null curve, for which $\dd s^2=0$ by definition. Nevertheless, the mass-less case can be considered a limit of the massive case. Let $\mathcal{O}$ be an observer and $\mathcal{P}$ a particle with mass $m$ and four-momentum~$\fvect{p}$. Suppose that $\mathcal{P}$ passes close to $\mathcal{O}$, so that we can use FNCs $(X^\alpha)$. Then everything happens as in Minkowski, and
\begin{equation}
E\e{free}^2 = (p^0)^2 = m^2 + \delta_{ab} p^a p^b \ .
\end{equation}
In the ultra-relativistic regime, i.e., if the energy~$E\e{free}$ of $\mathcal{P}$ is much larger than its rest-mass energy, we have $(p^0)^2 \approx \delta_{ab} p^a p^b$. In this regime, the particle moves almost at light-speed, and we can compare it to a photon. The corresponding four-momentum reads $\hbar\fvect{k}$, where
\begin{equation}
(k^\alpha)=(\omega,\vec{k})
\qquad \text{with} \quad
\begin{system}
\omega &: \text{cyclic frequency}\\
\vec{k} &: \text{wave-vector},
\end{system}
\end{equation}
is the photon's wave-four vector. Since, for $E\e{free}\rightarrow \infty$, we have $\fvect{p}=m\fvect{v}\rightarrow \hbar\fvect{k}$, and since for any value of $E\e{free}$ the trajectory of $\mathcal{P}$ satisfies $p^\nu \nabla_\nu p^\mu = 0$, we conclude that
\begin{empheq}[box=\fbox]{equation}\label{eq:null_geodesic}
k^\nu \nabla_\nu k^\mu = 0 \ .
\end{empheq}
The wave four-vector plays here the role of a four-velocity, in the sense that it is tangent to the photon's world-line. The main difference with the massive case is that this tangent vector is null,
\begin{equation}
\fvect{k}\cdot\fvect{k} = k^\mu k_\mu = 0 \ ,
\end{equation}
photons are thus following \emph{null geodesics} of space-time.

Another difference with the massive case is that one cannot write $k^\mu=\dd x^\mu/\dd \tau$, since there is no proper time along a null curve. Instead, one writes $\dd x^\mu/\dd\lambda$, where $\lambda$ is an affine parameter on the photon's world-line. In terms of $\lambda$, eq.~\eqref{eq:null_geodesic} can be rewritten as
\begin{equation}
\Ddf{k^\mu}{\lambda}
=
\ddf{k^\mu}{\lambda} 
	+ \Gamma\indices{^\mu_\nu_\rho} k^\nu k^\rho 
= \ddf[2]{x^\mu}{\lambda} 
	+ \Gamma\indices{^\mu_\nu_\rho} \ddf{x^\mu}{\lambda}
															\ddf{x^\nu}{\lambda} 
= 0 \ .
\end{equation}

\begin{exercise}\label{ex:frequency_photon}
Let us interpret $\lambda$ physically. Suppose that a photon passes by an observer~$\mathcal{O}$ with four-velocity~$\fvect{u}$. Show that, in the observer's frame, between~$\lambda$ to $\lambda+\dd\lambda$, the photon has moved by a distance~$\dd\ell=\omega\dd\lambda$, where $\omega$ is the cyclic frequency of the photon as measured by $\mathcal{O}$.
\end{exercise} 

\subsection{Physics in curved space-time}

Let us close this section by sketching how one uses the Einstein equivalence principle to incorporate gravity into the laws of physics in four dimensions.

\paragraph{Mechanics in curved space-time} The geodesic equation characterising free fall, $\Dd u^\mu / \dd\tau = 0$, has exactly the same form as the analogue of Newton's equation in Minkowski space-time, eq.~\eqref{eq:relativistic_Newton_2}, for $F^\mu=0$. The analogy goes even further: in the presence of gravitation, the equation of motion of a particle in the presence of gravity reads
\begin{equation}
\Ddf{p^\mu}{\tau} = F^\mu \ ,
\end{equation}
where the only difference with sec.~\ref{sec:physics_four_dimensions} is that the metric is now a general~$g_{\mu\nu}$, and not necessarily the Minkowski metric~$f_{\mu\nu}$. If the four-force derives from a potential~$U$, and that we write the above equation explicitly, we find
\begin{equation}\label{eq:eom_curved_space-time}
\ddf{}{\tau}\pac{ (m+U) u^\mu }
+ (m+U) \Gamma\indices{^\mu_\nu_\rho} u^\nu u^\rho
= -\partial^\mu U \ .
\end{equation}
The first term on the left-hand side contains the acceleration of the particle, and the second term with Christoffel symbols now contains not only fictitious forces, but also gravity. This shows that gravity can be considered a fictitious force: its effect only appears in frames that are not freely falling, i.e. non-inertial frames. Just like in the Minkowski case, eq.~\eqref{eq:eom_curved_space-time} derives from an action principle, with
\begin{equation}
S = -\int_A^B (m+U) \, \dd\tau \ .
\end{equation}

\paragraph{Minimal coupling} Consider a matter field~$\mat{\psi}$. This field can stand for a scalar field, like the Higgs boson or the Nordstr\"{o}m field, but also for a spinor field, like fermions, or for a vector field like the photon, etc. Suppose that, \emph{in the absence of gravity}, where space-time is described by the Minkowski metric, the classical dynamics of this field is ruled by an action of the form
\begin{equation}
S[\mat{\psi}] = \int \mathcal{L}(\mat{\psi},\partial_\alpha\mat{\psi})\;\dd^4 X \ ,
\end{equation}
where $\dd^4 X\define \dd X^0 \dd X^1 \dd X^2 \dd X^3$, and it is understood that $(X^\alpha)$ are ICCs. The integrand~$\mathcal{L}$ is called the \emph{Lagrangian density} of the field, and it is assumed to depend only on $\psi$ and its first derivatives. This is the case, for example, of the Lagrangian of the whole standard model of particle physics. The action~$S$ can be rewritten in an arbitrary coordinate system~$(x^\mu)$ as follows:
\begin{enumerate}
\item The partial derivative~$\partial_\alpha\mat{\psi}$ must be replaced with a covariant derivative~$\nabla_\mu\mat{\psi}$. If $\mat{\psi}$ is a scalar field, it does not change anything, but if it is, e.g., a vector field, we have seen that the covariant derivative ensures a correct behaviour with respect to coordinate transformations.
\item Change the element of space-time~$\dd^4 X$ accordingly. Indeed, we know that for any change of variable~$X^\alpha\rightarrow x^\mu$ in an integral, the differential element must be multiplied by the absolute value of the Jacobian of the transformation:
\begin{equation}
\dd^4 X = \abs{\det \pac{\pd{X^\alpha}{x^\mu}}} \dd^4 x \ .
\end{equation}
\end{enumerate}

\begin{exercise}
Using the expression~\eqref{eq:Minkowski_explicit} of the Minkowski metric, show that
\begin{equation}
\abs{\det \pac{\pd{X^\alpha}{x^\mu}}} = \sqrt{ -\det\pac{f_{\mu\nu}} } \ .
\end{equation}
The determinant of the metric $\det\pac{f_{\mu\nu}}$ is usually denoted simply~$f$, for short.
\end{exercise}
Summarising, \emph{in the absence of gravity}, the action of~$\mat{\psi}$ reads
\begin{equation}
S[\mat{\psi}] = \int
							\mathcal{L}(\mat{\psi}, \nabla_\mu\mat{\psi},f_{\mu\nu})
							\; \sqrt{-f} \, \dd^4 x \ ,
\end{equation}
where we specified the dependence in the Minkowski metric~$f_{\mu\nu}$ because, as $\mathcal{L}$ is a scalar, if it depends on $\nabla_\mu\mat{\psi}$ somewhere, we need something to contract indices.

The minimal change that we can make to this action, in order to incorporate gravity, consists in replacing the Minkowski metric~$f_{\mu\nu}$ by a general~$g_{\mu\nu}$ accounting for the distortions of space-time. We are therefore left with
\begin{empheq}[box=\fbox]{equation}
S[\mat{\psi},\mat{g}]
= \int
	\mathcal{L}(\mat{\psi}, \nabla_\mu\mat{\psi},g_{\mu\nu})
	\; \sqrt{-g} \, \dd^4 x \ ,
\end{empheq}
so that, in the case where the effects of gravity are negligible ($g_{\mu\nu}\approx f_{\mu\nu}$), we recover the dynamics of the action we started from. This defines the \emph{minimal coupling} between $\mat{\psi}$ and gravitation. It is minimal because, in principle we could have added other terms to $S$, which would also vanish for $g_{\mu\nu} = f_{\mu\nu}$; for example, terms depending on the Riemann curvature tensor:
\begin{equation}
\mathcal{L}
(\mat{\psi}, \nabla_\mu\mat{\psi},g_{\mu\nu}, R_{\mu\nu\rho\sigma}, \ldots)\ .
\end{equation}
However, this would \emph{violate the Einstein equivalence principle}. Indeed, let $\mathcal{O}$ be a freely-falling observer, and $\mathcal{T}$ a narrow space-time ``tube'' around her world-line. Within this tube, we can use FNCs $(X^\alpha)$ such that~$g_{\alpha\beta}=\eta_{\alpha\beta}$ and $\nabla_\alpha=\partial_\alpha$. However, even with this coordinate system, $R_{\alpha\beta\gamma\delta}\not=0$ in general. Thus, the dynamics of $\mat{\psi}$ in $\mathcal{T}$ would explicitly depend on the local curvature of space-time, regardless of how narrow~$\mathcal{T}$ is. In other words, the results of an experiment using the physics of $\mat{\psi}$ would depend on where and when it is carried out, and on the velocity of the experimentalist who performs it.

\paragraph{Example of electrodynamics} The minimal-coupling prescription can be applied to electrodynamics. The fundamental field of electromagnetism is the four-vector potential~$(A_{\alpha})=(-V,\vec{A})$, where $V$ denotes the electrostatic potential and $\vec{A}$ the vector potential. Those potentials are related to the electric and magnetic fields via
\begin{align}
\vec{E} &= -\partial_t \vec{A} - \vec{\nabla} V \ , \\
\vec{B} &= \vec{\nabla} \times \vec{A} \ ,
\end{align}
which can be gathered in the antisymmetric Faraday tensor
\begin{equation}
F_{\alpha\beta} = \partial_\alpha A_\beta - \partial_\beta A_\alpha
\qquad\text{with}\quad
[F_{\alpha\beta}]
=
\begin{bmatrix}
0 & -E^1 & -E^2 & -E^3 \\
E^1 & 0 & B^3 & -B^2 \\
E^2 & -B^3 & 0 & B^1 \\
E^3 & B^2 & -B^1 & 0
\end{bmatrix}.
\end{equation}
With such notation, Maxwell's equations read~$\partial_\alpha F^{\alpha\beta} = 4\pi J^\beta$, where $(J^\alpha)=(\rho\e{e}, \vec{J}\e{e})$ denotes the electric four-current; $\rho\e{e}$ is the electric charge density, while $\vec{J}\e{e}$ is the electric current density. This equation derives from an action with Lagrangian density
\begin{equation}
\mathcal{L} = -\frac{1}{16\pi} \, F^{\alpha\beta} F_{\alpha\beta}
						+ A_\alpha J^\alpha \ .
\end{equation}

Applying the minimal coupling prescription, we thus obtain the action of electrodynamics in the presence of gravitation,
\begin{equation}\label{eq:Maxwell_action}
S[A_\mu, g_{\mu\nu}]
= \int
		\pac{ -\frac{1}{16\pi} \, g^{\mu\rho} g^{\nu\sigma}
					F_{\mu\nu} F_{\rho\sigma}
				+ A_\mu J^\mu
				} \sqrt{-g}\,\dd^4 x \ .
\end{equation}
The fact that any field naturally couples to the metric in this way is responsible for the \emph{universality} of gravitation: it affects everything, and, in turn, is affected by everything.

\begin{exercise} Taking the variation of eq.~\eqref{eq:Maxwell_action} with respect to $A_\mu$, show that the field equation for electrodynamics in the presence of gravity reads
\begin{equation}
\nabla_\mu F^{\mu\nu} = 4\pi J^\nu.
\end{equation}
\textit{Hint}: Prove and use the identity $\partial_\mu (\sqrt{-g} \, F^{\mu\nu})=\sqrt{-g}\,\nabla_\mu F^{\mu\nu}$.
\end{exercise}

\section{Matter tells space-time how to curve}

The previous lecture concerned the \emph{passive} aspect of gravitation, namely, how physics undergoes the effect of an external gravity field, encoded in the geometry of space-time. We now address its \emph{active} side, namely, how this geometry is generated.

\subsection{Energy-momentum tensor}

Just like the Poisson equation of Newtonian gravitation relates the gravitational field~$\Phi$ to the matter mass density~$\rho$, we would like to have an equation relating the metric~$g_{\mu\nu}$ to the energetic properties of matter. Moreover, since the laws of physics are coordinate-independent, the field equation of GR must be covariant: they must take either a scalar, a vector, or a tensor form.

\paragraph{Why a tensor?} As seen in \S~\ref{sec:differential_geometry}, all the geometric quantities that can be constructed from the metric have an even number of indices ($g_{\mu\nu}, R_{\mu\nu\rho\sigma}, \ldots$); therefore, we need to construct a field related to the energy of matter which is, \textit{a minima}, a scalar, and if it does not work, a tensor with two indices, or four, six, etc.

We have seen that the energy of a particle cannot be separated from its momentum. Both notions are encapsulated in its four-momentum~$\fvect{p}$. This suggests that we cannot construct directly a scalar field that would describe the energy of a set of particles: it has to be, at least, a vector. This, combined with the previous geometric argument, encourages us to build a tensor field using $\fvect{p}$.

\paragraph{Point particles} Consider a single point particle, assumed for simplicity be massive ($m\not=0$), with four-momentum~$\fvect{p}$, and whose world-line is described by $Y^\alpha(t)$ in the FNC system of an arbitrary observer\footnote{We use $X^0=t$, because we want to keep the notation~$T$ for the energy-momentum tensor}~$(X^\alpha)=(t,X^a)$. A tensor field built from two occurrences of $p^\alpha$ could be, for example,
\begin{equation}\label{eq:energy-momentum_first_attempt}
T^{\alpha\beta}(t, X^c)
= \frac{p^\alpha p^\beta}{m} \, \delta\e{D}^{(3)}[X^c-Y^c(t)] \qquad \text{(first attempt)}.
\end{equation}
In the above, the three-dimensional Dirac ``function'' $\delta\e{D}^{(3)}$ ensures that $T^{\alpha\beta}(t, X^c)=0$ if $(t,X^c)$ is not on the word-line of the particle; besides, we divided by the mass~$m$ so that the result has the dimension of a mass per unit volume, like~$\rho$.

The issue with this first attempt is that $T^{\alpha\beta}$ does not transform as a tensor under Lorentz boosts. This is because the Dirac function~$\delta\e{D}^{(3)}$ is not a scalar. Suppose one performs a Lorentz boost~$X^\alpha \rightarrow \tilde{X}^\beta = B\indices{^\beta_\alpha} X^\alpha$, then
\begin{equation}
\delta\e{D}^{(3)}(X^a)
= \frac{\dd^3 \tilde{X}}{\dd^3 X} \, \delta\e{D}^{(3)}(\tilde{X}^b)
= |\det[B\indices{^a_b}]| \, \delta\e{D}^{(3)}(\tilde{X}^b)
= \gamma \, \delta\e{D}^{(3)}(\tilde{X}^b) \ .
\end{equation}
The Lorentz factor that appears above can be understood as an effect of the relativistic contraction of lengths. We can circumvent this problem by replacing, in eq.~\eqref{eq:energy-momentum_first_attempt}, $m$ by $p^0$, whose transformation under boosts compensates for the transformation of $\delta^{(3)}\e{D}$. With this replacement, and for a set of $N$ particles following the world-lines~$Y_n^\alpha(t)$, we have
\begin{empheq}[box=\fbox]{equation}
\label{eq:energy-momentum_N_particles_inertial}
T^{\alpha\beta}(t,X^c)
= \sum_{n=1}^N
	\frac{p_n^\alpha p_n^\beta}{p^0_n} \, \delta\e{D}^{(3)}[X^c-Y_n^c(t)] \ .
\end{empheq}
This is called the \emph{energy-momentum tensor} (or stress-energy tensor) of the system of $N$ point particles, in a local inertial frame. We can finally rewrite it in an explicitly coordinate-independent way, by turning the three-dimensional Dirac function into a four-dimensional one. For that purpose, we can introduce an integration along the particles' world-lines~$y^\rho_n(\lambda)$, parametrised by $\lambda$, so that
\begin{equation}\label{eq:energy-momentum_N_particles}
T^{\mu\nu}(x^\rho)
=
 \sum_{n=1}^N \int
	\frac{p^\mu_n p^\nu_n}{p^0_n} \, \ddf{x^0}{\lambda} \,
			\frac{\delta\e{D}^{(4)}[x^\rho-y^\rho_n(\lambda)]}{\sqrt{-g}}
		 \; \dd\lambda \ ,
\end{equation}
where $\dd x^0/\dd\lambda$ is here to ensure the correct normalisation of the Dirac function, whose temporal part concerns $x^0$, while integration is performed over $\lambda$.

\begin{exercise}
Show that $T^{\mu\nu}$, as defined in eq.~\eqref{eq:energy-momentum_N_particles}, behaves as a tensor under general coordinate transformations. Check that eq.~\eqref{eq:energy-momentum_N_particles_inertial} is recovered with FNCs.
\end{exercise}

Equation~\eqref{eq:energy-momentum_N_particles} has the advantage of being valid even if $m\not=0$. In the massive case, it can be put under a more aesthetic form, by choosing~$\lambda=\tau_n$ for each integral; indeed,
\begin{equation}
\frac{1}{p^0_n} \ddf{x^0}{\tau_n} 
= \frac{1}{m_n u^0_n} \ddf{x^0}{\tau_n}
= \frac{1}{m_n} \ ,
\end{equation}
and hence
\begin{equation}\label{eq:energy-momentum_N_massive_particles}
T^{\mu\nu}(x^\rho)
=
 \sum_{n=1}^N m_n \int u^\mu_n u^\nu_n \, 
			\frac{\delta\e{D}^{(4)}[x^\rho-y^\rho_n(\tau)]}{\sqrt{-g}}
		 \; \dd\tau \ .
\end{equation}

\paragraph{Physical interpretation} It is interesting to explore the physical meaning of the tensor~$T^{\alpha\beta}$ as given in eq.~\eqref{eq:energy-momentum_N_particles_inertial}. Let us start with its $[00]$ component, which reads
\begin{equation}
T^{00}(t,X^c)
= \sum_{n=1}^N p_n^0 \, \delta\e{D}^{(3)}[X^c-Y_n^c(t)]
= \sum_{n=1}^N E_n \, \delta\e{D}^{(3)}[X^c-Y_n^c(t)]
\define \rho \ .
\end{equation}
This quantity represents the \emph{energy density} of the system of $N$ particles, usually denoted~$\rho$, despite the fact that it does not only contain the rest-mass energy but also the kinetic energy of the particles. Furthermore, if the particles were experiencing any non-gravitational potential energy~$U$, then the latter would also count in $E_n$.

The $[0a]$ components read
\begin{equation}
T^{0a}(t,X^c)
= \sum_{n=1}^N p_n^a \, \delta\e{D}^{(3)}[X^c-Y_n^c(t)] \ ,
\end{equation}
which represents the \emph{momentum density} of the system. Alternatively, since $p_n^a = E_n v_n^a$, where $v_n^a$ is the velocity of the particle $n$, $T^{0a}$ can also be seen as the \emph{energy flux density} in the direction $X^a$. For a small surface $\dd A$ with unit normal $\vec{n}$, the energy carried by the particles going through this surface in the direction of $\vec{n}$ during $\dd t$ is~$\dd E = T^{0 a} n_a \dd A \, \dd t$.

\begin{figure}[h!]
\centering
\input{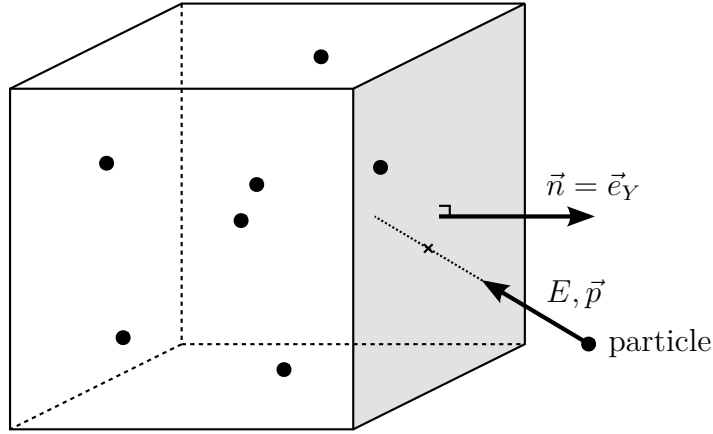}
\caption{We consider a small element of volume~$\dd V=\dd X \dd Y \dd Z$. During $\dd t$, particles get in and out. When  a particle enters through the right face, its energy contributes to $-T^{0Y}$, and its momentum~$\vec{p}=(p^a)$ to $-T^{Ya}$. The sign would be positive if the particle were exitting.}
\label{fig:energy-momentum}
\end{figure}

Finally, the component $[ab]$ is
\begin{equation}
T^{ab}(t,X^c)
= \sum_{n=1}^N v_n^a p_n^b \, \delta\e{D}^{(3)}[X^c-Y_n^c(t)] \ ,
\end{equation}
and thus represents the \emph{momentum flux density} in the direction $X^a$ projected on $X^b$, or vice-versa since $T^{ab}=T^{ba}$. For a small surface $\dd A$ with unit normal $\vec{n}$, the amount of momentum carried by the particles crossing the surface in the direction of $\vec{n}$ during $\dd t$ is~$\dd\vec{P}=T^{ab} n_a \vec{\partial}_b \, \dd A \, \dd t$. This is summarised in fig.~\ref{fig:energy-momentum}.

\paragraph{Perfect fluid} Consider a subset~$N_{\mathcal{D}}$ of our $N$ particles, localised in a small spatial domain~$\mathcal{D}$ with volume~$V_{\mathcal{D}}$, and let us assume that the local inertial frame corresponding to the coordinates $(X^\alpha)$ coincides with the barycentric frame of this subset, i.e. the rest-frame of its centre of mass. We would like to analyse the effective behaviour of $T^{\alpha\beta}$, once smoothed over~$\mathcal{D}$. We have already seen that $T^{00}$ represents the energy density~$\rho$ of the system. More precisely, for the domain~$\mathcal{D}$, we have
\begin{equation}
\ev{T^{00}}_{\mathcal{D}}
\define \frac{1}{V_{\mathcal{D}}} \int_{\mathcal{D}} T^{00}(t,X^c) \, \dd^3 X
= \frac{E_{\mathcal{D}}}{V_{\mathcal{D}}}
\define \rho_{\mathcal{D}} \ .
\end{equation}
Regarding $T^{0a}$, we find
\begin{equation}
\ev{T^{0a}}_{\mathcal{D}}
\define \frac{1}{V_{\mathcal{D}}} \int_{\mathcal{D}} T^{0a}(t,X^c) \, \dd^3 X
= \frac{1}{V_{\mathcal{D}}} \sum_{n\in\mathcal{D}} p^a_n
= 0
\end{equation}
in the barycentric frame. Finally, for the $[ab]$ component,
\begin{equation}
\ev{T^{ab}}_{\mathcal{D}}
\define \frac{1}{V_{\mathcal{D}}} \int_{\mathcal{D}} T^{ab}(t,X^c) \, \dd^3 X
= \frac{1}{V_{\mathcal{D}}} \sum_{n\in\mathcal{D}} 
	\gamma_n m_n v^a_n v^b_n \ .
\end{equation}
In the barycentric frame, if $a\not= b$, we can consider~$v^a$ and $v^b$ as independent random variables, with the same distribution if we assume that the system is isotropic; therefore,
\begin{equation}
\frac{1}{V_{\mathcal{D}}}\sum_{n\in\mathcal{D}} \gamma_n m_n v^a_n v^b_n
= \frac{N_{\mathcal{D}}}{V_{\mathcal{D}}}
	\frac{\ev{\gamma m v^2}}{3} \, \delta^{ab}
\define P_{\mathcal{D}} \, \delta^{ab} \ ,
\end{equation}
where $P_{\mathcal{D}}$ represents the \emph{kinetic pressure} of the particles in $\mathcal{D}$. Summarising,
\begin{equation}\label{eq:energy-momentum_mesoscopic}
\begin{system}
\ev{T^{00}}_{\mathcal{D}} &= \rho_{\mathcal{D}} \\
\ev{T^{0a}}_{\mathcal{D}} &= 0 \\
\ev{T^{ab}}_{\mathcal{D}} &= P_{\mathcal{D}} \, \delta^{ab}
\end{system}
\qquad \text{that is} \quad
\ev{T^{\alpha\beta}}_{\mathcal{D}}
= \rho_{\mathcal{D}} \, u^\alpha_{\mathcal{D}} u^\beta_{\mathcal{D}} 
	+ P_{\mathcal{D}} 
		(\eta^{\alpha\beta} + u^\alpha_{\mathcal{D}} u^\beta_{\mathcal{D}}) \ ,
\end{equation}
if $\fvect{u}_{\mathcal{D}}$ represents the four-velocity of the barycentric frame of $\mathcal{D}$. Since this domain is, in fact, arbitrary, we understand that eq.~\eqref{eq:energy-momentum_mesoscopic} describes the mesoscopic behaviour the system of $N$ particles. When their mutual interaction and the non-diagonal part of $\ev{T^{ab}}_{\mathcal{D}}$ is negligible, we say that the system behaves as a \emph{perfect fluid}, and its energy-momentum tensor is modelled by
\begin{empheq}[box=\fbox]{equation}
T^{\mu\nu} = \rho \, u^\mu u^\nu + P (g^{\mu\nu} + u^\mu u^\nu) ,
\end{empheq}
where $\fvect{u}$ is the local four-velocity of the fluid.

\paragraph{Relation with the action} The general expression of the energy-momentum tensor of a matter species actually derives from its action. Let us derive this particular relationship in the case of a single point particle with mass $m$. We have seen that the action of this particle is
\begin{equation}
S = -m \int \dd\tau 
	= -m\int \sqrt{-g_{\mu\nu} \dot{y}^\mu \dot{y}^\nu} \; \dd\lambda \ ,
\end{equation}
with $\dot{y}^\mu \define \dd y^\mu/\dd\lambda$, $\lambda$ being an arbitrary parameter on the world-line~$y^\mu(\lambda)$ of the particle.

This action can be rewritten as an integral over space-time, by introducing a Dirac delta function peaked on the particle's trajectory,
\begin{align}
S
&= -m \int \dd\lambda \int \dd^4 x \;
	\delta^{(4)}\e{D}[x^\rho-y^\rho(\lambda)]\,
	\sqrt{-g_{\mu\nu} \dot{y}^\mu \dot{y}^\nu}\\
&= -m \int \dd^4 x \int \dd\lambda \;
	\delta^{(4)}\e{D}[x^\rho-y^\rho(\lambda)]\,
	\sqrt{-g_{\mu\nu} \dot{y}^\mu \dot{y}^\nu} \ .
\end{align}
Varying this action with respect to the metric, we find
\begin{align}
\delta S 
&= 
\int \dd^4 x
\paac{ 
\int \dd\lambda \; 
	\frac{m \dot{y}^\mu \dot{y}^\nu\, 
	\delta^{(4)}\e{D}[x^\rho-y^\rho(\lambda)]}
			{2\sqrt{-g_{\mu\nu} \dot{y}^\mu \dot{y}^\nu}} 
} \delta g_{\mu\nu} \\
&= \int \dd^4 x
\paac{ 
\frac{1}{2} m
\int \dd\tau \; 
	u^\mu u^\nu \, \delta^{(4)}\e{D}[x^\rho-y^\rho(\tau)]
} \delta g_{\mu\nu} \ ,
\end{align}
where we changed integration variable from $\lambda$ to $\tau$ in the second line. We recognise in the curly brackets something that really looks like the energy-momentum tensor~\eqref{eq:energy-momentum_N_massive_particles}, for $N=1$; more precisely,
\begin{empheq}[box=\fbox]{equation}\label{eq:energy-momentum_general}
T^{\mu\nu} = \frac{2}{\sqrt{-g}} \, \frac{\delta S}{\delta g_{\mu\nu}} \ .
\end{empheq}
Equation~\eqref{eq:energy-momentum_general} is actually the general definition of the energy-momentum of a matter species. Once the action is known, $T^{\mu\nu}$ follows by functional derivation.

\subsection{Einstein's equation}

\paragraph{The equation of relativistic gravitation} The equation of the Einstein-Fokker reformulation of Nordstr\"{o}m's gravity was $R=24\pi G T$, where $T$ is the trace of the energy-momentum tensor of matter. This equation does not produce the correct law of gravitation; the one that does was derived by Einstein in 1915, and reads
\begin{empheq}[box=\fbox]{equation}\label{eq:Einstein}
R_{\mu\nu} - \frac{1}{2} \, R g_{\mu\nu} = 8\pi G \, T_{\mu\nu} \ .
\end{empheq}
It is naturally called \emph{Einstein's equation}, or the Einstein field equation. Its trace yields
\begin{equation}
R = -8\pi G\,T,
\end{equation}
which should be noted to differ from Nordstr\"{o}m's theory. Substituting the above in the original formulation of Einstein's equation yields
\begin{equation}
R_{\mu\nu} = 8\pi G \pa{ T_{\mu\nu} - \frac{1}{2} \, T g_{\mu\nu}} ,
\end{equation}
which is a useful expression. It shows in particular that in vacuum ($T_{\mu\nu}=0$) space-time is Ricci-flat ($R_{\mu\nu}=0$).

Einstein's equation is a non-linear system of 10 coupled partial differential equations for 10 functions ($g_{\mu\nu}$) of 4 variables ($x^\mu$). Non-linearity comes from the fact that the Ricci tensor involves the inverse of the metric, which is a non-linear operation, and products of the Christoffel symbols. As a consequence, contrary to many theories of physics (including Newtonian gravitation), Einstein's gravitation does not satisfy the superposition principle: if one doubles the amount of energy in the Universe, the metric does not get multiplied by two. However, Ricci curvature does.

Einstein's equation tells us that the Ricci curvature of space-time is locally ruled by the density of energy and momentum of matter. This is an important fact, which distinguishes it from Newton's gravity: \emph{not only mass actively gravitates, but any form of energy}. In particular, a hot gas, which has more energy than a cold gas, is heavier. A light beam, which contains energy and momentum, also curves space-time around it, and hence produces gravitational attraction.

\paragraph{The cosmological constant} Another term can be added to Einstein's equation without changing its essential properties,
\begin{equation}
R_{\mu\nu} - \frac{1}{2} \, R g_{\mu\nu} + \Lambda g_{\mu\nu}
= 8\pi G \, T_{\mu\nu} .
\end{equation}
where $\Lambda$ is called the \emph{cosmological constant}, and adds a constant Ricci curvature to space-time. Its net effect is a repulsive gravitational force that grows linearly with distance. The cosmological constant was introduced by Einstein in 1917, when he proposed the very first relativistic cosmological model~\cite{1917SPAW.......142E}. The role of $\Lambda$ was to counter-balance the attractive nature of gravity, and describe a Universe in agreement with Einstein's philosophical prior: a homogeneous, isotropic, eternal, and static Universe~\cite{1917SPAW.......142E}. The discovery of the expansion of the Universe by Hubble in 1929~\cite{1929PNAS...15..168H} led Einstein to refer to the cosmological constant as the ``biggest blunder of [his] life''~\footnote{According to George Gamow in his autobiography~\cite{Gamow}.}. Yet, $\Lambda$ is today the best way to explain the current \emph{acceleration} of the expansion of the Universe, discovered 70 years after Hubble's observations~\cite{1998AJ....116.1009R, 1999ApJ...517..565P}. Note that the cosmological constant is not a strictly relativistic concept: in Newtonian physics, it can be added to the Poisson equation as $\Delta\Phi+\Lambda=4\pi G\rho$.

\paragraph{Conservation of energy and momentum} The left-hand side of eq.~\eqref{eq:Einstein} is called the Einstein tensor. Its standard notation is $G_{\mu\nu}$, but along with other relativists I personally dislike this notation, since there is already a $G$ in Einstein's equation, referring to Newton's constant. We will therefore denote it
\begin{equation}
E_{\mu\nu} \define R_{\mu\nu} - \frac{1}{2} R g_{\mu\nu}.
\end{equation}

\begin{exercise}
Using the Bianchi identity~\eqref{eq:Bianchi}, show that the covariant divergence of the Einstein tensor vanishes, $\nabla_\mu E^{\mu\nu} = 0$.
\end{exercise}

When applied to the Einstein's equation, this relation yields
\begin{equation}
\nabla_\mu T^{\mu\nu} = 0,
\end{equation}
which corresponds to the local conservation of energy and momentum. To understand this, consider a local inertial frame $(X^\alpha)$ and a small spatial domain~$\mathcal{D}$. In that frame, the Christoffel symbols can be considered to vanish over~$\mathcal{D}$, and the equation reads
\begin{equation}
0 = \partial_\alpha T^{\alpha\beta}
= \partial_T T^{0\beta} + \partial_a T^{a\beta},
\end{equation}
which we can integrate over~$\mathcal{D}$ to get
\begin{equation}\label{eq:integrated_cons_energy}
\partial_T \int_\mathcal{D} T^{0\beta} \; \dd V
= -\int_{\partial\mathcal{D}} T^{a\beta} \; \dd A_a
\end{equation}
after applying the Green-Ostrogradski divergence theorem. For $\beta=0$, this corresponds to the conservation of energy. Indeed, we have seen that $T^{00}=\rho$ represents the energy density, while $T^{a0}=\Pi^a$ is the energy flux density, hence eq.~\eqref{eq:integrated_cons_energy} becomes
\begin{equation}
\partial_T E_\mathcal{D} = - \int_{\partial\mathcal{D}} \vec{\Pi}\cdot \dd \vec{A} \ ,
\end{equation}
which tells us that the variation of the energy inside~$\mathcal{D}$ is exactly equal to the energy entering through its boundary. For $\beta=b$, $T^{0b}=\Pi^b$ shall now be interpreted as a momentum density, so that its integral is the total momentum~$\vec{P}_{\mathcal{D}}$ inside~$\mathcal{D}$. Thus, eq.~\eqref{eq:integrated_cons_energy} reads
\begin{equation}
\partial_T P^b_{\mathcal{D}} = - \int_{\partial\mathcal{D}} T^{ab}\, \dd A_a \ ,
\end{equation}
which, like for energy, tells us that the variation of the momentum inside~$\mathcal{D}$ is equal to the momentum entering in it through its boundary.

\medskip

\noindent\textit{Remark.} Thanks to a mathematical property of the Riemann tensor, namely the Bianchi identity, Einstein's equation is consistent with the local conservation of energy and momentum. It is then a matter of taste what one should consider the most fundamental---is Einstein's equation a fundamental law of nature, which implies energy-momentum conservation; or is the latter more fundamental, and Einstein's equation is \emph{forced} to respect it, like any alternative theory of gravity should?

\begin{exercise}
Show that the conservation of energy and momentum~$\nabla_\mu T^{\mu\nu}=0$ of a perfect fluid leads to the following set of equations:
\begin{align}
u^\mu \nabla_\mu \rho + (\rho+P) \nabla_\mu u^\mu &= 0, \\
(\rho + P) u^\nu \nabla_\nu u^\mu 
+ (g^{\mu\nu}+ u^\mu u^\nu) \nabla_\nu P
&= 0.
\end{align}
Show that they can be interpreted as the continuity and Euler equations of hydrodynamics. Where is gravity in these equations?
\end{exercise}

\subsection{Action principle for gravitation}

\paragraph{Einstein-Hilbert action} Just like mechanics or field theory, relativistic gravitation can be formulated in terms of an action. The Einstein-Hilbert action is defined as
\begin{empheq}[box=\fbox]{equation}
S\e{EH}[\mat{g}]
= \frac{1}{16\pi G}
	\int \dd^4 x \,\sqrt{-g} \; R \ ,
\end{empheq}
where $R$ is the Ricci scalar, and $g$ denotes the determinant of the matrix~$[g_{\mu\nu}]$. One could add a cosmological constant term~$S_\Lambda$ to this action, as
\begin{equation}
S_{\Lambda}[\mat{g}]
\define -\frac{1}{8\pi G} \int \dd^4 x \, \sqrt{-g}\;\Lambda \ .
\end{equation}
We will show that the functional derivative of $S\e{g} \define S\e{EH}+S_\Lambda$ with respect to the metric corresponds to~$E_{\mu\nu}+\Lambda g_{\mu\nu}$.

\paragraph{Deriving Einstein's equation} Consider a region~$\mathcal{D}$ of space-time with metric~$g_{\mu\nu}$, and let us change this metric by an amount~$\delta g_{\mu\nu}$, such that~$\delta g_{\mu\nu}=0$ on the boundary~$\partial\mathcal{D}$ of $\mathcal{D}$. We first write~$R=g^{\mu\nu}R_{\mu\nu}$, so that
\begin{equation}\label{eq:variation_Einstein-Hilbert}
16\pi \delta S\e{g}
= \int_{\mathcal{D}} \dd^4 x \, \sqrt{-g} \,
	\pac{
			\frac{\delta \sqrt{-g}}{\sqrt{-g}} \, (R-2\Lambda)
			+ \delta g^{\mu\nu} R_{\mu\nu}
			+ g^{\mu\nu} \delta R_{\mu\nu}
			} .
\end{equation}

\begin{exercise}\label{ex:metric_determinant}
Let $\mat{M}$ be an invertible matrix, whose components are slightly varied by an amount $\delta\mat{M}$. The determinant of $\mat{M}+\delta\mat{M}$ can then be written as
\begin{equation}
\det(\mat{M}+\delta\mat{M}) = \det\mat{M} \det\pa{\mat{1}+\mat{M}^{-1}\delta\mat{M}},
\end{equation}
where we used that~$\det(\mat{A}\mat{B})=\det\mat{A}\det\mat{B}$. Expanding the above at first order, show that
\begin{equation}
\delta \det{\mat{M}} 
\define \det(\mat{M}+\delta\mat{M}) - \det\mat{M}
= \det\mat{M} \, \tr(\mat{M}^{-1}\delta\mat{M}).
\end{equation}
Applying this general result to the metric, conclude that
\begin{equation}
\frac{\delta\sqrt{-g}}{\sqrt{-g}} = \frac{1}{2} \, g^{\mu\nu} \delta g_{\mu\nu} \ .
\end{equation}
\end{exercise}

Since $g^{\mu\nu}$ is the inverse of $g_{\mu\nu}$, their variations are not independent. More precisely, considering the variation of $g^{\mu\rho}g_{\rho\nu}=\delta^\mu_\nu$, we get
\begin{equation}
\delta g^{\mu\rho} g_{\rho\nu} + g^{\mu\rho} \delta g_{\rho\nu} = 0 \ ,
\end{equation}
which we contract again with the inverse metric to get
\begin{equation}
\delta g^{\mu\nu} = - g^{\mu\rho} g^{\nu\sigma} \delta g_{\rho\sigma} \ .
\end{equation}

Combining the first two terms of the integrand of eq.~\eqref{eq:variation_Einstein-Hilbert}, and leaving the third term aside, we find
\begin{equation}
16\pi G \, \delta S\e{g}
= \int 
	\underbrace{
						\pa{ \frac{1}{2} \, R \, g^{\mu\nu} - R^{\mu\nu} 
								- \Lambda g^{\mu\nu} }
						}_{-E^{\mu\nu}-\Lambda g^{\mu\nu}}
	\delta g_{\mu\nu}
 	\sqrt{-g} \, \dd^4 x
 	+ \underbrace{
 							\int g^{\mu\nu} \delta R_{\mu\nu} \, \sqrt{-g} \, \dd^4 x 
 							}_{\define \delta B}
 		\ ,
\end{equation}
where we have recognised the Einstein tensor in the first integral. Let us now show that the second integral, $\delta B$, vanishes. The trick consists in using FNCs~$(X^\alpha)$, such that the Christoffel symbols vanish, and we are left with
\begin{equation}
\delta R_{\alpha\beta}
= \delta R\indices{^\gamma_\alpha_\gamma_\beta}
= \delta\Gamma\indices{^\gamma_\alpha_\beta_,_\gamma}
	- \delta\Gamma\indices{^\gamma_\alpha_\gamma_,_\beta}.
\end{equation}

\begin{exercise}
Show that, under an arbitrary coordinate transformation $(x^\mu)\rightarrow(y^\alpha)$, the Christoffel symbols transform as
\begin{equation}
\Gamma\indices{^\alpha_\beta_\gamma}
= \pd{y^\alpha}{x^\mu}
	\frac{\partial^2 x^\mu}{\partial y^\beta \partial y^\gamma}
	+ \pd{y^\alpha}{x^\mu} \pd{x^\nu}{y^\beta} \pd{x^\rho}{y^\gamma}
		 \, \Gamma\indices{^\mu_\nu_\rho},
\end{equation}
and conclude that the components of the variation $\delta\Gamma\indices{^\mu_\nu_\rho}$ transform as a tensor, even though the Christoffel symbols themselves do not.
\end{exercise}

Since $\delta \Gamma\indices{^\mu_\nu_\rho}$ behaves like a tensor, we can define its covariant derivative, which coincides with its partial derivative in inertial coordinates. Thus,
\begin{equation}
\delta R_{\alpha\beta} = 
\delta\Gamma\indices{^\gamma_\alpha_\beta_;_\gamma}
- \delta\Gamma\indices{^\gamma_\alpha_\gamma_;_\beta} \ ,
\end{equation}
which is a tensor equation (all its terms behave as tensors), so it is valid in any coordinate system, and not only in the FNCs used to get it. In $\delta B$,
\begin{equation}
g^{\mu\nu} \delta R_{\mu\nu} 
= g^{\mu\nu} \pa{ \delta\Gamma\indices{^\rho_\mu_\nu_;_\rho}
								- \delta\Gamma\indices{^\rho_\mu_\rho_;_\nu} }
= \nabla_\rho \pa{ g^{\mu\nu} \delta\Gamma\indices{^\rho_\mu_\nu} 
								- g^{\mu\rho} \delta\Gamma\indices{^\nu_\mu_\nu} }
\define \nabla_\rho V^\rho,
\end{equation}
where we have used that the covariant derivative of the metric vanishes, and we have exchanged the names of $\nu$ and $\rho$ in the second equality.

\begin{exercise}\label{ex:boundary_Hilbert}
For any vector field~$(V^\mu)$, demonstrate the identity
\begin{equation}
\sqrt{-g} \, \nabla_\mu V^\mu = \partial_\mu \pa{\sqrt{-g} \, V^\mu}
\end{equation}
and conclude that any integral of the form
\begin{equation}
\int_{\mathcal{D}} \dd^4 x \, \sqrt{-g} \; \nabla_\mu V^\mu
\end{equation}
is actually an integral of $V^\mu$ over the boundary~$\partial\mathcal{D}$.
\end{exercise}
From exercise~\ref{ex:boundary_Hilbert}, we conclude that $\delta B$ is a boundary term,
\begin{equation}
\delta B
= \int\dd^4 x\,\sqrt{-g} \; g^{\mu\nu} \delta R_{\mu\nu} 
= \int_{\partial\mathcal{D}} \dd \Sigma_\rho 
			\pa{ g^{\mu\nu} \delta\Gamma\indices{^\rho_\mu_\nu} 
					- g^{\mu\rho} \delta\Gamma\indices{^\nu_\mu_\nu} } .
\end{equation}
We can get rid of this term by imposing that, on $\mathcal{D}$, $\delta g_{\mu\nu,\rho}=0$, along with $\delta g_{\mu\nu}=0$, which what is usually assumed when the Lagrangian density of an action depends on the second derivatives of the field. Another approach consists in adding a counter-term in the definition of the Einstein-Hilbert action, which kills~$\delta B$. Under those conditions, we found
\begin{equation}
\frac{\delta S\e{g}}{\delta g_{\mu\nu}} 
= -\frac{\sqrt{-g}}{16\pi G} \pa{ E^{\mu\nu} + \Lambda g^{\mu\nu} } .
\end{equation}

\paragraph{Action formulation: everything at once} Let us summarise everything by putting together the action of gravitation~$S\e{g}$ with the action~$S\e{m}$ of all the matter fields~$\mat{\psi}_1,\ldots,\mat{\psi}_n$ of the standard model of particle physics, which are minimally coupled to gravity. The total action reads
\begin{equation}
S[\mat{\psi}_1, \ldots \mat{\psi}_N, \mat{g}] = S\e{m}[\mat{\psi}_1, \ldots \mat{\psi}_N,\mat{g}] + S\e{g}[\mat{g}] \ .
\end{equation}
On the one hand, the variation of $S$ with respect to $\mat{\psi}_n$ yields the equation of motion for the corresponding matter field, which takes the effect of gravity in to account. On the other hand, the variation of $S$ with respect to the metric yields
\begin{align}
0
&= \frac{\delta S\e{m}}{\delta g_{\mu\nu}}
	+  \frac{\delta S\e{g}}{\delta g_{\mu\nu}} \\
&= \frac{\sqrt{-g}}{2} \, T^{\mu\nu} 
	- \frac{\sqrt{-g}}{16\pi G} \pa{ E^{\mu\nu} + \Lambda g^{\mu\nu} } \\
&= \frac{\sqrt{-g}}{16\pi G}
	 \pa{8\pi G T^{\mu\nu} -  E^{\mu\nu} - \Lambda g^{\mu\nu}} .
\end{align}
which is Einstein's equation, in the presence of a cosmological constant, and where 
\begin{equation}
T^{\mu\nu} \define \frac{2}{\sqrt{-g}} \frac{\delta S\e{m}}{\delta g_{\mu\nu}}
\end{equation}
is the total energy-momentum tensor of matter.

\section*{Newton versus Einstein}
\addcontentsline{toc}{section}{Newton versus Einstein}

The first two chapters of this course have reviewed Newton's and Einstein's theories of gravity. We have seen in detail how conceptually different these two approaches are. Table~\ref{tab:comparison_Newton_Einstein} summarises these differences.

\renewcommand{\arraystretch}{1.5}
\begin{table}[h!]
\centering
\begin{tabular}{r|cc|}
\cline{2-3}
	& {\bfseries Newton}
	& {\bfseries Einstein} \\
\cline{2-3}
Space and time
	& absolute
	& relative \\
Inertia quantified by
	& mass
	& energy \\
Nature of gravity
	& force
	& space-time geometry\\
Fundamental field
	& gravitational potential $\Phi$
	& space-time metric $g_{\mu\nu}$ \\
Gravitational acceleration
	& $g^i = -\partial^i\Phi$
	& $-\Gamma\indices{^\mu_\nu_\rho}u^\nu u^\rho$ \\
Equivalence principle ensured by
	& $m\e{in}=m\e{pg}$
	& minimal coupling \\
Free fall
	& $\displaystyle{\Ddf{p^i}{t} = m g^i}$
	& $\displaystyle{\Ddf{p^\mu}{\tau} = 0}$ \\[4mm]
Mechanics
	& $\displaystyle{\Ddf{p^i}{t} = m g^i + F^i}$
	& $\displaystyle{\Ddf{p^\mu}{\tau} = F^\mu}$ \\
Source of gravity
	& mass
	& energy and momentum \\
Field equation
	& $\Delta\Phi+\Lambda = 4\pi G \rho$
	& $E_{\mu\nu} + \Lambda g_{\mu\nu} = 8\pi G T_{\mu\nu}$\\
Gravitation propagates
	& instantaneously
	& at the speed of light \\
Gravitational waves
	& no
	& yes \\
Mathematical features
	& 3D, scalar, linear
	& 4D, tensorial, non-linear\\
\cline{2-3}
\end{tabular}
\caption{Comparison between Newton's and Einstein's theories of gravitation.}
\label{tab:comparison_Newton_Einstein}
\end{table}

\chapter{The general-relativistic world}
\label{chap:GR_world}

\lettrine{T}{he} previous chapter of this course was dedicated to the construction of a relativistic theory of gravitation. In this third and last chapter, we will review some of the main real-world new features of this theory, such as gravitational time dilation, gravitational waves, and black holes.

\minitoc

\newpage

\section{Weak gravitational fields}
\label{sec:weak_fields}

General relativity (GR) is, today, the best description of gravity that we dispose of. In particular, it is \emph{better} than Newtonian gravity. This does not mean, however, that Newton's theory is absolutely wrong; on the contrary, we have seen in the first chapter that it provides an excellent description of nature in our daily experience. Just like Galilean kinematics is a limit of special relativity when velocities are sub-luminal, Newtonian gravity should be a limit of GR in some regime. That is the regime of weak gravitational fields.

\subsection{Linearised Einstein's equation}

\paragraph{Definition of a weak field} Space-time will be said to be in the weak-field regime if its metric is nearly Minkowskian, i.e. if there exists a coordinate system $\{x^\mu\}$ such that
\begin{equation}\label{eq:quasi_Minkowskian_metric}
g_{\mu\nu} = \eta_{\mu\nu} + h_{\mu\nu}
\end{equation}
in the whole region under consideration. This last remark is important. We have seen in the last chapter that, by virtue of local flatness, eq.~\eqref{eq:quasi_Minkowskian_metric} can always be satisfied in a small region of space-time. In that sense, any gravitational field is \emph{locally} weak, but not necessarily \emph{globally}. The quantity $h_{\mu\nu}$ is called the \emph{metric perturbation}, as it quantifies the departure from Minkowski.

\paragraph{Linearising Einstein's equation} Any non-linear equation can be made approximately linear by considering only first-order perturbations about one of its solutions. Here we consider small perturbations about the Minkowski space-time. As the Minkowski metric has a vanishing Einstein tensor, expanding $\fvect{E}[\fvect{g}]$ about $\fvect{\eta}$ at first order in $\fvect{h}$ should yield
\begin{equation}
\fvect{E}[\fvect{\eta}+\fvect{h}] = \fvect{\mathcal{D}} \fvect{h} + \mathcal{O}(\fvect{h}^2) \ ,
\end{equation}
where $\fvect{\mathcal{D}}$ is a linear differential operator to be determined. Neglecting the second-order terms leads us to the linearised Einstein's equation $\fvect{\mathcal{D}}\fvect{h} = 8\pi G \fvect{T}$.

In order to derive the explicit expression of $\fvect{\mathcal{D}}\fvect{h}$, we start with expanding the Christoffel symbols at first order in $\fvect{h}$,
\begin{align}
\Gamma\indices{^\rho_\mu_\nu}
&= \frac{1}{2}  g^{\rho\sigma} \pa{ g_{\sigma\mu,\nu} + g_{\sigma\nu,\mu} - g_{\mu\nu,\sigma} } \\
&= \frac{1}{2}  g^{\rho\sigma} \pa{ h_{\sigma\mu,\nu} + h_{\sigma\nu,\mu} - h_{\mu\nu,\sigma} } 
\qquad \text{since }\eta_{\mu\nu}=\cst\\
&= \frac{1}{2}  \eta^{\rho\sigma} \pa{ h_{\sigma\mu,\nu} + h_{\sigma\nu,\mu} - h_{\mu\nu,\sigma} } + \mathcal{O}(\fvect{h}^2) \qquad \text{since }g^{\mu\nu}=\eta^{\mu\nu}+\mathcal{O}(\fvect{h}) .
\end{align}
We can then calculate the Ricci tensor at the same order,
\begin{align}
R_{\mu\nu}
&= \Gamma\indices{^\rho_\mu_\nu_,_\rho}
	- \Gamma\indices{^\rho_\rho_\mu_,_\nu}
	\overbrace{
	+ \Gamma\indices{^\rho_\sigma_\rho}
		\Gamma\indices{^\sigma_\mu_\nu}
	- \Gamma\indices{^\rho_\sigma_\mu}
		\Gamma\indices{^\sigma_\rho_\nu}
		}^{\mathcal{O}(\fvect{h}^2)} \\
&= \frac{1}{2} \eta^{\rho\sigma}\pa{ h_{\sigma\mu,\nu\rho} + h_{\sigma\nu,\mu\rho} - h_{\mu\nu,\sigma\rho} - h_{\sigma\mu,\rho\nu} - h_{\sigma\rho,\mu\nu} + h_{\mu\rho,\sigma\nu}} + \mathcal{O}(\fvect{h}^2) \\
&= \frac{1}{2} \pa{ h\indices{^\rho_\nu_,_\mu_\rho} - \Box h_{\mu\nu} - h_{,\mu\nu} + h\indices{_\mu^\rho_,_\rho_\nu} } + \mathcal{O}(\fvect{h}^2) \ ,
\end{align}
where $\Box\define \eta^{\mu\nu}\partial_\mu\partial_\nu$ and $h=h^\mu_\mu=\eta^{\mu\nu} h_{\mu\nu}$ is the trace of $\fvect{h}$.

Combining $R_{\mu\nu}$ with its trace to build the Einstein tensor~$E_{\mu\nu}$, and dropping quadratic terms, finally yields the \emph{linearised Einstein's equation}
\begin{equation}\label{eq:linearised_Einstein}
\Box h_{\mu\nu} + h_{,\mu\nu} - h\indices{^\rho_\mu_,_\rho_\nu} - h\indices{^\rho_\nu_,_\rho_\mu} - \pa{ \Box h - h\indices{^\rho^\sigma_,_\rho_\sigma} } \eta_{\mu\nu}
= -16\pi G T_{\mu\nu} \ ,
\end{equation}
where the left-hand side is $-2\fvect{\mathcal{D}}\fvect{h}$, which we aimed to determine.

\paragraph{Trace-reversed perturbation} Equation~\eqref{eq:linearised_Einstein} is more conveniently handled with
\begin{equation}
\gamma_{\mu\nu} \define h_{\mu\nu} - \frac{1}{2} h \eta_{\mu\nu} \ ,
\end{equation}
which can be dubbed trace-reversed metric perturbation, instead of $h_{\mu\nu}$. Note that the above relation is inverted as $h_{\mu\nu}=\gamma_{\mu\nu}-\gamma \eta_{\mu\nu}/2$.

\begin{exercise}
Show that, in terms of $\gamma_{\mu\nu}$, eq.~\eqref{eq:linearised_Einstein} reads
\begin{equation}\label{eq:linearised_Einstein_gamma}
\Box \gamma_{\mu\nu} + \gamma\indices{_\rho_\sigma^,^\rho^\sigma} \eta_{\mu\nu} - \gamma\indices{_\mu_\rho_,_\nu^\rho} - \gamma\indices{_\nu_\rho_,_\mu^\rho}
= -16\pi G T_{\mu\nu} \ .
\end{equation}
\end{exercise}

\paragraph{Gauge freedom} A very important thing about the metric perturbation $h_{\mu\nu}$ (or $\gamma_{\mu\nu}$) is that it is \emph{not unique} for a given space-time. It actually depends on the particular coordinate system that was used to define the Minkowskian background.

This ambiguity, called \emph{gauge freedom}, is a general feature of pertubative schemes. Let us take a concrete example. The surface of a football is approximately spherical: its radius is almost constant. Departures from sphericity can be described perturbatively as $r(\theta,\ph)=R+h(\theta,\ph)$, where $h\ll R$. But clearly there is no unique way to define $R$ and $h$: I can choose $R$ to be the radius~$R_1$ of the ball at the junction between two pentagons, or alternatively $R_2>R_1	$ its radius at the centre of one of the pentagons. This yields two different definitions for the perturbation, $r = R_1+h_1=R_2+h_2$.

\begin{figure}[h!]
\centering
\input{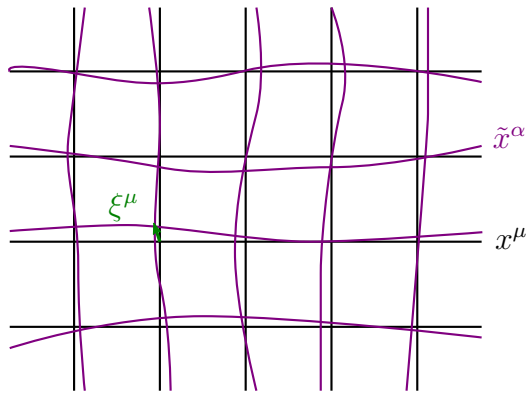}
\caption{Two coordinate systems $(x^\mu)$ and $(\tilde{x}^\alpha)$ related by an infinitesimal transformation.}
\label{fig:gauge_transformation}
\end{figure}

Let us examine what happens to the metric as we perform an infinitesimal coordinate transformation~$x^\mu\rightarrow \tilde{x}^\mu = x^\mu - \xi^\mu(x^\nu)$, where $\xi^\mu\ll 1$ (see fig.~\ref{fig:gauge_transformation}). Because the metric is a tensor, we have
\begin{align}
\tilde{g}_{\alpha\beta}(\tilde{x}^\gamma)
&= \frac{\partial x^\mu}{\partial \tilde{x}^\alpha}
		\frac{\partial x^\nu}{\partial \tilde{x}^\beta} \,
		g_{\mu\nu}[x^\rho(\tilde{x}^\gamma)] \\
&= \pa{ \delta^\mu_\alpha + \xi\indices{^\mu_,_\alpha} }
	\pa{ \delta^\nu_\beta + \xi\indices{^\nu_,_\beta} }
	\pac{ \eta_{\mu\nu} + h_{\mu\nu}(\tilde{x}^\rho+\xi^\rho) } \\
&= \eta_{\alpha\beta}
		+ h_{\alpha\beta}(\tilde{x}^\gamma)
		+ \xi_{\alpha,\beta} + \xi_{\beta,\alpha}
		+ \ldots \\
&= \eta_{\alpha\beta} + \tilde{h}_{\alpha\beta}(\tilde{x}^\gamma) \ ,
\end{align}
with, at linear order
\begin{empheq}[box=\fbox]{equation}
\tilde{h}_{\mu\nu}=h_{\mu\nu} + 2\xi_{(\mu,\nu)} \ .
\end{empheq}
Thus, in the slightly distorted coordinate system~$(\tilde{x}^\alpha)$, the metric perturbation is no longer $h_{\mu\nu}$, but $\tilde{h}_{\mu\nu}$. There is no reason to prefer the former over the latter: both perturbations describe the same space-time; simply, they do it in a different way.

\begin{exercise}\label{ex:gauge_independence_Riemann}
Show that the Riemann tensor is gauge independent, namely, that for any gauge transformation $\tilde{h}_{\mu\nu}=h_{\mu\nu}+2\xi_{(\mu,\nu)}$, we have
\begin{equation}
\tilde{R}_{\mu\nu\rho\sigma} = R_{\mu\nu\rho\sigma} \ .
\end{equation}
\end{exercise}

This is structurally similar to what happens in electrodynamics: the electromagnetic field~$F_{\mu\nu}$ remains invariant under a gauge transformation of the potential~$A_\mu$.

\paragraph{Harmonic gauge} The gauge freedom allows us to impose additional conditions on the metric perturbation without affecting its actual nature. Taking again the football example, we can always choose $R$ such that the average radius perturbation $h$ is zero, without changing the shape of the ball. In electrodynamics, one can always impose the Lorenz gauge $\nabla_\mu A^\mu=0$ without affecting the electromagnetic field.

The harmonic gauge, also called Hilbert or De Donder gauge, is the gravitational analogue of the Lorenz gauge, and corresponds to imposing
\begin{empheq}[box=\fbox]{equation}\label{eq:harmonic_gauge}
\gamma\indices{_\mu_\nu^,^\nu} = 0 \ .
\end{empheq}

\begin{exercise}
Show that it is always possible to impose the condition~\eqref{eq:harmonic_gauge}; namely, show that if $\gamma_{\mu\nu}$ does not satisfy it, then one can find a gauge transformation $h_{\mu\nu}\rightarrow\tilde{h}_{\mu\nu}$ such that the corresponding $\tilde{\gamma}_{\mu\nu}$ does.
\end{exercise}

In the harmonic gauge, three of the four terms on the left-hand side of eq.~\eqref{eq:linearised_Einstein_gamma} drop, and we are left with
\begin{empheq}[box=\fbox]{equation}\label{eq:linearised_Einstein_harmonic}
\Box \gamma_{\mu\nu} = -16\pi G T_{\mu\nu} \ .
\end{empheq}

\subsection{Newtonian regime}

\paragraph{Gravitational potential} Let us assume that matter is non-relativistic, i.e., that it is made of particles moving slowly compared to the speed of light in the coordinate system~($x^\mu$). In that case the dominant component of the energy-momentum tensor is the rest-mass energy density $T_{00}=\rho$. Specifically, if $v\ll 1$ is the typical velocity of the sources, then
\begin{equation}
\rho=T_{00} \gg T_{0a} \sim v T_{00} \gg T_{ab} \sim v^2 T_{00} \ ,
\end{equation}
so that we can neglect $T_{0a}, T_{ab}$ in the following. In that case, eq.~\eqref{eq:linearised_Einstein_harmonic} reduces to
\begin{align}
\Box\gamma_{00} &= -16\pi G \rho
\label{eq:box_gamma_00}\\
\Box\gamma_{0a} &= \Box\gamma_{ab} = 0 \ .
\end{align}
Homogeneous solutions correspond to gravitational waves, which are the subject of \S~\ref{sec:gw}. For now, we drop such contributions and consider the particular solution $\gamma_{0a}=\gamma_{ab}=0$; besides, we solve eq.~\eqref{eq:box_gamma_00} using the well-known \emph{Green function} of the $\Box$ operator,
\begin{equation}\label{eq:gamma_00_result}
\gamma_{00}(t, \vec{x})
= 4 G \int \frac{\rho(t-||\vec{x}-\vec{y}||, \vec{y})}{||\vec{x}-\vec{y}||} \: \dd^3 y \ ,
\end{equation}
where $||\vec{x}-\vec{y}||$ denotes the Euclidean distance between points with Cartesian coordinates\footnote{We are facing, here, a notation subtlety: $(x^a)$ are Cartesian coordinates, because the spatial part of the metric is approximately $\delta_{ab}$, but we cannot denote them with capital letters ($X^a$), because these are reserved to FNCs $(X^\alpha)=(\tau,X^a)$.} $x^a, y^a$. Equation~\eqref{eq:gamma_00_result} is reminiscent of expression~\eqref{eq:Nordstrom_field} of Nordstr\"{o}m's field, except for a factor $-4$. It is thus natural to introduce the notation
\begin{equation}
\gamma_{00} = -4\Phi \ ,
\end{equation}
where $\Phi$ shall be interpreted as the gravitational potential.

\paragraph{Metric} Going back to the actual metric perturbation~$h_{\mu\nu}=\gamma_{\mu\nu}-\gamma \eta_{\mu\nu}/2$, and using $\gamma=-\gamma_{00}=4\Phi$, we find
\begin{align}
h_{00}
&= \gamma_{00} - \frac{1}{2}\gamma\eta_{00}
= -2\Phi \\
h_{0a}
&= \gamma_{0i} - \frac{1}{2}\gamma\eta_{0i} = 0 \\
h_{ab}
&= \gamma_{ab} - \frac{1}{2}\gamma\eta_{ab}
= -2\Phi \, \delta_{ab} \ ,
\end{align}
so that the line element reads
\begin{empheq}[box=\fbox]{equation}\label{eq:Newtonian_metric}
\dd s^2 = -(1+2\Phi) \dd t^2 + (1-2\Phi) \delta_{ab} \dd x^a \dd x^b
\end{empheq}
for weak gravitational fields in the Newtonian regime.

\paragraph{Motion} Let us analyse the motion of a massive non-relativistic particle in a space-time described by eq.~\eqref{eq:Newtonian_metric}. The equation of motion is
\begin{equation}
\ddf{p^\mu}{\tau} +
\Gamma\indices{^\mu_\nu_\rho} p^\nu u^\rho
= F^\mu \ ,
\end{equation}
with $p^\mu = m u^\mu$. Since the particle is non-relativistic, we can write $(u^\mu)\approx (1, v^a)$, and expand the equation of motion at lowest order in $v^a, \Phi \ll 1$. In particular, we have
\begin{align}
\ddf{t}{\tau} = u^0 &= 1 + \mathcal{O}(v^2) \ ,\\
\Gamma\indices{^\mu_\nu_\rho} p^\nu u^\rho
&= m \Gamma\indices{^\mu_0_0} + \mathcal{O}(v) \ .
\end{align}
For $\mu= a$ (spatial index), the Christoffel symbols read
\begin{equation}
\Gamma\indices{^a_0_0}
= \frac{1}{2} \delta^{ab} (h_{b0,0} + h_{b0,0} - h_{00,b})
= \partial^a\Phi \ ,
\end{equation}
whence
\begin{empheq}[box=\fbox]{equation}
\ddf{p^a}{t} = -m \partial^a \Phi + F^a \ ,
\end{empheq}
which is equivalent to Newton's second law of mechanics in the presence of gravity.

\begin{exercise}
Study the case of mass-less particles ($m=0$).
\end{exercise}

\begin{exercise}
Show that $R_{0a0b}=\Phi_{,ab}$. Compare with the expression of the tidal tensor of Newtonian gravity, defined in \S~\ref{subsec:tides}. Just like tidal forces cannot be eliminated by working in a freely-falling frame, Riemann curvature is the residual gravitational effect appearing in FNCs, see \S~\ref{subsec:geodesic_motion}.
\end{exercise}

\subsection{Gravitational dilation of time}
\label{subsec:gravitational_dilation_time}

\begin{minipage}{13cm}
\paragraph{Age of twins} Two twin sisters, Alexandra and Biki, have lived together until their majority, when they leave the parental house (event $L$). After that, each one lives her own life; they travel at different speeds and experience different gravitational potentials, before meeting again (event $M$). Between $L$ and $M$, Alexandra and Biki thus followed different world-lines in space-time. The respective proper time measured by each sister between $L$ and $M$ reads
\begin{equation}
\Delta\tau_{LM}
= \int_L^M \dd\tau
= \int_L^M \sqrt{-g_{\mu\nu} \ddf{x^\mu}{t} \ddf{x^\nu}{t}} \; \dd t \ ,
\end{equation}
where the integral is calculated along her own world-line.
\end{minipage}
\hfill
\begin{minipage}{2cm}
\begingroup%
  \makeatletter%
  \providecommand\color[2][]{%
    \errmessage{(Inkscape) Color is used for the text in Inkscape, but the package 'color.sty' is not loaded}%
    \renewcommand\color[2][]{}%
  }%
  \providecommand\transparent[1]{%
    \errmessage{(Inkscape) Transparency is used (non-zero) for the text in Inkscape, but the package 'transparent.sty' is not loaded}%
    \renewcommand\transparent[1]{}%
  }%
  \providecommand\rotatebox[2]{#2}%
  \ifx\svgwidth\undefined%
    \setlength{\unitlength}{55.86385498bp}%
    \ifx\svgscale\undefined%
      \relax%
    \else%
      \setlength{\unitlength}{\unitlength * \real{\svgscale}}%
    \fi%
  \else%
    \setlength{\unitlength}{\svgwidth}%
  \fi%
  \global\let\svgwidth\undefined%
  \global\let\svgscale\undefined%
  \makeatother%
  \begin{picture}(1,2.54808672)%
    \put(0,0){\includegraphics[width=\unitlength,page=1]{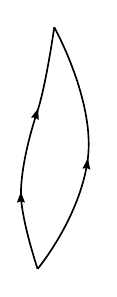}}%
    \put(0.26772285,0.01745315){\color[rgb]{0,0,0}\makebox(0,0)[lb]{\smash{$L$}}}%
    \put(0.36126888,2.38362439){\color[rgb]{0,0,0}\makebox(0,0)[lb]{\smash{$M$}}}%
    \put(-0.00179007,1.24733265){\color[rgb]{0,0,0}\makebox(0,0)[lb]{\smash{A}}}%
    \put(0.85970356,1.07680175){\color[rgb]{0,0,0}\makebox(0,0)[lb]{\smash{B}}}%
  \end{picture}%
\endgroup%

\end{minipage}

For $v,\Phi\ll 1$, we have
\begin{align}
-g_{\mu\nu} \ddf{x^\mu}{t} \ddf{x^\nu}{t} 
&= -g_{00} - g_{ab} v^a v^a \\
&= (1+2\Phi) - (1-2\Phi) \delta_{ab} v^a v^b \\
&= 1+2\Phi - v^2 + \mathcal{O}(v^2\Phi) \ ,
\end{align}
whence, \emph{at leading order} in $v, \Phi$,
\begin{empheq}[box=\fbox]{equation}
\Delta\tau = \int_L^M \pa{1-\frac{v^2}{2} + \Phi} \dd t \ .
\end{empheq}
In other words, the twin who, on average, travels faster and experiences stronger gravitational fields (recall that $\Phi<0$) is younger than the other when they meet at $M$.

\begin{exercise}
Suppose that Alexandra stays at home, in Amsterdam, while Biki flies to Douala, stays there 10 hours, and comes back. We assume that her plane flies with constant velocity $v=1000\U{km/h}$, and constant altitude of $12\U{km}$. Both Alexandra and Biki have identical watches, and when Biki is back to Amsterdam, Alexandra's watch indicates that 24 hours have elapsed since Biki's departure. What is the duration indicated on Biki's watch?
\end{exercise}

\noindent
\begin{minipage}{12cm}
\paragraph{Gravitational redshift} Loosely speaking, the above shows that gravitation slows down the passage of time. This also affects frequency measurements, an effect called \emph{gravitational redshift}. Consider an emitter~$\mathcal{E}$ sending a photon (event $E$) to an observer~$\mathcal{O}$, who receives it at $O$. The photon travels along a null geodesic whose tangent vector is $k^\mu=\dd x^\mu/\dd\lambda$, the wave four-vector. We have seen in exercise~\ref{ex:frequency_photon} that the cyclic frequency of a photon as measured by an observer is the projection of $\fvect{k}$ onto the observer's four-velocity~$\fvect{u}$,
\begin{equation}
\omega\e{em} = - (u_\mu k^\mu)_E \ ,
\qquad
\omega\e{obs} = -(u_\mu k^\mu)_O \ .
\end{equation}
\end{minipage}
\hspace*{0.5cm}
\begin{minipage}{2cm}
\begingroup%
  \makeatletter%
  \providecommand\color[2][]{%
    \errmessage{(Inkscape) Color is used for the text in Inkscape, but the package 'color.sty' is not loaded}%
    \renewcommand\color[2][]{}%
  }%
  \providecommand\transparent[1]{%
    \errmessage{(Inkscape) Transparency is used (non-zero) for the text in Inkscape, but the package 'transparent.sty' is not loaded}%
    \renewcommand\transparent[1]{}%
  }%
  \providecommand\rotatebox[2]{#2}%
  \ifx\svgwidth\undefined%
    \setlength{\unitlength}{145.71446533bp}%
    \ifx\svgscale\undefined%
      \relax%
    \else%
      \setlength{\unitlength}{\unitlength * \real{\svgscale}}%
    \fi%
  \else%
    \setlength{\unitlength}{\svgwidth}%
  \fi%
  \global\let\svgwidth\undefined%
  \global\let\svgscale\undefined%
  \makeatother%
  \begin{picture}(1,0.80030924)%
    \put(0,0){\includegraphics[width=\unitlength,page=1]{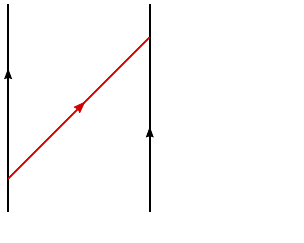}}%
    \put(0.04888418,0.14527522){\color[rgb]{0,0,0}\makebox(0,0)[lb]{\smash{$E$}}}%
    \put(0.51162347,0.65219756){\color[rgb]{0,0,0}\makebox(0,0)[lb]{\smash{$O$}}}%
    \put(0,0){\includegraphics[width=\unitlength,page=2]{emitter_observer.pdf}}%
    \put(-0.00463235,0.01895831){\color[rgb]{0,0,0}\makebox(0,0)[lb]{\smash{$\mathcal{E}$}}}%
    \put(0.4665509,0.02060362){\color[rgb]{0,0,0}\makebox(0,0)[lb]{\smash{$\mathcal{O}$}}}%
    \put(0.04278596,0.35276963){\color[rgb]{0,0.50196078,0}\makebox(0,0)[lb]{\smash{$\fvect{u}_E$}}}%
    \put(0.36539021,0.72586676){\color[rgb]{0,0.50196078,0}\makebox(0,0)[lb]{\smash{$\fvect{u}_O$}}}%
  \end{picture}%
\endgroup%

\end{minipage}

\medskip

Let us assume that both $\mathcal{E}$ and $\mathcal{O}$ are \emph{at rest} in the coordinate system $(x^\mu)$. Then their four-velocity reads, at leading order in $\Phi$,
\begin{equation}\label{eq:four-velocities_emitter_observer}
(u^\mu_E) = ( 1-\Phi_E , \vec{0} ) \ ,
\qquad
(u^\mu_O) = ( 1-\Phi_O , \vec{0} ) \ ,
\end{equation}
so that
\begin{equation}
\omega\e{em} = (1+\Phi_E) k^0_E \ ,
\qquad
\omega\e{obs} = (1+\Phi_O) k^0_O \ .
\end{equation}

\begin{exercise}
Check that the expressions~\eqref{eq:four-velocities_emitter_observer} are normalised, i.e. $\fvect{u}\cdot\fvect{u}=-1$, at leading order in $\Phi$.
\end{exercise}

The last step consists in determining $k^0_E, k^0_O$. We have seen in \S~\ref{subsec:geodesic_motion} that the null geodesic equation derives from the Lagrangian
\begin{equation}
L
= g_{\mu\nu} k^\mu k^\nu
= -(1+2\Phi) (k^0)^2
	+ (1-2\Phi) \delta_{ab} k^a k^b \ .
\end{equation}
Using the time component, $\mu=0$, we conclude that, in a static potential,
\begin{equation}
0 = \ddf{}{\lambda} \pd{L}{k^0} - \pd{L}{t}
= -\ddf{}{\lambda} \pac{ (1+2\Phi) k^0 } \ ,
\qquad \text{i.e.} \qquad
(1+2\Phi) k^0=\cst \ .
\end{equation}
Therefore,
\begin{empheq}[box=\fbox]{equation}
\frac{\omega\e{em}}{\omega\e{obs}}
= \frac{(1+\Phi_E) k^0\e{em}}{(1+\Phi_O) k^0\e{obs}}
\approx \frac{1+\Phi_O}{1+\Phi_E}
\approx 1 + \Phi_O - \Phi_E \ .
\end{empheq}
If the emitter lies within a deeper gravitational potential than the observer ($\Phi_E<\Phi_O$), then the latter sees a reduced frequency, i.e. a redder light---whence the name gravitational redshift. In the opposite situation ($\Phi_O<\Phi_E$), light is blue-shifted. Everything happens as if the photon were loosing energy climbing up, and gaining energy rolling down.


\section{Gravitational waves}
\label{sec:gw}

Newton's theory gives a rather rigid picture of gravity: the gravitational field instantly adapts to the motion of matter, and cannot propagate freely. Things are different in GR, where gravitational potentials are retarded, and which allows the existence of gravitational waves (hereafter GWs). After an intense experimental effort, such waves were finally detected by the Laser Interferometer Gravitational Observatory (LIGO) on the 14th of September 2015~\cite{Abbott:2016blz}, followed by ten other events from 2015 to 2017 (see e.g. the list of GW observations on \href{https://en.wikipedia.org/wiki/List_of_gravitational_wave_observations}{Wikipedia}). For their decisive contribution to this breakthrough, R.~Weiss, K.~Thorne, and B.~Barish shared the 2017 Nobel Prize in Physics.

We have seen in \S~\ref{sec:weak_fields} that the linearised Einstein equation reads $\Box \gamma_{\mu\nu} = -16\pi G T_{\mu\nu}$. In vacuum ($T_{\mu\nu}=0$), this becomes
\begin{equation}
\Box \gamma_{\mu\nu} = 0 \ ,
\end{equation}
which has propagating solutions. Just like electromagnetic waves are vacuum solutions of Maxwell's equations, GWs are vacuum solutions of Einstein's equation.

\subsection{Transverse trace-less gauge}

\paragraph{Trace-less gauge} In vacuum, the gauge freedom allows us to set the trace of the metric perturbation to zero, $h=\gamma=0$.

\begin{exercise}
Show that, under a gauge transformation for $h_{\mu\nu}$, the trace-reversed metric perturbation~$\gamma_{\mu\nu}$ transforms as
\begin{align}
\gamma_{\mu\nu}
&\rightarrow
\tilde{\gamma}_{\mu\nu} = \gamma_{\mu\nu} + \xi_{\mu,\nu} + \xi_{\nu,\mu} - \xi\indices{^\rho_,_\rho} \eta_{\mu\nu}  \ ,\\
\text{and thus}, \quad
\gamma &\rightarrow \tilde{\gamma} = \gamma - 2 \xi\indices{^\mu_,_\mu} \ .
\end{align}
\end{exercise}

From the above exercise, we conclude that, if $\gamma_{\mu\nu}$ has a non-vanishing trace~$\gamma$, then we can perform a gauge transformation with $\xi^\mu$ such that $\xi\indices{^\mu_,_\mu}=\gamma/2$ in order to eliminate it. Therefore, we can assume without loss of generality that $\gamma=0$ in the following; this is known as the \emph{trace-less gauge}. In that gauge, there is no difference between the original metric perturbation and the trace-reversed perturbation,
\begin{empheq}[box=\fbox]{equation}
\gamma_{\mu\nu} = h_{\mu\nu} \ .
\end{empheq}

\noindent\textit{Remark.} One must be careful, when enforcing the trace-less gauge, not to break the harmonic gauge, i.e., not to end up with $\gamma\indices{_\mu_\nu^,^\nu}\not=0$. Under a gauge transformation,
\begin{equation}
\gamma\indices{_\mu_\nu^,^\nu}
\rightarrow \gamma\indices{_\mu_\nu^,^\nu} + \Box \xi_\mu \ ,
\end{equation}
so if the harmonic gauge was initially satisfied, we just have to ensure that $\Box \xi_\mu=0$. This constraint can be satisfied simultaneously with the trace-killer $\xi\indices{^\mu_,_\mu}=\gamma/2$. This is easier to see in Fourier space,
\begin{equation}
\xi^\mu(x^\nu)
= \int \frac{\dd^4 k}{(2\pi)^4} \; \ex{ \i k_\nu x^\nu } \, \hat{\xi}^\mu(k_\nu) \ ,
\end{equation}
in terms of which
\begin{align}
\text{eliminate trace:}
&\qquad \xi\indices{^\mu_,_\mu} = \frac{1}{2}\gamma
\longleftrightarrow \i k_\mu \hat{\xi}^\mu = \frac{1}{2}\gamma \ ,\\
\text{preserve harmonic gauge:}
&\qquad \Box\xi^\mu = 0
\longleftrightarrow - k_\nu k^\nu \hat{\xi}^\mu = 0 \ .
\end{align}
These are clearly independent conditions on the vector field~$\xi^\mu$.

\paragraph{Plane waves} The general solution of $\Box h_{\mu\nu}=0$ is a superposition of plane waves
\begin{equation}\label{eq:plane_GW}
h_{\mu\nu} = H_{\mu\nu} \ex{\i k_\rho x^\rho}
						+ \cc \ ,
\end{equation}
where $H_{\mu\nu}\in \mathbb{C}$ is a constant called the \emph{polarisation tensor}, $k^\rho$ is the wave four-vector, and \cc\xspace means ``complex conjugate''. In the remainder of this section, we will analyse the properties of such plane waves. In terms of $H_{\mu\nu}$ and $k^\mu$, the wave equation and the two gauge conditions are equivalent to
\begin{align}
\Box h_{\mu\nu}=0
&\Longleftrightarrow k^\mu k_\mu = 0 \ ,\\
h\indices{_\mu_\nu^,^\nu} = 0 
&\Longleftrightarrow k^\mu H_{\mu\nu} = 0 \ ,\\
h^\mu_\mu = 0 &\Longleftrightarrow H^\mu_\mu = 0 \ .
\end{align}

\paragraph{Transverse gauge} We have not entirely exhausted the gauge freedom yet. Suppose, without any loss of generality, that the GW propagates in the $z=x^3$ direction, then $(k^\mu)=(\omega,0,0,\omega)$, and $k^\mu H_{\mu\nu}=0$ implies $H_{00}+H_{03}=0$.

\begin{exercise}\label{ex:transverse_gauge}
Consider a gauge transformation where $\xi^\mu$ takes the form
\begin{equation}
\xi^\mu = \Xi^\mu \ex{\i k_\nu x^\nu} + \cc \ ,
\end{equation}
where $\Xi^\mu$ is a constant amplitude and $k^\mu$ is the same wave four-vector as the GW.
\begin{itemize}
\item What are the requirements on $\Xi_\mu$ such that this transformation preserves both the harmonic and trace-less gauges?
\item Show that it is possible to impose $H_{0\mu}=H_{3\mu}=0$ with this transformation.
\end{itemize}
\end{exercise}

The condition enforced by exercise~\ref{ex:transverse_gauge} is called the \emph{transverse gauge}. Together with the trace-less gauge, they define the transverse trace-less (TT) gauge, in which the only non-vanishing components of $H_{\mu\nu}$ are $H_{11}\define H_{+}$, $H_{22}=-H_+$, and $H_{12}=H_{21}\define H_\times$,
\begin{equation}
[H_{\mu\nu}]
=
\begin{bmatrix}
0 & 0 & 0 & 0 \\
0 & H_+ & H_\times & 0 \\
0 & H_\times & -H_+ & 0 \\
0 & 0 & 0 & 0
\end{bmatrix} .
\end{equation}
The two parameters~$H_+, H_\times \in \mathbb{C}$ are the complex amplitudes of the two \emph{polarisations} of a GW. Thus, just like electromagnetic waves, GWs have two independent polarisations.

\subsection{Effect on matter and detection}

In the previous paragraph, we made a number of mathematical transformations in order to derive the simplest form of a GW, but it is hard to keep track of its actual physical meaning. Einstein himself, who first suggested their existence in 1916, changed his opinion several times: are GWs real, or just an artefact of some particular coordinate choice, just like the gravitational force?

\paragraph{Riemann tensor of a GW} In the previous chapters, we insisted on the fact that while the gravitational acceleration can be eliminated in a freely-falling frame, tidal forces cannot; the latter are genuine gravitational effects, encoded in the space-time curvature. The best way to assess the existence and meaning of GWs thus consists in calculating their contribution to the Riemann tensor.

At linear order in the metric perturbation,
\begin{align}
R_{\mu\nu\rho\sigma}
&= \Gamma_{\mu\nu\sigma,\rho} - \Gamma_{\mu\nu\rho,\sigma} \\
&= \frac{1}{2} \pa{ h_{\mu\nu,\sigma\rho} + h_{\mu\sigma,\nu\rho} - h_{\nu\sigma,\mu\rho} - h_{\mu\nu,\rho\sigma} - h_{\mu\rho,\nu\sigma} + h_{\nu\rho,\mu\sigma} } \\
&= \frac{1}{2} \pa{ h_{\mu\sigma,\nu\rho} - h_{\nu\sigma,\mu\rho} - h_{\mu\rho,\nu\sigma} + h_{\nu\rho,\mu\sigma} } \\
&= \frac{1}{2} \pa{ -k_\nu k_\rho H_{\mu\sigma} + k_\mu k_\rho H_{\nu\sigma} + k_\nu k_\sigma H_{\mu\rho} - k_\mu k_\sigma H_{\nu\rho}} \ex{\i k_\lambda x^\lambda} + \cc ,
\label{eq:Riemann_GW}
\end{align}
where in the last line we used the expression~\eqref{eq:plane_GW} of the GW. We see that $R_{\mu\nu\rho\sigma}\not= 0$ in general, which indicates that GWs produce tidal forces.

\paragraph{Tidal forces of a GW} In order to describe those forces, it is convenient to work in the frame of a freely-falling observer, described by FNCs~$(X^\alpha)$---see \S~\ref{subsec:geodesic_motion}. In the vicinity of the observer ($X^a=0$), the metric reads
\begin{align}
g_{00} &= -1-R_{0a0b} X^a X^b + \ldots
\label{eq:Fermi_g_00}\\
g_{0a} &= -\frac{1}{3} (R_{0bac} + R_{0cab}) X^b X^c + \ldots
\label{eq:Fermi_g_0a} \\
g_{ab} &= \delta_{ab} -\frac{1}{3} (R_{acbd} + R_{adbc}) X^c X^d + \ldots
\label{eq:Fermi_g_ab}
\end{align}

How do tidal forces appear in that frame? The equation of motion of a non-relativistic particle is
\begin{equation}
0 = \Ddf{p^a}{\tau} - F^a
\approx
\ddf{p^a}{\tau} + m \Gamma\indices{^a_0_0} - F^a \ ,
\end{equation}
where $F^a$ is the sum of all non-gravitational forces applied on the particle. Using eqs.~\eqref{eq:Fermi_g_00} and \eqref{eq:Fermi_g_0a}, we can express the Christoffel symbol as
\begin{align}
\Gamma\indices{^a_0_0}
&= \frac{1}{2} \delta^{ab} \pa{ 2 g_{b0,0} - g_{00,b} } \\
&= -\frac{2}{3} (R\indices{_0_b^a_c} + R\indices{_0_c^a_b})_{,0} X^b X^c
		+ R\indices{_0^a_0_b} X^b \ .
\label{eq:Christoffel_FNC}
\end{align}

If the observer is moving slowly with respect to the coordinate system $(x^\mu)$, then the FNCs can be considered a particular gauge, because they express the metric as a perturbation with respect to $\eta_{\mu\nu}$. We have seen in exercise~\ref{ex:gauge_independence_Riemann} that the Riemann tensor is gauge independent; thus, its expression in Fermi normal coordinates is the same as its expression~\eqref{eq:Riemann_GW} in the TT gauge. In particular, we see that the two terms of eq.~\eqref{eq:Christoffel_FNC} behave like
\begin{align}
\frac{2}{3} (R\indices{_0_b^a_c} + R\indices{_0_c^a_b})_{,0} X^b X^c
&\sim \partial\partial\partial \fvect{h} \,  |\fvect{X}|^2 
\sim |\fvect{h}| \, \omega^3 |X|^2 \, \\
R\indices{_0^a_0_b} X^b
&\sim \partial\partial \fvect{h} \,  |\fvect{X}|
\sim |\fvect{h}| \, \omega^2 |X| \ .
\end{align}
Assuming that the wavelength~$\lambda=2\pi/\omega$ of the GW is much larger than the distance~$|\fvect{X}|$ between the particle and the origin of the coordinate system, we conclude that the first term on the right-hand side of eq.~\eqref{eq:Christoffel_FNC} can be neglected. Hence,
\begin{align}
\Gamma\indices{^a_0_0}(\tau, \vec{X})
&\approx R\indices{_0^a_0_b}(\tau, \vec{0}) X^b \\
&= \frac{1}{2} \omega^2 H^a_b X^b \ex{\i\omega[z(\tau,\vec{0})-t(\tau,\vec{0})]} + \cc \\
&\approx \frac{1}{2} \omega^2 H^a_b X^b \ex{-\i\omega\tau} +\cc
\end{align}
In the last line, we used the fact that the TT-gauge coordinates~$(x^\mu)$ and the FNCs $(X^\alpha)$ are related by a gauge transformation; their difference is of the same order of magnitude as $H_{\mu\nu}$. In the end, the equation of motion of the particle in the freely-falling frame reads
\begin{empheq}[box=\fbox]{equation}\label{eq:eom_particle_GW}
\ddf{p^a}{\tau} = F^a + \frac{1}{2} m\omega^2 H^a_b X^b \ex{-\i\omega\tau} + \cc
\end{empheq}
where the second term is the tidal force~$F^a\e{GW}$ due to the GW.

\begin{figure}[h!]
\centering
\includegraphics[width=\columnwidth]{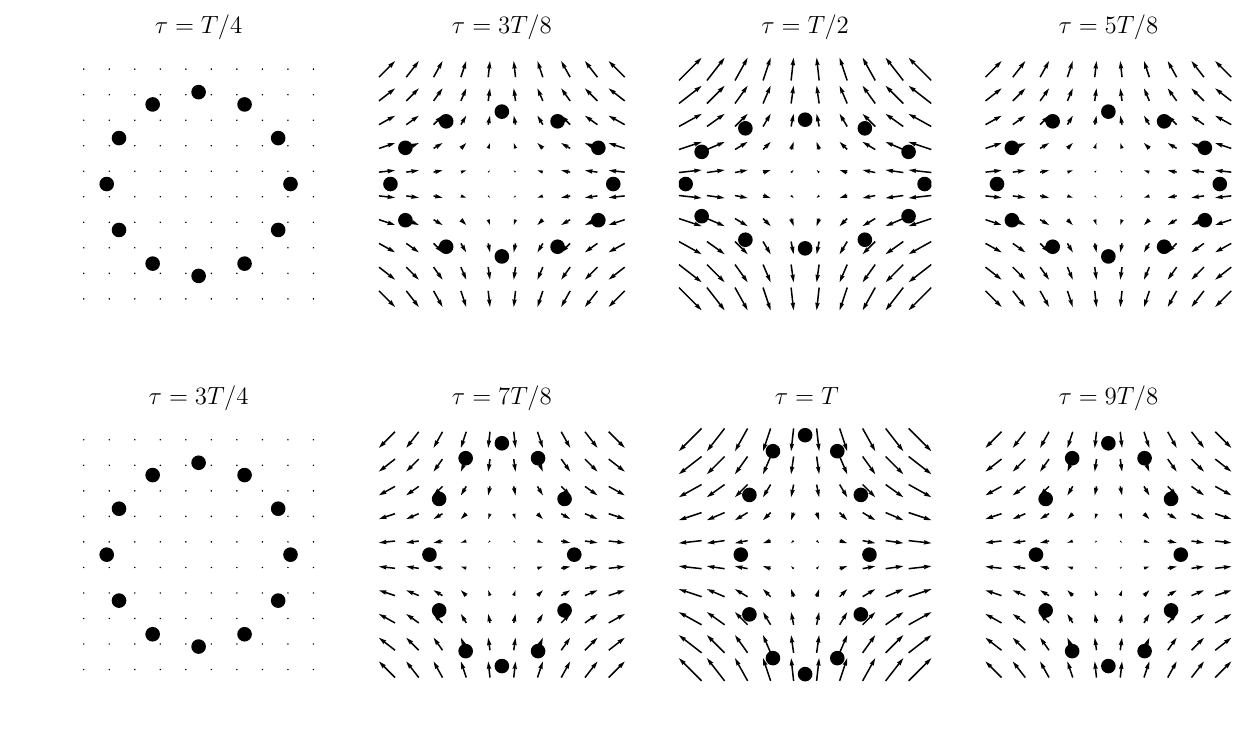}
\caption{Tidal forces, in the plane $OXY$, created by a GW with $H_+=0.3$, $H_\times=0$ and propagating along $Z$. Eight different steps of a period $T=2\pi/\omega$ are represented, as well as the effect of the GW on a ring of test particles, represented by black disks.}
\label{fig:GW_plus}
\end{figure}

\paragraph{Effect on matter} The impact of a GW on matter is more conveniently visualised if we consider the two polarisations $H_+, H_\times$ independently. Let us first suppose that $H_\times=0$. The tidal forces being orthogonal to $Z$, we can study what happens in the plane $Z=0$. Then, modulo a redefinition of the origin of time~$\tau$, we can assume that $H_+\in\mathbb{R}_+$, so that
\begin{align}
F^X\e{GW} &= m \omega^2 H_+ X \cos\omega\tau \ ,\\
F^Y\e{GW} &= -m \omega^2 H_+ Y \cos\omega\tau \ .
\end{align}
Figure~\ref{fig:GW_plus} represents this force field at different times~$\tau$. It also represents the effect of this force on a ring of test particles, i.e. particles subject to gravity only. Applying eq.~\eqref{eq:eom_particle_GW} for $F^a=0$, we find $a^b=m\omega^2 H^a_b X^b \cos\omega \tau$, that is to say
\begin{align}
\label{eq:eom_test_particle_GW_X}
\ddot{X} &= \omega^2 H_+ X\cos\omega\tau \ , \\
\label{eq:eom_test_particle_GW_Y}
\ddot{Y} &= -\omega^2 H_+ Y\cos\omega\tau \ ,
\end{align}
for each particle. If the amplitude of the GW is small, $H_+\ll 1$, which is the case in reality, then we can write $X^a(\tau)=X_0^a+\delta X^a(\tau)$, with $|\delta\vec{X}| \ll |\vec{X}_0|$. For particles at rest at $\tau=0$, and working at leading order in $\delta X^a$, eqs.~\eqref{eq:eom_test_particle_GW_X} and \eqref{eq:eom_test_particle_GW_Y} are integrated as
\begin{align}\label{eq:displacement_X_GW}
\delta X(\tau) &\approx -H_+ X_0 \cos\omega\tau \ ,\\
\delta Y(\tau) &\approx H_+ Y_0 \cos\omega\tau \ ,
\end{align}
which is what appears in fig.~\ref{fig:GW_plus}.

\begin{figure}[h!]
\centering
\includegraphics[width=\columnwidth]{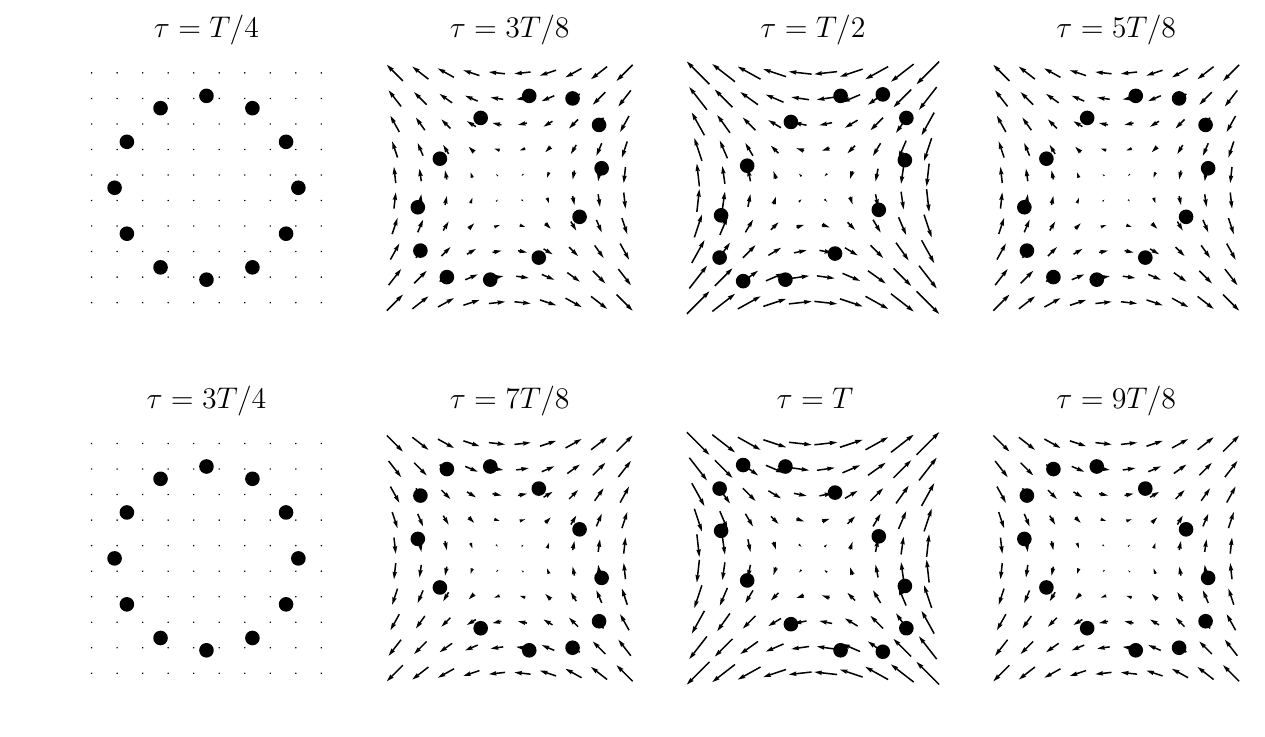}
\caption{Same as fig.~\ref{fig:GW_plus}, but with $H_\times=0.3$ and $H_+=0$.}
\label{fig:GW_cross}
\end{figure}

The case $(H_\times>0$, $H_+=0)$ is analysed similarly, and its effect on a ring of particles is depicted in fig.~\ref{fig:GW_cross}. Comparing figs.~\ref{fig:GW_plus} and \ref{fig:GW_cross}, it becomes pretty clear why the two polarisations are respectively denoted $H_+$, $H_\times$.

\begin{exercise}
Write a Python code generating a GIF animation representing the motion of a ring of particles under the effect of a GW, for any $H_+, H_\times\in \mathbb{C}$. The case $H_\times=\i H_+$ is called \emph{circular polarisation}; do you understand why?
\end{exercise}

\paragraph{Detection by interferometry}

The amplitude of GWs, even when due to spectacularly violent phenomena such as the collision of two black holes, is extremely small. For instance, the peak amplitude of the first event ever detected, called GW150914, was $|\fvect{h}|\sim 10^{-21}$. Following, e.g., eq.~\eqref{eq:displacement_X_GW}, this means that the associated displacement between two freely falling particles separated by a distance $X_0=1000\U{km}$ would be on the order of $\delta X \sim |\fvect{h}|X_0 \sim 10^{-15}\U{m}$, which is the size of an atomic nucleus.

\begin{figure}[h!]
\centering
\includegraphics[width=0.46\columnwidth]{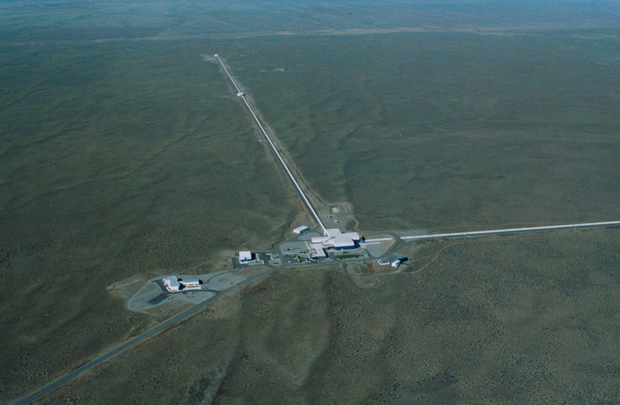}
\hfill
\includegraphics[width=0.53\columnwidth]{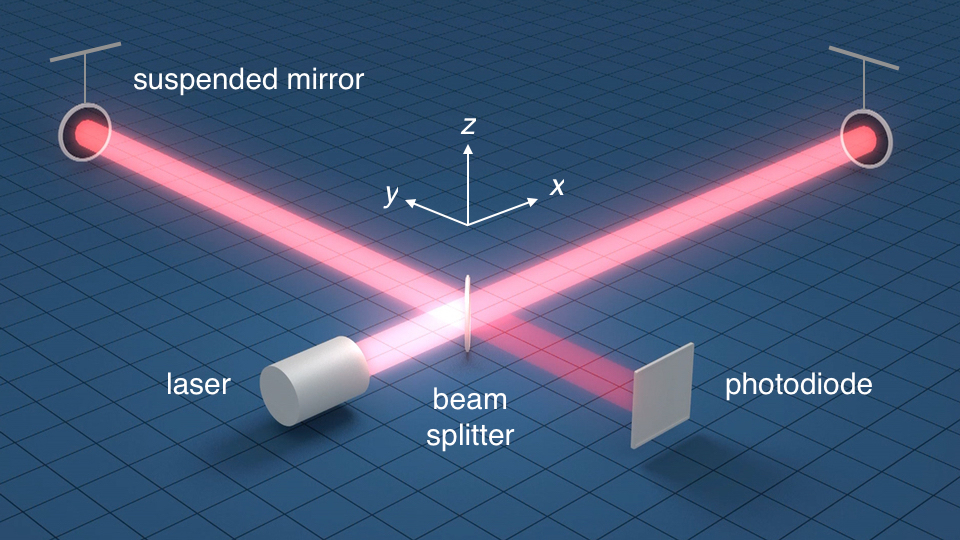}
\caption{\textit{Left panel}: LIGO, Hanford site (USA). The two arms of the interferometer are about four-kilometre long. \textit{Right panel}: Schematic view of the interferometer. A laser beam is split in two, each half-beam is reflected by a suspended mirror, both are recombined, and the resulting superposition is measured by a photo-diode. Adapted from \href{https://www.ligo.caltech.edu}{\tt https://www.ligo.caltech.edu}.}
\label{fig:LIGO}
\end{figure}

The only way to measure such a tiny displacement consists in exploiting luminous interferences. This is the method employed by the American \textit{Laser Interferometer Gravitational-wave Observatory} (LIGO, see fig.~\ref{fig:LIGO}), the European \textit{Virgo}, the near-future Japanese \textit{Kamioka Gravitational Wave Detector} (Kagra) or the Indian indIGO, and the future space mission \textit{Laser Interferometer Space Antenna} (LISA).

The general method is the following. A laser beam is split in two perpendicular directions, called the arms of the interferometer. Each half-beam is then reflected by a suspended mirror at the end of its arm, and the reflected half-beams are finally recombined. The interference between the beams is measured with a very sensitive photo-detector. Let us set the origin~$O$ of the reference frame at the beam splitter; when a GW passes through the interferometer, the associated tidal forces push or pull the suspended mirrors with respect to $O$, thereby increasing or reducing the effective length of each arm, which affects the interference pattern. This produces a very particular time-dependent signal measured by the photo-detector, which allows experimentalists to detect the GW.

\subsection{Production of gravitational waves}

Just like electromagnetic waves are produced by moving electric charges, GWs are produced by moving forms of energy. More precisely, GWs are produced whenever the quadrupolar moment of a distribution of energy evolves non-linearly with time. The goal of this last paragraph is to derive the so-called \emph{quadrupole formula} describing the production of GWs.

\paragraph{Post-Minkowskian expansion} We start again from the linearised Einstein's equation $\Box \gamma_{\mu\nu}=-16\pi G T_{\mu\nu}$, whose solution by the Green-function method yields
\begin{equation}
\gamma_{\mu\nu}(t,\vec{x})
= 4 G\int \frac{T_{\mu\nu}(t-||\vec{x}-\vec{y}||, \vec{y})}{||\vec{x}-\vec{y}||} \; \dd^3 y \ .
\end{equation}
Suppose that the above $T_{\mu\nu}$ is associated with matter that is well-localised in a small region~$\mathcal{R}$ of space, and that we are evaluating the metric at a distance $r$ much larger than that region. If the time-evolution of $T_{\mu\nu}$ is slow enough, then the retarded time $t-||\vec{x}-\vec{y}||$ is well approximated by $t-r$, and we have
\begin{equation}
\gamma_{\mu\nu}(t,\vec{x})
\approx \frac{4 G}{r} \int_{\mathcal{R}}
T_{\mu\nu}(t-r, \vec{y}) \; \dd^3 y \ ,
\end{equation}
that is,
\begin{align}
\gamma_{00} &= \frac{4G}{r} \int_{\mathcal{R}}
\rho \; \dd^3 y \ ,
\qquad \text{(gravitational potential)}\\
\gamma_{0a} &= \frac{4G}{r} \int_{\mathcal{R}}
\rho v_a \; \dd^3 y  \ ,
\qquad \text{(gravito-magnetism)}\\
\gamma_{ab} &= \frac{4G}{r} \int_{\mathcal{R}}
\rho v_a v_b\; \dd^3 y \ ,
\qquad \text{(gravitational waves)}\label{eq:GW_post_Minkowskian}
\end{align}
where $\rho$ is the matter energy density and $v^a$ its velocity field, modelled as a fluid. It is understood that the above integrands are evaluated at $(t-r,\vec{y})$. The idea consists in matching eq.~\eqref{eq:GW_post_Minkowskian} with the GW solution that we have investigated so far.

\paragraph{Quadrupole formula} At linear order in the metric perturbation,
\begin{equation}
0
= T\indices{^\mu^\nu_;_\nu}
= T\indices{^\mu^\nu_,_\nu} 
	+ \Gamma\indices{^\mu_\nu_\rho} T^{\rho\nu}
	+ \Gamma\indices{^\nu_\nu_\rho} T^{\mu\rho}
\approx T\indices{^\mu^\nu_,_\nu} \ ,
\end{equation}
that is
\begin{equation}\label{eq:energy-momentum_conservation_linear}
\partial_t T^{\mu 0} + \partial_a T^{\mu a} = 0 \ .
\end{equation}
Using the identity
$
(y^a T^{cb})_{,c}
= T^{ab} + y^a T\indices{^c^b_,_c},
$
we can rewrite the integral of eq.~\eqref{eq:GW_post_Minkowskian} as
\begin{align}
\int_{\mathcal{R}} T^{ab} \; \dd^3 y
&= \underbrace{\int_{\mathcal{R}}
							(y^a T^{cb})_{,c} \; \dd^3 y
						}_{0}
	 - \int_{\mathcal{R}} y^a T\indices{^c^b_,_c} \; \dd^3 y\\
&= - \frac{1}{2}
		\int_{\mathcal{R}}
			 \pa{ y^a T\indices{^c^b_,_c}
					+ y^b T\indices{^c^a_,_c} } 
				\dd^3 y
\qquad \text{because $T^{ab}$ is symmetric} \\
&= \frac{1}{2} \partial_t
		\int_{\mathcal{R}}
			 \pa{ y^a T\indices{^0^b}
					+ y^b T\indices{^0^a} } 
				\dd^3 y
\qquad \text{using \eqref{eq:energy-momentum_conservation_linear}.}
\end{align}
A similar operation, based on an integration by parts, can be performed a second time,
\begin{align}
\int_{\mathcal{R}}
	\pa{ y^a T\indices{^0^b}
			+ y^b T\indices{^0^a} } 
	\dd^3 y
&= \int_{\mathcal{R}}
	\pa{ y^a y^b T^{0c} }_{,c} \; \dd^3 y
		- \int_{\mathcal{R}}\pa{ y^a y^b  T^{0c} }_{,c} \; \dd^3 y \\
&= \partial_t \int_{\mathcal{R}} y^a y^b T^{00} \; \dd^3 y \ ,
\end{align}
so that finally
\begin{equation}\label{eq:gamma_ab_quadrupole}
\gamma^{ab}(t,\vec{x})
= \frac{4 G}{r} \int_{\mathcal{R}} T^{ab}(t-r,\vec{y}) \; \dd^3 y
= \frac{2 G}{r} \partial^2_t \int_{\mathcal{R}}
												 y^a y^b \rho(t-r,\vec{y}) \; \dd^3 y \ .
\end{equation}

After transforming eq.~\eqref{eq:gamma_ab_quadrupole} to the transverse trace-less gauge, we conclude that
\begin{empheq}[box=\fbox]{equation}\label{eq:quadrupole_formula}
h_{ab}\h{TT} = \frac{2 G}{3 r} \; P^{cd}_{ab} \ddot{Q}_{cd} \ ,
\end{empheq}
where $P^{cd}_{ab}$ is the projector orthogonally to the GW wave-vector, and
\begin{equation}
Q_{cd} = \int_{\mathcal{R}}
				(3 y^a y^b - \delta^{ab}\delta_{cd} y^c y^d) \rho
				\; \dd^3 y
\end{equation}
is the \emph{quadrupolar moment} of the energy distribution of matter within $\mathcal{R}$. Equation~\eqref{eq:quadrupole_formula} is known as the quadrupole formula\footnote{Although its result is correct, the standard derivation presented here is actually wrong. This is because the source of $h^{ij}$ is not only $T^{ij}$, but also the gravitational field itself, which has the same order of magnitude as $T^{ij}$. Hence, it is na\"{i}ve to calculate $h_{ij}$ by direct integration of $\Box \gamma_{ij}=-16\pi G T_{ij}$. I thank Guillaume Faye for letting me know about this issue. See ref.~\cite{Bonetti:2017hnb} for details.}. It shows that GWs can only be emitted by an accelerated quadrupole. As an anti-example, a  spherical mass distribution whose radius oscillates does not. However, a binary system of massive objects spiralling around each other has a non-zero $\ddot{\fvect{Q}}$, and hence emits GWs. Among the 11 GW events detected from 2015 to 2017, 10 were due to black hole mergers, and 1 to a neutron-star merger.

\section{The Schwarzschild black hole}

In the previous two sections, we have only explored some weak-field properties of the general theory of relativity. One could be curious about what happens when the metric strongly differs from Minkowski, and hence when the non-linearity of Einstein's equation starts to play an important role. \emph{Black holes} are an example of such strong gravitational field situations. In this lecture, we will focus on the simplest case, which is a single static, non-rotating, and non-electrically charged black hole.

\subsection{The Schwarzschild solution}

In January 1916, about one month after Einstein published his field equation, the German physicist Karl Schwarzschild found its very first exact solution~\cite{Schwarzschild:1916uq}, describing space-time surrounding a \emph{static} and \emph{spherically symmetric} massive object\footnote{Einstein himself seems to have been very surprised by this finding; he did not expect that one could actually find exact solutions to such a complicated equation. Not to mention that this happened during World War I, while Schwarzschild was serving in the German army.}.

\paragraph{Staticity} A space-time metric is said to be \emph{stationary} if there exists a coordinate system $(t,x^i)$ such that $\partial_t g_{\mu\nu}=0$,
\begin{equation}
\dd s^2
= g_{00}(x^k) \dd t^2 
+ 2g_{0i}(x^k) \dd t \dd x^i
+ g_{ij}(x^k) \dd x^i \dd x^j \ .
\end{equation}
It is said to be \emph{static} if, furthermore, it is invariant under the transformation $t\rightarrow -t$, which imposes $g_{0i}=0$. Hence,
\begin{equation}
\dd s^2
= g_{00}(x^k) \dd t^2 
+ g_{ij}(x^k) \dd x^i \dd x^j \ .
\end{equation}

\paragraph{Spherical symmetry} A metric is said to be \emph{spherically symmetric} if there exists a coordinate system $t, R, \theta, \ph$ such that, for $t=\cst$,
\begin{equation}
\dd s^2 = \dd R^2
+ g_{\theta\theta}(R) \pa{ \dd\theta^2 + \sin^2\theta \dd\ph^2 } .
\end{equation}
If we define $r=\sqrt{g_{\theta\theta}}$ as the new radial coordinate, then a static and spherically symmetric metric must read
\begin{equation}
\dd s^2 = g_{00}(r) \dd t^2 + g_{rr}(r) \dd r^2 + r^2 \pa{ \dd\theta^2 + \sin^2\theta \dd\ph^2 } .
\end{equation}
Since $g_{00}<0$ and $g_{rr}>0$, we can parametrise them as $g_{00}(r)=-\exp 2\nu(r)$ and $g_{rr}(r)=\exp2\lambda(r)$, where $\nu, \lambda$ are functions or $r$. The metric then reads
\begin{empheq}[box=\fbox]{equation}\label{eq:static_spherically-symmetric_metric}
\dd s^2 = -\ex{2\nu(r)}\dd t^2 + \ex{2\lambda(r)} \dd r^2
+ r^2 \pa{\dd\theta^2 + \sin^2\theta \dd\ph^2} .
\end{empheq}

\paragraph{Einstein's equation} We want to model, with a metric of the form~\eqref{eq:static_spherically-symmetric_metric}, the space-time geometry generated by a single massive body located at $r=0$, space being otherwise empty. In other words, $\forall r>0 \quad T_{\mu\nu}=0$, so that Einstein's equation is equivalent to $R_{\mu\nu}=0$ in that region.

\begin{exercise}
Show that the Ricci tensor of the metric~\eqref{eq:static_spherically-symmetric_metric} reads
\begin{align}
\label{eq:R_tt}
R_{tt}
&= \ex{2(\nu-\lambda)} \pac{ \nu'' + (\nu')^2 - \nu'\lambda' + \frac{2\nu'}{r} } ,\\
\label{eq:R_rr}
R_{rr}
&= -\nu'' - (\nu')^2 + \nu'\lambda' + \frac{2\lambda'}{r} \ ,\\
\label{eq:R_thth}
R_{\theta\theta}
&= 1+ \ex{-2\lambda} \pac{ r(\lambda'-\nu') - 1 } , \\
\label{eq:R_phph}
R_{\ph\ph}
&= R_{\theta\theta} \sin^2\theta \ ,
\end{align}
where a prime denotes a derivative with respect to $r$, and the off-diagonal terms are all zero. Such calculations can be performed by hand, or with the use of a computer algebra system, such as Mathematica, Maple (with the Tensor package), or SageMath (with SageManifolds).
\end{exercise}

Combining eqs.~\eqref{eq:R_tt} and \eqref{eq:R_rr}, we find
\begin{equation}
0 = \ex{-2(\nu-\lambda)} R_{tt} + R_{rr} = \frac{2}{r}(\nu'+\lambda') \ ,
\end{equation}
that is, $\nu(r)+\lambda(r)=C=\cst$. This constant can always be absorbed in a rescaling of the time coordinate, in the sense that
\begin{equation}
\ex{2\nu} \dd t^2
= \ex{-2\lambda} \pa{\ex{C}\dd t}^2
\rightarrow \ex{-2\lambda} \dd t^2
\end{equation}
under the transformation $t\rightarrow \ex{C}t$. Thus, we can consider without loss of generality that $C=0$ and $\lambda=-\nu$. Equation~\eqref{eq:R_thth} then becomes, in terms of $\nu(r)$ only,
\begin{equation}
1 = \ex{2\nu} \pa{2r\nu'+1} = \pa{ r\ex{2\nu} }',
\end{equation}
whence
\begin{equation}
\ex{2\nu} = 1 - \frac{r\e{S}}{r} \ ,
\end{equation}
where $r\e{S}$ is a constant to be determined. We have obtained the Schwarzschild metric
\begin{empheq}[box=\fbox]{equation}\label{eq:Schwarzschild_Droste}
\dd s^2 =
- \pa{ 1-\frac{r\e{S}}{r} } \dd t^2
+ \pa{ 1-\frac{r\e{S}}{r} }^{-1} \dd r^2
+ r^2 \pa{ \dd\theta^2 + \sin^2\theta \dd\ph^2} .
\end{empheq}
In fact, the above expression of the metric is the one that was independently derived the Dutch physicist Johannes Droste, later the same year 1916~\cite{1917KNAB...19..447D}. In his original article, Schwarzschild was using another coordinate system whose origin was located at $r=r\e{S}$, which made the results look much more complicated. Thus, eq.~\eqref{eq:Schwarzschild_Droste} should be referred to as the Schwarzschild metric in Droste coordinates.

It is customary to introduce the notation
\begin{equation}
A(r) \define 1 - \frac{r\e{S}}{r} \ ,
\qquad
\dd\Omega^2 \define \dd\theta^2 + \sin^2\theta \dd\ph^2 ,
\end{equation}
so that eq.~\eqref{eq:Schwarzschild_Droste} simply reads $\dd s^2 = -A(r) \dd t^2 + A^{-1}(r) \dd r^2 + r^2 \dd\Omega^2$.

\paragraph{Determining $\boldsymbol{r\e{S}}$} The quantity $r\e{S}$ is the only characteristic length scale of the problem. Far away from the massive body at $r=0$, i.e. for $r\gg r\e{S}$, we should recover the weak-field metric. In particular, we expect to find
\begin{equation}
g_{00}(r\gg r\e{S}) = -(1+2\Phi) ,
\end{equation}
where $\Phi=-GM/r$ is the Newtonian gravitational potential created by the massive object. We immediately identify
\begin{empheq}[box=\fbox]{equation}
r\e{S} = 2 GM,
\end{empheq}
where $M$ is the mass of the central body. If we were restoring the missing $c$ factors, this would become $r\e{S}=2GM/c^2$. This quantity is known as the \emph{Schwarzschild radius}.

\subsection{Geodesics}
\label{subsec:Schwarzschild_geodesics}

In order to explore the physics of the Schwarzschild geometry, it is useful to determine the trajectories of freely-falling particles, i.e. the geodesics of that space-time.

\paragraph{Geodesic equation and conserved quantities} The action producing the geodesic motion of massive and mass-less particles is proportional to
\begin{equation}\label{eq:action_geodesic}
s[x^\mu] = -\int\sqrt{|g_{\mu\nu} \dot{x}^\mu\dot{x}^\nu|} \; \dd\lambda \ ,
\end{equation}
with $\dot{x}^\mu\define \dd x^\mu/\dd\lambda$. If $\lambda$ is an affine parameter, then
\begin{equation}
g_{\mu\nu} \dot{x}^\mu\dot{x}^\nu
= \eps
\define
\begin{cases}
-1 & \text{for the time-like case ($\lambda=\tau$)}, \\
0 & \text{for the null case}.
\end{cases}
\end{equation}
In both cases, $\eps^2=\eps$, and hence we can remove the square-root of the integrand of eq.~\eqref{eq:action_geodesic}. In other words, the Lagrangian can be considered to be
\begin{align}
L
&= g_{\mu\nu} \dot{x}^\mu \dot{x}^\nu \\
&=-A(r) \, \dot{t}^2 + A^{-1}(r) \, \dot{r}^2
	+ r^2 \pa{ \dot{\theta}^2 + \sin^2\theta \dot{\ph}^2 }.
	\label{eq:Lagrangian_geodesics_Schwarzschild}
\end{align}

\begin{exercise}
Applying the Euler-Lagrange equation to the Lagrangian~\eqref{eq:Lagrangian_geodesics_Schwarzschild}, show that there exist two constants of motion $E,L$ such that
\begin{align}
A(r) \dot{t} &= E \ ,\\
\label{eq:theta_dot_Schwarzschild}
(r^2 \dot{\theta})\dot{}
&= r^2 \sin\theta \cos\theta \dot{\ph}^2 ,\\
\label{eq:phi_dot_Schwarzschild}
r^2 \sin^2\theta \dot{\ph} &= L \ .
\end{align}
These constants are related to the conservation of energy and angular momentum.
\end{exercise}

Combining eqs.~\eqref{eq:theta_dot_Schwarzschild} and \eqref{eq:phi_dot_Schwarzschild}, we find $(r^2 \dot{\theta})\dot{}=(L/r)^2 \cos\theta/\sin^3\theta$; multiplying this equation by $2r^2\dot{\theta}$ and integrating the result, we get
\begin{equation}
\pa{r^2 \dot{\theta}}^2 + \frac{L^2}{\sin^2\theta} = \cst.
\end{equation}
If we set the coordinate system such that, initially, $\theta=\pi/2$, $\dot{\theta}=0$, then the constant is $L^2$, and we conclude that $(r^2\dot{\theta})^2 + (L/\tan\theta)^2=0$. When the sum of two positive quantities vanishes, both quantities must be zero, so $\theta=\pi/2$ for the whole trajectory. This is analogous to the Keplerian problem of \S~\ref{subsec:orbits_planets}. Without any loss of generality, we can thus consider $\theta=\pi/2$ from now on. The full set of equations describing geodesic motion in the Schwarzschild space-time is, therefore,
\begin{empheq}[box=\fbox]{align}
A(r) \dot{t} &= E \\
\theta &= \pi/2 \\
r^2\dot{\ph} &= L \\
\label{eq:geodesic_Schwarzschild_r}
\frac{1}{A(r)} \pa{ \dot{r}^2 - E^2 }& + \frac{L^2}{r^2} 
= \eps \ .
\end{empheq}

\begin{figure}[h!]
\centering
\includegraphics[width=0.75\columnwidth]{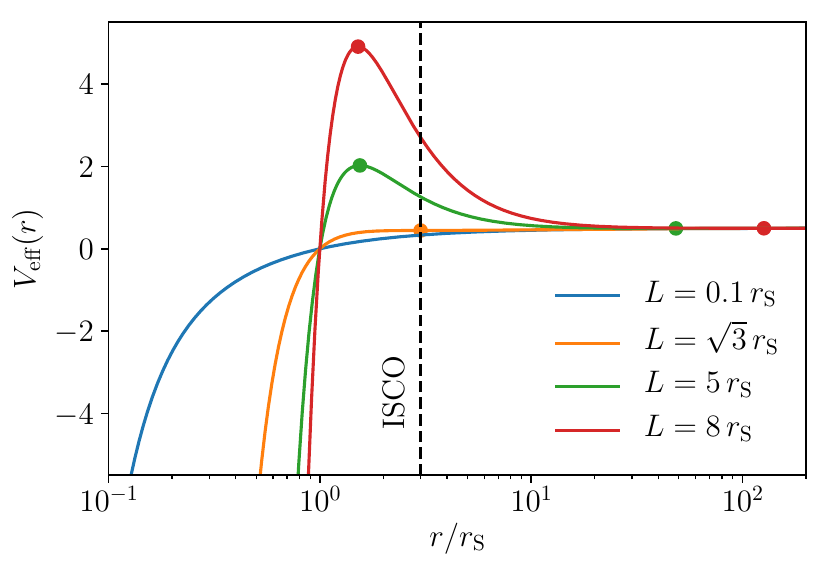}
\caption{Effective potential~$V\e{eff}(r)$ for massive particles ($\eps=-1$) and different values of $L$. The positions of circular orbits, when they exist, are indicated with disks. For $L>\sqrt{3}r\e{S}$, there exist one stable and one unstable orbit. They merge into the ISCO for $L=\sqrt{3}r\e{S}$.}
\label{fig:circular_orbits_Schwarzschild}
\end{figure}

\paragraph{Circular orbits} The equation of motion~\eqref{eq:geodesic_Schwarzschild_r} for $r$ can be rewritten
\begin{equation}
\frac{\dot{r}^2}{2} + V\e{eff}(r) = \frac{E^2}{2} \ ,
\qquad \text{with} \quad
V\e{eff}(r) \define
\frac{A(r)}{2} \pac{ \pa{\frac{L}{r}}^2 - \eps }
\end{equation}
playing the role of an effective potential. Circular orbits ($r=\cst$) are possible if $V'\e{eff}=0$. They are stable if $V\e{eff}''>0$. The effective potential is illustrated in fig.~\ref{fig:circular_orbits_Schwarzschild}.

\begin{exercise}
Show that the radius~$r$ of any circular orbit satisfies
\begin{equation}\label{eq:circular_radius}
-\eps GM r^2 - L^2 r + 3GM L^2 = 0 \ .
\end{equation}
\end{exercise}
For photons ($\eps=0$), eq.~\eqref{eq:circular_radius} is linear, thus it admits a single solution $r=3GM$. At that distance, the gravitational field of the central massive body is strong enough to allow light to orbit around it. However, this orbit in unstable: $V''(3GM)=-L^2/(3GM)^4<0$, hence it cannot be observed in reality.

For massive particles ($\eps=-1$), eq.~\eqref{eq:circular_radius} is quadratic, with discriminant $\Delta=L^2(L^2-3r\e{S}^2)$. There are three possibilities:
\begin{enumerate}
\item If $L^2>3r\e{S}^2$, eq.~\eqref{eq:circular_radius} has two solutions
\begin{equation}
r_{\pm}
= \frac{L}{r\e{S}} \pa{ L \pm \sqrt{L^2-3r\e{S}^2} } ,
\end{equation}
corresponding to one stable ($r_+$) and one unstable ($r_-$) orbit. For $L\gg r\e{S}$, the stable orbit $r_+\approx 2L^2/r\e{S}$ corresponds to the Newtonian limit, while $r_-\approx 3GM$ is an unstable relativistic orbit.
\item If $L^2=3r\e{S}^2$, the two solutions~$r_\pm$ merge into $r\e{ISCO}=6GM$, known as the \emph{innermost stable circular orbit} (ISCO).
\item If $L^2<3r\e{S}^2$, there is no circular orbit: the particle does not have enough angular momentum to keep away from the central massive object, and spirals towards the centre $r=0$. This is a strictly relativistic prediction; Newtonian gravitation does not have such a feature.
\end{enumerate}

\paragraph{Radial free fall} If $L=0$, then $\dot{\ph}=0$, which corresponds to a radial free fall. For photons, the equation of motion is simply~$\dot{r}^2=E^2$. For massive particles, it reads
\begin{equation}\label{eq:radial_fall_Newton-like}
\frac{1}{2} \dot{r}^2 - \frac{GM}{r} = \frac{E^2-1}{2} \ ,
\end{equation}
which is exactly the same as its Newtonian counterpart, if $(E^2-1)/2$ is interpreted as the total energy of the particle per unit mass.

It is important to notice that eq.~\eqref{eq:radial_fall_Newton-like} involves $\dot{r}\define \dd r/\dd\tau$, but $\tau$ is not really the time that an exterior observer, watching the particle fall, would use. Consider a static observer in a space station very far from the central mass ($r\e{obs}\gg r\e{S}$). The proper time of such an observer is then $\dd\tau\e{obs}=\sqrt{A(r\e{obs})} \dd t \approx \dd t$ since $A(r\e{obs})\approx 1$. If this observer watches a particle fall towards the central mass, then she sees a trajectory $r(t)$ such that
\begin{equation}
\ddf{r}{t} = \frac{\dot{r}}{\dot{t}} = A(r) \sqrt{1 - \frac{A(r)}{E^2}} \rightarrow 0 \quad \text{for }r \rightarrow r\e{S}.
\end{equation}
Hence, the particle will appear to slow down as it approaches the sphere $r=r\e{S}$, and the observer never actually sees it crossing its surface. This is an extreme illustration of the \emph{gravitational dilation of time} discussed in \S~\ref{subsec:gravitational_dilation_time}.

\begin{exercise}\label{ex:time_to-singularity}
Consider a particle starting a radial free fall at $r_0>r\e{S}$ with no initial velocity ($\dot{r}=0$). Determine the time~$\tau$ that the particle takes to reach $r=0$ as measured in its own frame. Is it finite or infinite?
\end{exercise}

\subsection{Event horizon and black hole}

\paragraph{Singularity at $\boldsymbol{r\e{S}}$?} A quick look at the expression~\eqref{eq:Schwarzschild_Droste} of the Schwarzschild metric suffices to notice that something wrong happens for $r=r\e{S}$. The infinite dilation of time mentioned above is one of its manifestations. When, in 1922, Einstein presented the Schwarzschild solution\footnote{Schwarzschild did not have the chance to participate to the lively debate provoked by his solution, because he died in May 1916.} at the Collège de France (Paris), he was obviously aware of that problem. At that time, many mathematicians and physicists considered it as a proof that Einstein's theory could not be correct. On the other hand, several alternative coordinate systems were proposed by Painlevé, Gullstrand, Eddington, Finkelstein, Lemaître, Robertson, Synge, Kruskal, Szekeres, and Novikov, for which the metric appears to be well-behaved for $r=r\e{S}$. It took about 40 years for this debate to be closed, and definitely understand that the apparent singularity at $r=r\e{S}$ was actually a feature of the Droste coordinates. An observer radially falling towards $r=0$ does not experience anything particular when reaching $r=r\e{S}$. However, when this surface is crossed, one can never come back to the region $r>r\e{S}$, as we will see in a few paragraphs.

\begin{exercise}
Show that the \emph{Kretschmann scalar}, defined as $K \define R^{\mu\nu\rho\sigma}R_{\mu\nu\rho\sigma}$ reads
\begin{equation}
K = \frac{12 r\e{S}^2}{r^6}
\end{equation}
for the Schwarzschild metric. Conclude that there is no curvature singularity at $r=r\e{S}$,  but that there is one at $r=0$.
\end{exercise}

\paragraph{Kruskal-Szekeres coordinates} The detailed structure of the Schwarzschild space-time can be explored using the \emph{Kruskal-Szekeres} coordinate system $(T,R,\theta,\ph)$~\cite{Kruskal:1959vx, Szekeres:1960gm}. We leave the two angular coordinate unchanged, and define new time and radial coordinates
\begin{align}
T &\define
\sqrt{\abs{\frac{r}{r\e{S}}-1}}
\,\exp\pa{\frac{r}{2r\e{S}}}\,
\sinh\pa{\frac{t}{2 r\e{S}}} ,
\label{eq:Kruskal-Szekeres_T}\\
R &\define
\sqrt{\abs{\frac{r}{r\e{S}}-1}}\,
\exp\pa{\frac{r}{2r\e{S}}}\,
\cosh\pa{\frac{t}{2 r\e{S}}} ;
\label{eq:Kruskal-Szekeres_R}
\end{align}
these imply, in particular,
\begin{align}
\pa{\frac{r}{r\e{S}} - 1} \exp\pa{\frac{r}{r\e{S}}}
&= R^2-T^2 ,\\
\tanh\pa{ \frac{t}{2 r\e{S}} } &= \frac{T}{R} \ .
\end{align}

\begin{exercise}
Show that the Schwarzschild metric in Kruskal-Szekeres coordinates reads
\begin{equation}
\dd s^2 = \frac{4 r\e{S}^3}{r} \, \ex{-r/r\e{S}}
				\pa{ -\dd T^2 + \dd R^2 } + r^2 \dd\Omega^2,
\end{equation}
where it is understood that $r=r(T,R)$, implicitly defined by eqs.~\eqref{eq:Kruskal-Szekeres_T} and \eqref{eq:Kruskal-Szekeres_R}. Conclude that the metric is indeed regular at $r=r\e{S}$.
\end{exercise}

\begin{figure}[h!]
\centering
\input{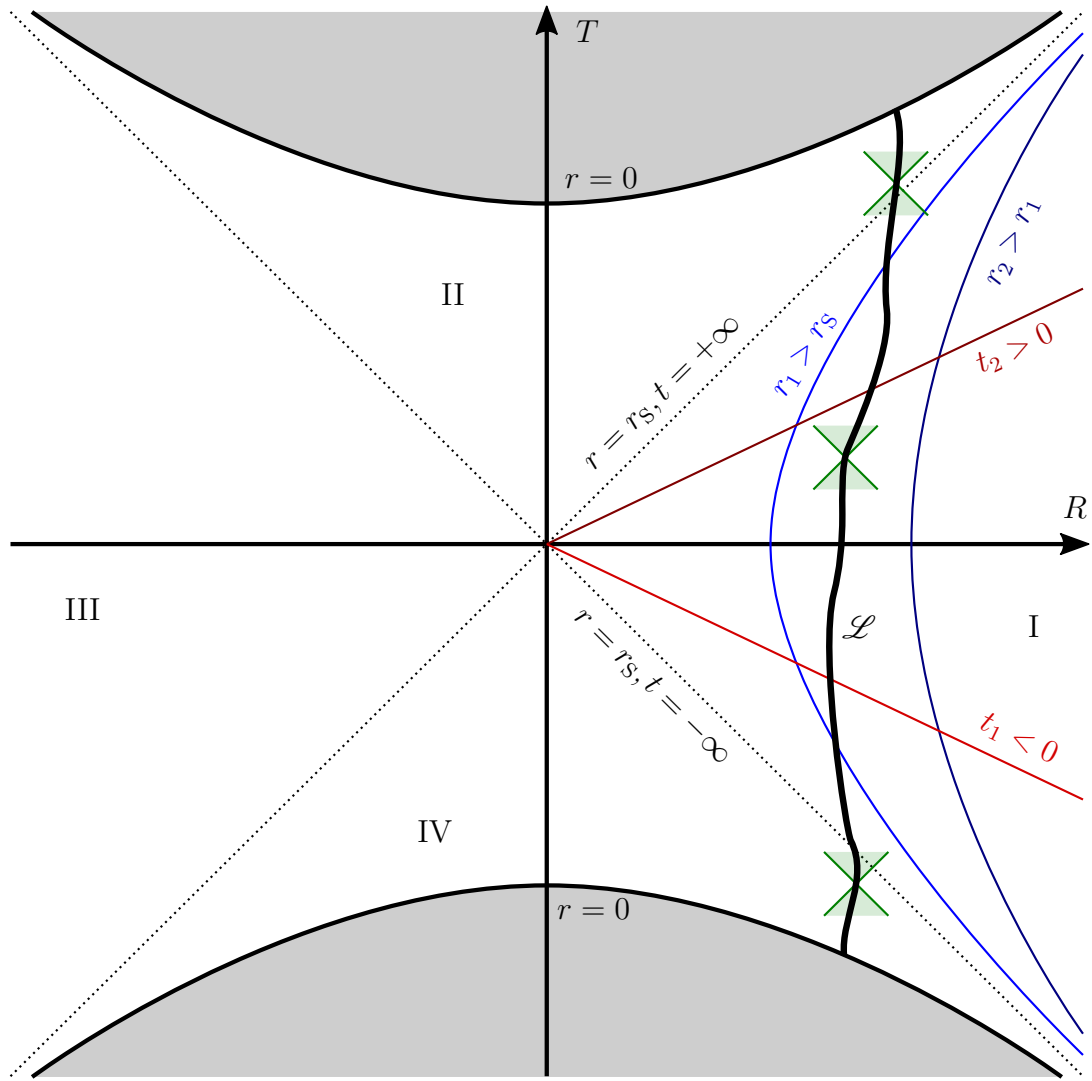}
\caption{Kruskal diagram of the Schwarschild space-time. The axes $T,R$ indicate Kruskal-Szekeres coordinates. The two gray regions are excluded, their contour indicating the central singularity $r=0$. Dotted lines represent the event horizon of the black hole, and split the diagram into four regions: exterior (I), black interior (II), parallel exterior (III), and white interior (IV). The thick black curve is the world-line of a particle emitted and reabsorbed by the black hole, along which three local light-cones are indicated in green. Blue lines represent $r=\cst$ world-lines, while red lines represent $t=\cst$ hyper-surfaces.}
\label{fig:Kruskal_diagram}
\end{figure}

An important feature of Kruskal-Szekeres coordinates is that they trivialise radial null geodesics. Indeed, radial null curves ($\dd s^2=0$ with $\dd\Omega^2=0$) are simply given by
\begin{equation}
\dd T = \pm \dd R \ .
\end{equation}
Due to the spherical symmetry of the Schwarzschild space-time, these are also geodesics, so that radial light rays are simply straight lines in the $(T,R)$ plane. Table~\ref{tab:Droste-Kruskal} draws a correspondence between the Droste and Kruskal-Szekeres coordinates for various locations. The full structure of the Schwarzschild space-time can then be represented in the \emph{Kruskal diagram} (fig.~\ref{fig:Kruskal_diagram}), which consists of the plane $(T,R)$.

\begin{table}[h!]
\centering
\begin{tabular}{rcc}
Location & Droste & Kruskal-Szekeres \\ 
\hline 
static particle & $r=\cst$ & $R^2-T^2=\cst$ \\ 
horizon & $r=r\e{S}$ & $R^2-T^2=0 \implies t=\pm\infty$ \\ 
singularity & $r=0$ & $R^2-T^2=-1$ \\ 
spatial slice & $t=\cst$ & $T=R\times\cst$ \\ 
\end{tabular}
\caption{Correspondence between Droste and Kruskal-Szekeres coordinates for various elements of the Schwarzschild space-time.}
\label{tab:Droste-Kruskal}
\end{table}

\paragraph{Event horizon} We are now ready to understand why the Schwarzschild space-time describes a \emph{black hole}. Let us focus on the regions labelled I and II in the Kruskal diagram. Region I is the part that is well described by the Droste coordinates $(t,r)$; it represents the \emph{exterior of the black hole}, $r>r\e{S}$. In this region, particles can be accelerated so as to maintain $r=\cst$, because the associated hyperbolas are time-like curves. This region is not fundamentally different from the exterior of any massive body.

Now consider a particle following the time-like curve $\mathscr{L}$ upwards. In the upper part, the particle moves towards the centre $r=0$. When the particle crosses the line $T=R$ ($r=r\e{S}$), it enters region II, which is the \emph{interior of the black hole}. From that point, we see that its causal future can only lead to the singularity at $r=0$. The particle cannot get out of region II, nor send any message to the exterior, because region I is now entirely space-like for the particle. This is why this region is a black hole: nothing can get out of it, not even light. No information can ever propagate from the interior (II) to the exterior (I).

The surface $r=r\e{S}$ is called the \emph{event horizon} of the black hole. Note that, in terms of the time coordinate $t$, the particle never actually reaches the horizon, because of the extreme time dilation mentioned at the end of \S~\ref{subsec:Schwarzschild_geodesics}. It is not the case from the point of view of the particle itself (see exercise~\ref{ex:time_to-singularity}).

\paragraph{White hole and parallel Universe} The other two regions of the Schwarzschild space-time (III and IV) could not have been revealed without the Kruskal-Szekeres coordinate system. Region IV is the \emph{interior of a white hole}: contrary to the interior of the black hole, the causal future of any particle in that region lies at the exterior ($r>r\e{S}$, region I). Taken as a whole, $\mathscr{L}$ depicts the entire world-line of a particle emitted from the interior, which is then re-absorbed by the black hole.

Region III is even more intriguing. It represents another exterior for the white/black hole (with $R<0$) which is causally disconnected from region I. It is sometimes coined as a \emph{parallel Universe}, which people in region I cannot interact with.

\paragraph{Diving into a black hole?} This is not precisely a good idea. Any observer crossing the horizon of a sufficiently large\footnote{The following reasoning only applies if the Schwarzschild radius~$r\e{S}$ is larger than the observer's body. If not, it can still chop a part of his body.} black hole is bound to reach the singularity in a finite amount of time. At $r=0$, curvature diverges, hence the observer gets radially stretched by very intense tidal forces. Technically speaking, this process is known as \emph{spaghettification}.

\subsection{Black holes in nature}

Whenever a certain amount of matter collapses under the effect of gravity, if nothing prevents this collapse, then the final state is a black hole. Specifically, if some matter distribution~$M$ is concentrated in a sphere whose radius is smaller than $r\e{S}=2GM/c^2$, then it is a black hole. A good order of magnitude to keep in mind is the Schwarzschild radius of the Sun, $r\e{S}=2GM_\odot/c^2=3\U{km}$. It means that if the whole mass of the Sun were concentrated in a ball with a radius of $3\U{km}$, then it would be a black hole. For comparison, the Sun's actual radius is $R_\odot=7\times 10^8\U{m}$.

Black holes are sometimes pictured as scary objects that absorb everything in their neighbourhood. It is not really the case. Although nothing can escape from the interior region of a black hole, it is not that easy to enter this region at all, because its cross-section ($\sim r\e{S}^2$) is generally very small. Any object moving towards a black hole with an impact parameter larger than a few $r\e{S}$ would actually orbit around it, just like the planets of the Solar system orbit around the Sun.

\begin{wrapfigure}{r}{0.3\columnwidth}
\centering
\includegraphics[width=0.3\columnwidth]{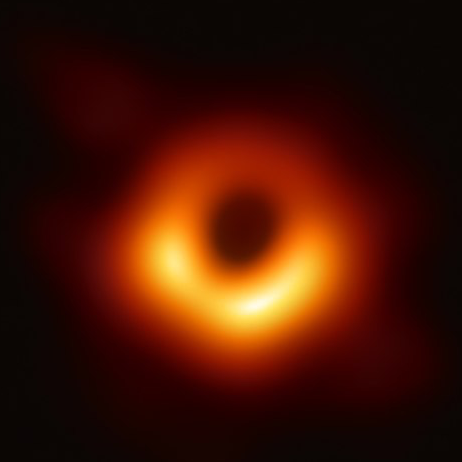}
\caption{Image of M87* taken by the Event Horizon Telescope. The orange halo is the accretion disk, seen in radio frequencies. The central black region is the \emph{shadow} of the black hole, whose radius is approximately $2.6\times r\e{S}$.}
\label{fig:M87}
\end{wrapfigure}

We believe nowadays that most galaxies have a \emph{super-massive black hole} at their centre, although their origin is not yet fully understood. In our own Milky Way resides Sagittarius~A\textsuperscript{*} (SgrA\textsuperscript{*}), a relatively quiet super-massive black hole with mass $M\approx 4.3\times 10^6 M_\odot$. Its Schwarzschild radius thus approaches 12 million kilometres, which is approximately 30 times the distance between the Earth and the Moon. Another, now famous, example, is the super-massive black hole at the centre of the Messier 87 (M87) galaxy, a super-giant elliptical galaxy located more than 50 billion light-years away from us. M87\textsuperscript{*} was indeed the first black hole ever \emph{directly} observed by a telescope (see fig.~\ref{fig:M87}), the Event Horizon Telescope~\cite{Akiyama:2019cqa}.

In other galaxies, the central black hole is less quiet. Black holes are usually surrounded by an accretion disk: a disk of very hot gas, part of which is progressively absorbed by the black hole. When accretion is very rapid, the extreme temperature reached in the disk makes it extremely bright; so bright that these objects were initially thought to be stars of our own galaxy, while they can actually be a billion time further away. This confusion led astronomers to call such galaxies-with-a-greedy-black-hole \emph{quasars} (for quasi-stars), or quasi-stellar objects (QSO).

Besides super-massive black holes, there is a range of masses for other black holes in nature. Pretty common ones are the so-called \emph{stellar black holes}, which are the final product of stellar evolution for very massive stars. There is currently a fascinating debate about the origin of the black hole mergers that produced the GWs observed by the LIGO/Virgo collaboration. These black holes, with masses of a few to tens of solar masses, are more massive than what most stellar models tend to predict. More speculatively, they could be \emph{primordial black holes}, formed at the very early stages of our Universe from the collapse of very dense regions, mostly made of light. Shall they actually exist, these primordial black holes could represent a part of the mysterious dark matter.

%
%
%

\chapter*{Conclusion}
\addstarredchapter{Conclusion}

\lettrine{A}{s} we have now reached the end of our journey, let me emphasise that it was far from being comprehensive. There would be so much more to say about the general-relativistic world, and more formally about the theory of relativity itself. I wish that we could have covered the key experimental tests of relativity on Earth and in the Solar system, such as the Pound \& Rebka experiment, the relativistic precession of Mercury's perihelion, or the Shapiro time delay of light propagation. I wish that we could have explored gravito-magnetic phenomena, such as the precession of gyroscope in the gravity field of the Earth, or the spinning Kerr black holes. Not to mention the role of relativistic gravitation in astrophysics, such as in neutron stars and pulsars, or in the physics of the early Universe. Fortunately, all these topics are covered in several excellent textbooks, such as the references of these notes.

Is there any reason to try to go beyond GR? From the strictly experimental perspective, not really, for Einstein's theory is one of the most successfully tested in physics. The weak equivalence principle is confirmed with a precision of a part in $10^{14}$, while post-Newtonian parameters agree with the predictions of GR with an accuracy of a part in at least $10^4$, up to $10^{20}$~\cite{Will:2014kxa}. With the detection, in 2017, of the combined gravitational-wave and gamma-ray signals emitted by a binary neutron-star merger~\cite{TheLIGOScientific:2017qsa, Goldstein:2017mmi}, the speed of gravitational information was found to match the speed of light, as predicted by GR, with a precision of $10^{-15}$. We can also mention the 2018 analysis of the orbit of stars about the Sgr~A\textsuperscript{*} super-massive black hole, in excellent agreement with GR~\cite{refId0}. Einstein's theory therefore successfully passed the numerous and diverse tests to which it was submitted.

Some may argue that modern cosmology hints towards gravitational phenomena beyond GR. This idea stems from the two great mysteries of dark matter and dark energy. Dark matter, on the one hand, is the name given to the mass apparently missing from all structures in the Universe, from galaxies to galaxy clusters and the large-scale cosmic web. There is an overwhelming amount of observational evidence for such an anomaly, which is nonetheless very well modelled by a new form of matter, only interacting with itself and normal matter via gravitation. Could it be that there is no such dark matter, and that we just misunderstand how gravity works on astronomical scales? While such a scenario is possible, the most recent results of that field of research do not favour it. It turns out to be extremely difficult to build a model of gravitation which would explain the whole dark-matter phenomenology~\cite{2011IJMPD..20.2749D}. Therefore, dark matter seems to regard particle physics rather than gravitation. Dark energy, on the other hand, is the unknown phenomenon causing the current acceleration of cosmic expansion, which is extremely puzzling in a Universe where gravity is attractive! As mentioned in Chapter~\ref{chap:relativity}, adding a cosmological constant~$\Lambda$ to Einstein's equations easily solves that issue. Of course, one may propose more involved extensions of GR, like the so-called self-accelerating models. However, most of these models predict that gravitational and electromagnetic waves have different propagation speed~\cite{Lombriser:2016yzn}, which is now excluded.

In fact, the only compelling reasons to go beyond GR are strictly theoretical. The first one is the vacuum-energy problem.\footnote{In the scientific literature, this issue is better known as the ``cosmological-constant problem''. I do not particularly cherish this denomination, because it tends to generate confusion with the dark-energy issue in cosmology. These are distinct questions.} You have learned in this course that all forms of energy gravitates; hence this should also include the energy of quantum vacuum, which would behave as a cosmological constant~$\Lambda\e{vac}$. The problem is that a na\"{i}ve estimate of $\Lambda\e{vac}$ exceeds the measured value of $\Lambda$ by many orders of magnitude. Thus, there must exist an unknown (but probably quantum) mechanism cancelling that huge gravitational effect of vacuum. This leads us to our second big theoretical question: can gravitation be quantized? In the standard model of particle physics, the electromagnetic, weak, and strong interactions are intrinsically quantum. Why would gravitation be an exception indeed? Various programmes were undertaken to address this question, such as super-gravity and string theory with a field-theoretic approach, or loop quantum gravity with a canonical approach. So far, none of them resulted into a complete theory of quantum gravity, nor did they produce any falsifiable prediction to novel experimental tests.

Could it be that we have taken the wrong path? Could it be that gravitation cannot be quantized? There is something truly remarkable with the development of gravitation: it has always been sitting by its own, somehow disconnected from the rest of physics. All the other physical entities and concepts have experienced, at some point of their history, unification or fragmentation. For example, the attraction between magnets merged with lightning bolts, visible light, and radio waves within Maxwell's electromagnetism, which then merged with radioactivity into the electroweak fundamental interaction. On the contrary, all materials were progressively understood to have molecular and atomic substructure. Macroscopic concepts like temperature and pressure were understood to result from the statistical and collective effect of that substructure. Despite their name, atoms were decomposed into electrons and nucleons, themselves made of quarks held together by the strong interaction. Who knows where the division ends?

Nothing truly comparable ever happened to gravitation. Admittedly, it merged with the fabric of space and time, but never with any another physical phenomenon. At the end of his life, Einstein desperately tried and failed to unify electromagnetism and gravitation into a unique theory. What if, contrary to electromagnetism, weak, and strong interactions, gravitation should not be merged with something else, but rather fragmented just like matter? What if gravitation was not fundamental but rather emerged from a microscopic substructure, just like temperature and pressure? As eccentric as it may seem, this idea of an emergent gravity is backed by curious coincidences, such as the fact that the Einstein equation can be reformulated in thermodynamical terms~\cite{Jacobson:1995ab}, which led to a modest but continuous research effort during the last couple of decades.

Perhaps a new Einstein will arise to solve all these questions at once. Or perhaps will we need many Bohr, De Broglie, Dirac, Feynman, Heisenberg, Pauli, and Sch\"{o}dinger to work together and solve them separately. Perhaps, this time, they will be Africans and Asians rather than Europeans and Americans. Or perhaps they will be artificial intelligences. Perhaps the human civilisation will not manage to overcome its more immediate challenges  before solving these fascinating puzzles. I hope it will, and that you will be on board.

\phantomsection
\addstarredchapter{References}
\bibliography{bibliography_AIMS}
\bibliographystyle{JHEP}

\end{document}